\documentclass[nofootinbib,preprint, prX]{revtex4-1} 

\usepackage{amsmath}
\usepackage{mathrsfs}
\usepackage{amssymb}		
\usepackage{graphicx}		
\usepackage{graphics}
\usepackage[small,compact]{titlesec}
\usepackage{xspace}
\usepackage{natbib}
\usepackage{units}
\usepackage[justification=raggedright]{caption}
\usepackage{subfig}

\newcommand{\vect}[1]{\vec{\mathbf{#1}}}
\newcommand{\vectS}[1]{\vec{\boldsymbol{#1}}}

\newcommand{\propagator}[2]{\frac{i}{#1-\frac{\vect{#2}^{2}}{2M_{N}}+i\epsilon}}

\newcommand{\EFT}{$\mathrm{EFT}_{\not{\pi}}$\xspace}
\newcommand{\SixJ}[6]{\left\{\begin{array}{ccc} #1 & #2 & #3\\ #4 & #5 & #6 \end{array}\right\}}
\newcommand{\CG}[6]{C_{#1,#3,#5}^{#2,#4,#6}}
\newcommand{\srutTypeOne}[1]{\vrule width0pt height0pt depth #1\relax}
\newcommand{\comment}[1]{}

\begin{document}

\title{Fully Perturbative Calculation of $nd$ Scattering to Next-to-next-to-leading-order}

\author{Jared Vanasse}
\email{jjv9@phy.duke.edu}
\affiliation{\emph{Department of Physics,\\
Duke University,\\
Durham, NC 27708}
}

\date{\today}

\begin{abstract}
We introduce a new technique to calculate perturbative corrections to neutron-deuteron ($nd$) scattering that does not require calculation of the full off-shell scattering amplitude. Its relation to the more familiar partial-resummation technique is explained.  Also included is a calculation of the SD-mixing term that occurs at next-to-next-to-leading-order (NNLO) in pionless effective field theory (\EFT).  Using the new technique with the SD-mixing term a complete strictly perturbative phase-shift analysis of $nd$ scattering is performed up to NNLO including eigenphases and mixing angles. This is compared to potential model calculations and good agreement is found with the eigenphases and some of the mixing angles at low energies.
\end{abstract}

\keywords{latex-community, revtex4, aps, papers}

\maketitle

\section{Introduction}
	At sufficiently low energies ($E\lesssim m_{\pi}^{2}/M_{N}$), where pions can be integrated out, nuclear physics can be described by pionless effective field theory(\EFT) in which only nucleon degrees of freedom as well as possible external currents appear. (For a review of \EFT see \cite{Beane:2000fx}.)  This approach has been used quite
	successfully to calculate nucleon-nucleon ($NN$) scattering \cite{Chen:1999tn,Ando:2004mm,Ando:2007fh,Kong:1999sf}, electromagnetic form factors of the deuteron, and the neutron proton capture process \cite{Chen:1999bg,Rupak:1999rk,Ando:2005cz} in the two-body sector.  It has also been used to study $NN$ parity-violation \cite{Schindler:2009wd,Phillips:2008hn,Shin:2009hi} and neutrino-deuteron processes \cite{Kong:2000px,Butler:2000zp,Ando:2008va,Chen:2012hm}.  In the three-body sector calculations have been carried out for the bound state properties of Helium-3 \cite{Ando:2010wq}, parity-violation in neutron-deuteron ($nd$) interactions \cite{Griesshammer:2011md,Vanasse:2011nd}, and $nd$ scattering \cite{Bedaque:1998mb,Bedaque:1999ve,Gabbiani:1999yv,Bedaque:2002yg,Griesshammer:2004pe} as well as proton deuteron ($pd$) scattering \cite{Rupak:2001ci,Konig:2011yq} in which
	Coulomb forces were treated perturbatively.  However, in the three-body case for $nd$ scattering no strictly perturbative calculation has been carried out to NNLO since it apparently requires the full off-shell leading-order (LO) scattering
	amplitude, which is numerically expensive.  In order to calculate $nd$ and $pd$ scattering to higher orders the partially resummed approach \cite{Gabbiani:1999yv} has been used in which certain classes of diagrams are summed to all orders.  The calculation contains all the correct diagrams to
	the order one is working, but also contains certain higher order diagrams.  This method leads to good results for $nd$ scattering in both the effective range expansion (ERE) parametrization and Z parametrization \cite{Phillips:1999hh,Gabbiani:1999yv}.  However, certain problems occur in the quartet S-wave channel above the deuteron breakup threshold (DBT), namely the imaginary part of the phase shift at next-to-leading-order (NLO) and NNLO is negative and hence unphysical.  A possible way around this is to resum to all orders the effective range in the deuteron
	propagator and then use this deuteron propagator to calculate the NNLO quartet S-wave scattering amplitude \cite{Gabbiani:1999yv}.  This approach gives physical results for the phase shifts but introduces spurious poles in the deuteron propagator.  These poles occur for momenta greater than the cutoff of \EFT.  However, they will still introduce issues with the numerical solution \cite{Griesshammer:2004pe}.  A perturbative calculation using short range effective field theory (SREFT) for a trimer of Helium-4 atoms has been carried out by Ji and Phillips \cite{Ji:2012nj}.  They calculate the full off-shell leading-order scattering amplitude
	and use it to numerically solve all necessary Feynman diagrams up to and including NNLO.  This calculation is similar to $nd$ scattering in the doublet S-wave channel.  In this paper we propose a new method to calculate to arbitrary order
	perturbatively without the need to calculate the full off-shell LO scattering amplitude.  The method puts contributions into the inhomogeneous part of the integral equation that depend on amplitudes that are lower order than the order at which one is working.  At each order the integral equation has the same kernel but a different inhomogenous term.  The paper is organized as follows.  In section II we introduce the two-body Lagrangian and then in section III we calculate $nd$ scattering in the quartet channel to NNLO and introduce the new technique.  Section IV extends this to the doublet channel and in section V the SD-mixing terms for $nd$ scattering occurring at NNLO are calculated.  In section VI we describe our phase shift analysis conventions.  In section VII we analyze the results and make conclusions in section VIII.
\section{Two-Body Lagrangian}

\begin{figure}[hbt]
\includegraphics[width=140mm]{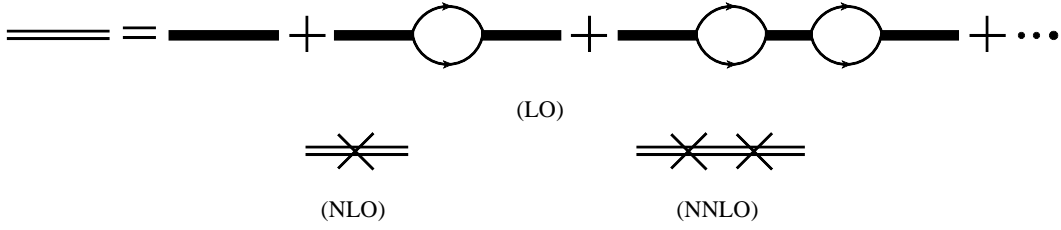}









\caption{\label{fig:LODeuteronPropagator} At LO the bare deuteron propagator $i/\Delta^{({}^{3}\!S_{1})}_{(-1)}$ is dressed by an infinite number of nucleon bubbles to give the LO dressed deuteron propagator.  At NLO the dressed deuteron propagator receives one effective range correction which comes from the deuteron kinetic term and the NLO correction $\Delta^{({}^{3}\!S_{1})}_{(0)}$.  Finally at NNLO the dressed deuteron propagator receives two such effective range corrections}
\end{figure}

The two-body Lagrangian in the auxiliary field formalism is given by 

\begin{align}
\label{eq:Lagrangian}
\mathcal{L}^{d}=\ &\hat{N}^{\dagger}\left(i\partial_{0}+\frac{\vect{\nabla}^2}{2M_{N}}\right)\hat{N}-\hat{t}_{i}^{\dagger}\left(i\partial_{0}+\frac{\vect{\nabla}^{2}}{4M_{N}}-\Delta^{({}^3\!S_{1})}_{(-1)}-\Delta^{({}^3\!S_{1})}_{(0)}\right)\hat{t}_{i}+y_{t}\left[\hat{t}_{i}^{\dagger}\hat{N} ^{T}P_{i}\hat{N} +h.c.\right]\\\nonumber
&-\hat{s}_{a}^{\dagger}\left(i\partial_{0}+\frac{\vect{\nabla}^{2}}{4M_{N}}-\Delta^{({}^1\!S_{0})}_{(-1)}-\Delta^{({}^1\!S_{0})}_{(0)}\right)\hat{s}_{a}+y_{s}\left[\hat{s}_{a}^{\dagger}\hat{N}^{T}\bar{P}_{a}\hat{N}+h.c.\right],
\end{align}

\noindent where $\hat{t}_{i}$ ($\hat{s}_{a}$) is the spin-triplet (spin-singlet) dibaryon field.  The auxiliary field Lagrangian can be shown to be equivalent to a Lagrangian containing just nucleon fields by integrating out the auxiliary fields and performing a field redefinition \cite{Bedaque:1999vb}.  The projector $P_{i}=\frac{1}{\sqrt{8}}\sigma_{2}\sigma_{i}\tau_{2}$ ($\bar{P}_{a}=\frac{1}{\sqrt{8}}\tau_{2}\tau_{a}\sigma_{2}$) projects out the spin-triplet iso-singlet (spin-singlet iso-triplet) combination of nuclei.  The term $\Delta^{({}^{3}\!S_{1})}_{(-1)}$ is sub-leading compared to $\Delta^{({}^{3}\!S_{1})}_{(0)}$.  Thus the LO bare deuteron propagator is simply given by $i/\Delta^{({}^{3}\!S_{1})}_{(-1)}$ and at LO is dressed by an infinite number of nucleon bubbles as in Fig \ref{fig:LODeuteronPropagator}.  This bubble sum can by solved explicitly by means of a geometric series yielding 

\begin{equation}
\label{eq:BubbleSum}
iD_{t}^{LO}(p_{0},\vect{p})=-\frac{4\pi i}{M_{N}y_{t}^2}\frac{1}{\frac{4\pi\Delta^{({}^3\!S_{1})}_{(-1)}}{M_{N}y_{t}^{2}}-\mu+\sqrt{\frac{\vect{p}^{2}}{4}-M_{N}p_{0}-i\epsilon}},
\end{equation}

\noindent where the $\mu$ dependence is obtained by using dimensional regularization with the power divergence subtraction scheme (PDS) \cite{Kaplan:1998tg} in which poles occurring in three dimensions are added back to our expressions.  The coefficients are then chosen such that the deuteron pole is in the right location, giving  

\begin{equation}
\label{eq:LOfit}
\frac{\Delta_{(-1)}^{{}^{3}\!S_{1}}}{y_{t}^{2}}=\frac{M_{N}}{4\pi}(\mu-\gamma_{t}),
\end{equation}
\vspace{.2cm}

\noindent where $\gamma_{t}=45.7025$ MeV is the deuteron binding momentum.  At NLO the deuteron propagator receives a correction from the deuteron kinetic term and the NLO correction $\Delta^{({}^{3}\!S_{1})}_{(0)}$, and at NNLO it receives two such corrections as shown in Fig.\ref{fig:LODeuteronPropagator}.  Thus the full deuteron propagator up to and including NNLO is given by 

\begin{align}
\label{eq:NNLOdeuteronprop}
iD_{t}^{NNLO}(p_{0},\vect{p})=iD_{t}^{LO}(p_{0},\vect{p})&\left(1+D_{t}^{LO}(p_{0},\vect{p})(\Delta_{(0)}^{({}^{3}\!S_{1})}+p_{0}-\frac{\vect{p}^{2}}{4M_{N}})\right.\\\nonumber
&\left.+(D_{t}^{LO}(p_{0},\vect{p}))^{2}(\Delta_{(0)}^{({}^{3}\!S_{1})}+p_{0}-\frac{\vect{p}^{2}}{4M_{N}})^{2}\right).
\end{align}

There exists two different approaches by which to fit the parameters $\Delta^{({}^{3}\!S_{1})}_{(-1)}$, $\Delta^{({}^{3}\!S_{1})}_{(0)}$, and $y_{t}$.  In the effective range expansion (ERE) parametrization one insures that the deuteron pole is given correctly and that the deuteron pole residue $Z_{t}=1/(1-\gamma_{t}\rho_{t})$ is given correctly in an expansion of the effective range about the deuteron pole, $\rho_{t}=1.65$ fm.  The other approach, termed the Z-parametrization \cite{Phillips:1999hh}, also fits to the deuteron pole, but fits to the deuteron pole residue exactly at NLO.  The deuteron propagator in both parametrizations has been used extensively throughout the literature and we merely quote the results for the deuteron propagator in the Z-parametrization \cite{Griesshammer:2004pe}, which is given by 

\begin{align}
\label{eq:DtpropNNLO}
&iD_{t}^{NNLO}(p_{0},\vect{p})=\frac{4\pi i}{M_{N}y_{t}^{2}}\frac{1}{\gamma_{t}-\sqrt{\frac{\vect{p}^{2}}{4}-M_{N}p_{0}-i\epsilon}}\\\nonumber
&\times\left[\underbrace{\srutTypeOne{.5cm} 1}_{\mathrm{LO}}+\underbrace{\frac{Z_{t}-1}{2\gamma_{t}}\left(\gamma_{t}+\sqrt{\frac{\vect{p}^{2}}{4}-M_{N}p_{0}-i\epsilon}\right)}_{\mathrm{NLO}}+\underbrace{\left(\frac{Z_{t}-1}{2\gamma_{t}}\right)^{2}\left(\frac{\vect{p}^{2}}{4}-M_{N}p_{0}-\gamma_{t}^{2}\right)}_{\mathrm{NNLO}}+\cdots\right],
\end{align}

\noindent and the resulting constraints on the coefficients which are

\begin{equation}
\label{eq:DtpropConstraints}
\frac{1}{y_{t}^{2}}=\frac{M_{N}^{2}}{8\pi\gamma_{t}}\frac{Z_{t}-1}{1+(Z_{t}-1)},\quad \Delta_{(-1)}^{({}^{3}\!S_{1})}=\frac{2\gamma_{t}Z_{t}}{M_{N}}\frac{\mu-\gamma_{t}}{Z_{t}-1},\quad\Delta_{(0)}^{({}^{3}\!S_{1})} =\frac{\gamma_{t}^{2}}{M_{N}}.
\end{equation}

The form of the ${}^{1}\!S_{0}$ dibaryon propagator is obtained analogously to that of the deuteron propagator.  However, in the ERE parametrization one expands about zero momentum, since there is no physical bound state pole.  Then one fits to obtain the correct scattering length.  At higher orders in the ERE one fits to the effective range.  For the Z-parametrization one insures that the virtual-bound state pole at LO in the ${}^{1}\!S_{0}$ channel is reproduced correctly.  At higher orders the correct residue about the virtual-bound state pole is reproduced and the pole is not changed.  Since there is no physical data for the virtual-bound state pole one must infer its location from scattering data, and doing so one finds the virtual-bound state binding momentum $\gamma_{s}=-7.8902$ MeV, and the residue about the pole $Z_{s}=.9015$ \cite{Griesshammer:2004pe}.  The ${}^{1}\!S_{0}$ dibaryon propagator in the Z-parametrization is given by

\begin{align}
\label{eq:DspropNNLO}
&iD_{s}^{NNLO}(p_{0},\vect{p})=\frac{4\pi i}{M_{N}y_{s}^{2}}\frac{1}{\gamma_{s}-\sqrt{\frac{\vect{p}^{2}}{4}-M_{N}p_{0}-i\epsilon}}\\\nonumber
&\times\left[\underbrace{\srutTypeOne{.5 cm}1}_{\mathrm{LO}}+\underbrace{\frac{Z_{s}-1}{2\gamma_{s}}\left(\gamma_{s}+\sqrt{\frac{\vect{p}^{2}}{4}-M_{N}p_{0}-i\epsilon}\right)}_{\mathrm{NLO}}+\underbrace{\left(\frac{Z_{s}-1}{2\gamma_{s}}\right)^{2}\left(\frac{\vect{p}^{2}}{4}-M_{N}p_{0}-\gamma_{s}^{2}\right)}_{\mathrm{NNLO}}+\cdots\right],
\end{align}

\noindent and the resulting constraints on the coefficients are 

\begin{equation}
\label{eq:DspropConstraints}
\frac{1}{y_{s}^{2}}=\frac{M_{N}^{2}}{8\pi\gamma_{s}}\frac{Z_{s}-1}{1+(Z_{s}-1)},\Delta_{(-1)}^{({}^{1}\!S_{0})}=\frac{2\gamma_{s}Z_{s}}{M_{N}}\frac{\mu-\gamma_{s}}{Z_{s}-1},\quad\Delta_{(0)}^{({}^{1}\!S_{0})} =\frac{\gamma_{s}^{2}}{M_{N}}.
\end{equation}

For the sake of later convenience we adopt the notation $D^{(n)}_{t}(p_{0},\vect{p})$ where $n=0,1,2,$... refers to LO, NLO, NNLO, and higher orders respectively.  Thus $D^{(0)}_{t}(p_{0},\vect{p})$ is the LO deuteron propagator and $D^{(2)}_{t}(p_{0},\vect{p})$ only picks out the NNLO piece of the deuteron propagator, the part in the brackets of Eq. (\ref{eq:DtpropNNLO}) that is under-braced with NNLO times the piece outside of the brackets.  An analogous notation is also used for the ${}^{1}\!S_{0}$ dibaryon propagator.

Finally we note that the deuteron wavefunction renormalization is given by the residue of the deuteron propagator about the deuteron pole  

\begin{equation}
\label{eq:ZD}
Z_{D}=\frac{8\pi\gamma_{t}}{M_{N}^{2}y_{t}^{2}}\left[\underbrace{\srutTypeOne{.1cm}1}_{\mathrm{LO}}+\underbrace{(Z_{t}-1)}_{\mathrm{NLO}}+\underbrace{\srutTypeOne{.1cm}0}_{\mathrm{NNLO}}+\cdots+\underbrace{\srutTypeOne{.1cm}0}_{\mathrm{N}^{n}\mathrm{LO}}+\cdots\right].
\end{equation}

\noindent In the Z-parametrization the deuteron pole residue is given exactly at NLO and therefore there are no NNLO or higher order corrections to the deuteron wavefunction renormalization. For later convenience we will denote $Z_{LO}$, $Z_{NLO}$, and $Z_{NNLO}$ as the part of the deuteron wavefunction renormalization occurring at each order in \EFT.  In particular $Z_{LO}=(8\pi\gamma_{t})/(M_{N}^{2}y_{t}^{2})$ , $Z_{NLO}=Z_{LO}(Z_{t}-1)$, and $Z_{NNLO}=0$.

\section{Quartet channel}
\begin{figure}[hbt]

\includegraphics[width=100mm]{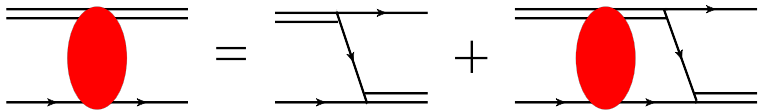}





\caption{\label{fig:LO_Quartet}(Color online)Integral equation for quartet channel at LO}
\end{figure}

In order to calculate $nd$ scattering in the quartet channel at LO in \EFT one must sum an infinite number of diagrams.  However, unlike the two-body case the diagrams do not factorize and cannot be solved analytically by means of a geometric series, rather the infinite sum of diagrams can be solved numerically by means of an integral equation represented in Fig. \ref{fig:LO_Quartet} and originally given by Skornyakov and Ter-Martirosian \cite{Skornyakov}.  The integral equation for the half off-shell scattering amplitude represented in Fig \ref{fig:LO_Quartet} is given by Eq. (\ref{eq:QuartetLOIntegralEq}) \cite{Gabbiani:1999yv}. 

\begin{align}
\label{eq:QuartetLOIntegralEq}
\left(it^{ji}\right)^{\beta b}_{\alpha
a}(\vect{k},\vect{p},h)=&\frac{y_{t}^{2}}{2}(\sigma^{i}\sigma^{j})^{\beta}_{\alpha}\delta^{b}_{a}\frac{i}{-\frac{\vect{k}^{2}}{4M_{N}}-\frac{\gamma_{t}^{2}}{M_{N}}+h-\frac{(\vect{k}+\vect{p})^{2}}{2M_{N}}+i\epsilon}+\\\nonumber
&+\frac{y_{t}^{2}}{2}(\sigma^{i}\sigma^{k})^{\beta}_{\gamma}\delta^{b}_{c}\int\!\frac{d^{4}q}{(2\pi)^{4}}(it^{jk})^{\gamma c}_{\alpha a}(\vect{k},\vect{q},h+q_{0})\\\nonumber
&\times iD_{t}^{(0)}\left(\frac{\vect{k}^{2}}{4M_{N}}-\frac{\gamma_{t}^{2}}{M_{N}}+h+q_{0},\vect{q}\right)\frac{i}{\frac{\vect{k}^{2}}{2M_{N}}-h-q_{0}-\frac{\vect{q}^{2}}{2M_{N}}+i\epsilon}\\\nonumber
&\times\frac{i}{-\frac{\vect{k}^{2}}{4M_{N}}-\frac{\gamma_{t}^{2}}{M_{N}}+2h+q_{0}-\frac{(\vect{q}+\vect{p})^ {2}}{2M_{N}}+i\epsilon}
\end{align}

\noindent where i (j) is the initial (final) deuteron polarization, $\alpha$ ($\beta$) the initial (final) nucleon spin, and $a$ ($b$) the initial (final) nucleon isospin.  The incoming momentum in the center of mass (c.m.) frame is $\vect{k}$ and the outgoing momentum in the c.m. frame is $\vect{p}$.  Finally the parameter $h=(\vect{k}^{2}-\vect{p}^{2})/2M_{N}$ measures the off-shellness of the outgoing particles.  Note that when $|\vect{p}|=|\vect{k}|$ the amplitude is full on shell and $h=0$. 

We now simplify Eq. (\ref{eq:QuartetLOIntegralEq}) by integrating over the energy and picking up a simple pole in the complex energy plane.  Then we project the spin in the quartet channel, set $h=(\vect{k}^{2}-\vect{p}^{2})/2M_{N}$,  and perform a partial wave decomposition of the amplitude yielding \cite{Gabbiani:1999yv}.

\begin{align}
\label{eq:QuartetLO}
t^{l}_{q}(k,p)=&-\frac{y_{t}^{2}M_{N}}{pk}Q_{l}\left(\frac{p^{2}+k^{2}-M_{N}E-i\epsilon}{pk}\right)\\\nonumber
&-\frac{2}{\pi}\int_{0}^{\Lambda}dq q^{2}t^{l}_{q}(k,q)\frac{1}{\gamma_{t}-\sqrt{\frac{3\vect{q}^{2}}{4}-M_{N}E-i\epsilon}}\frac{1}{qp}Q_{l}\left(\frac{p^{2}+q^{2}-M_{N}E-i\epsilon}{pq}\right),
\end{align}

\noindent where 

\begin{equation}
t^{l}_{q}(k,p)=2\int d\theta_{\hat{\mathbf{k}}\cdot\hat{\mathbf{p}}}P_{l}(\theta_{\hat{\mathbf{k}}\cdot\hat{\mathbf{p}}})(t^{ji})^{11}_{11}\left(\vect{k},\vect{p},\frac{\vect{k}^{2}-\vect{p}^{2}}{2M_{N}}\right)\delta_{(i,-(1+i2)/\sqrt{2})}\delta_{(j,-(1-i2)/\sqrt{2})},
\end{equation}

\noindent and

\begin{equation}
\label{eq:LegnedgreQ}
Q_{l}(a)=\frac{1}{2}\int_{-1}^{1}\!dx\frac{P_{l}(x)}{x+a},
\end{equation}

\noindent is equal to the Legendre polynomials of the second kind up to a factor of $(-1)^{l}$.  The energy appearing in Eq. (\ref{eq:QuartetLO}) is defined as $E=\frac{3\vect{k}^{2}}{4M_{N}}-\frac{\gamma_{t}^{2}}{M_{N}}$ (Note the subscript $q$ on the amplitude $t$ refers to the quartet channel.)  The partial wave decomposition changes the integral equation from three dimensions to one, but there is now an infinite set of integral equations.  However, at low energies only a few partial waves are needed to obtain sufficient convergence and for our purposes we calculate up to and including G-waves.  To solve the integral equations numerically we employ the Hetherington-Schick method \cite{Hetherington:1965zza,Brayshaw:1969ab,Ziegelmann}, in which one solves the integral equation along a contour rotated into the complex plane thereby avoiding the logarithmic singularities and the simple pole from the deuteron propagator.  This is valid as long as no singularities exist within the contour and this is satisfied for our equation.  Finally one can use the integral equation and the solution of the amplitude along the contour to solve for the amplitude along the real axis again provided there are no singularities within the resulting contour \cite{Aaron:1966zz}.

We note that in Eq. (\ref{eq:QuartetLO}) the integral ranges from 0 to $\Lambda$ instead of 0 to infinity.  The use of a cutoff for the integral serves two purposes.  First, it gives a cutoff regularization for potential divergences.  Second, since we will have to solve this integral equation numerically we have to impose some cutoff anyway since we cannot integrate to infinity.  One may be concerned that we are using two different regularization schemes here.  We are using dimensional regularization for the deuteron propagator and a cutoff regularization for the integral equation.  However, any effects from choosing different regularization schemes are of higher order \cite{Bedaque:1999ve}.  

\begin{figure}[hbt]

\includegraphics[width=40mm]{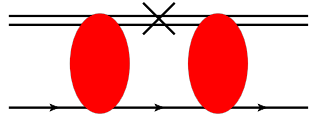}

\caption{\label{fig:NLOdiagram}(Color online)NLO diagram in quartet channel}
\end{figure}

\begin{figure}[hbt]

\includegraphics[width=100mm]{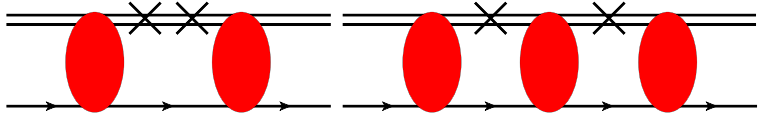}






\caption{\label{fig:NNLOdiagrams}(Color online)NNLO diagrams in quartet channel }
\end{figure}

At NLO the scattering amplitude is given by the diagram in Fig. \ref{fig:NLOdiagram} where a single insertion of the effective range correction is put between two half off-shell LO scattering amplitudes.  This diagram can be calculated by numerically integrating with the half off-shell LO scattering amplitude and has been done in Refs. \cite{Hammer:2001gh,Gabbiani:1999yv}.  At NNLO there are two diagrams one must calculate in Fig. \ref{fig:NNLOdiagrams}.  The first diagram simply sandwiches two effective range corrections between two half off-shell LO scattering amplitudes and can again be solved straightforwardly numerically.  However, the second diagram has an insertion of a full off-shell LO scattering amplitude and in principle cannot be solved without calculating the full off-shell LO scattering amplitude.  In order to circumvent the need to calculate the full off-shell LO scattering amplitude the partial-resummation technique has been employed by Bedaque et al. \cite{Gabbiani:1999yv}, which gives the correct diagrams up to NNLO but includes certain higher order diagrams and therefore is not strictly perturbative.  A fully perturbative calculation has been carried out by Ji and Phillips \cite{Ji:2012nj} by effectively calculating the full off-shell scattering amplitude.  However, their calculation was not for $nd$ scattering and was only for S-wave.  Now we will introduce a new approach to calculate the $nd$ scattering amplitudes to NNLO strictly perturbatively that avoids calculating the full off-shell LO scattering amplitude.


The full $nd$ scattering amplitude is in general given by 

\begin{align}
\label{eq:primary}
&t^{l}_{0,q}(k,p)+t^{l}_{1,q}(k,p)+t^{l}_{2,q}(k,p)=B^{l}_{0}(k,p)+B^{l}_{1}(k,p)+B^{l}_{2}(k,p)\\\nonumber
&\hspace{1cm}+(K^{l}_{0}(q,p,E)+K^{l}_{1}(q,p,E)+K^{l}_{2}(q,p,E))\otimes(t^{l}_{0,q}(q,k)+t^{l}_{1,q}(q,k)+t^{l}_{2,q}(q,k)),
\end{align}

\noindent where $t^{l}_{n,q}(k,p)$ is the scattering amplitude, $B^{l}_{n}(k,p)$ the inhomogeneous term, and $K^{l}_{n}(q,p,E)$ the kernel of the integral equation.  The subscripts $0$, $1$ and $2$ refer to LO, NLO, and NNLO parts respectively.  The symbol $\otimes$ represents an integration and is defined by 

\begin{equation}
\label{eq:otimes}
A(q)\otimes B(q)=\frac{2}{\pi}\int_{0}^{\Lambda}dqq^{2}A(q)B(q).
\end{equation}

\noindent The inhomogeneous part and kernel of the integral equation to each order in \EFT are defined by

\begin{equation}
\label{eq:inhom}
B^{l}_{0}(k,p)=-\frac{y_{t}^{2}M_{N}}{pk}Q_{l}\left(\frac{p^{2}+k^{2}-M_{N}E-i\epsilon}{pk}\right), B^{l}_{1}(k,p)=0, B^{l}_{2}(k,p)=0
\end{equation}

\begin{equation}
\label{eq:hom}
K^{l}_{n}(q,p,E)=\frac{M_{N}y_{t}^{2}}{4\pi}D_{t}^{(n)}\left(E-\frac{\vect{q}^{2}}{2M_{N}},\vect{q}\right)\frac{1}{qp}Q_{l}\left(\frac{q^{2}+p^{2}-M_{N}E-i\epsilon}{pq}\right)
\end{equation}

At NLO in the partial-resummation technique one keeps the inhomogeneous part, kernel, and amplitude up to and including NLO.  Denoting $t^{l}_{NLO}=t^{l}_{0,q}+t^{l}_{1,q}$  (Note for ease of discussion we have omitted the momentum dependence of all functions in the text) one obtains 

\begin{equation}
\label{eq:partialresum}
t^{l}_{NLO}(k,p)=B^{l}_{0}(k,p)+B^{l}_{1}(k,p)+(K^{l}_{0}(q,p,E)+K^{l}_{1}(q,p,E))\otimes t^{l}_{NLO}(k,q),
\end{equation}

\noindent for the NLO amplitude in the partial-resummation technique.  This technique again is not strictly perturbative  since the $K^{l}_{1}\otimes t^{l}_{1,q}$ term is NNLO.  Thus in order to make the calculation strictly perturbative one simply throws out the term $K^{l}_{1}\otimes t^{l}_{1,q}$ and obtains 

\begin{equation}
\label{eq:NLOamp}
t^{l}_{1,q}(k,p)=B^{l}_{1}(k,p)+K^{l}_{1}(q,p,E)\otimes t^{l}_{0,q}(k,q)+K^{l}_{0}(q,p,E)\otimes t^{l}_{1,q}(k,q),
\end{equation}

\noindent for the NLO piece of the scattering amplitude.  In this equation the term $K^{l}_{1}\otimes t^{l}_{0,q}$ is simply absorbed into the inhomogeneous part of an integral equation for $t^{l}_{1,q}$ since the amplitude $t^{l}_{0,q}$ is already calculated at LO.  The resulting kernel for the integral equation is $K^{l}_{0,q}$, the same as the kernel for the LO amplitude.  Now collecting all of the NNLO terms in Eq. (\ref{eq:primary}) one obtains 

\begin{equation}
\label{eq:NNLOamp}
t^{l}_{2,q}(k,p)=B^{l}_{2}(k,p)+K^{l}_{2}(q,p,E)\otimes t^{l}_{0,q}(k,q)+K^{l}_{1}(q,p,E)\otimes t^{l}_{1,q}(k,q)+K^{l}_{0}(q,p,E)\otimes t^{l}_{2,q}(k,q),
\end{equation}

\noindent for the NNLO piece of the scattering amplitude.  In this equation $K^{l}_{2}\otimes t^{l}_{0,q}$ and $K^{l}_{1}\otimes t^{l}_{1,q}$ are absorbed into the inhomogeneous part of an integral equation for $t^{l}_{2,q}$, since $t^{l}_{0.q}$ is already calculated at LO and $t^{l}_{1,q}$ is calculated in Eq (\ref{eq:NLOamp}).  Again the kernel for the integral equation of $t^{l}_{2,q}$ is given by $K^{l}_{0,q}$.  Thus at each order the kernel of the integral equation is always the same but the inhomogeneous part of the integral equation changes at each order and depends on lower order amplitudes.  Therefore, in order to calculate the NLO and NNLO amplitudes numerically one can simply use the kernel calculated for the LO scattering amplitude as it is now the kernel for all orders.  The Hetherington-Schick method can also be used to calculate the NLO and NNLO amplitudes.  Since $t^{l}_{0,q}$ and $K^{l}_{1,q}$ have no singularities in the contour used to perform the integration, the term $K^{l}_{1,q}\otimes t^{l}_{0,q}$ introduces no new singularities inside the contour.  Therefore, no new singularities are introduced in the contour for the $t^{l}_{1,q}$ scattering amplitude, and the same analysis holds at NNLO.

\section{Doublet Channel}
In the doublet channel two extra complications arise.  One comes from the fact that we now need to include the spin singlet dibaryon propagator, which gives two coupled integral equations.  The second arises from three-body forces, which only occur in the doublet S-wave channel starting at LO because there is no centrifugal barrier and the Pauli exclusion principle does not exclude a bound state in this channel.  Since the resulting kernel of the integral equation in the doublet S-wave channel is non-compact, the equation does not possess a unique solution (in the limit $\Lambda\to\infty$).  At finite cutoff this results in the solution varying greatly with the choice of cutoff.  This can be remedied by the insertion of a three-body force with an appropriate scale dependence on $\Lambda$.  The basic form of the three-body force up to NNLO is given by \cite{Bedaque:1998km,Bedaque:1999ve,Bedaque:2002yg}

\begin{equation}
{\mathcal{H}}(E,\Lambda)=\frac{2H_{0}^{LO}(\Lambda)}{\Lambda^{2}}+\frac{2H_{0}^{NLO}(\Lambda)}{\Lambda^{2}}+\frac{2H_{0}^{NNLO}(\Lambda)}{\Lambda^{2}}+\frac{2H_{2}^{NNLO}(\Lambda)}{\Lambda^{4}}(M_{N}E+\gamma_{t}^{2}).
\end{equation}

\begin{figure}[hbt]

\includegraphics[width=120mm]{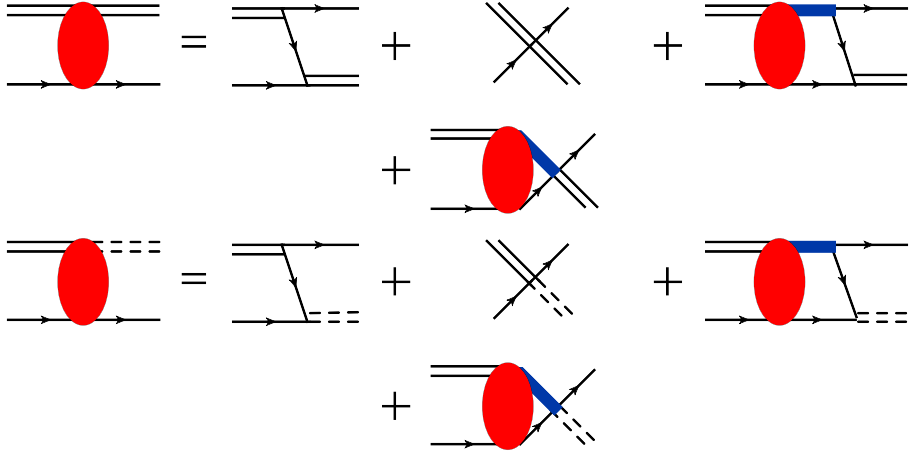}

\caption{\label{fig:LO_Integral_Equations_Doublet}(Color online)Integral Equations for doublet channel at LO.  Note diagrams with three-body forces only occur in doublet S-wave channel }
\end{figure}

The term $H_{0}$ is chosen at each order so that we get the correct scattering length of $a_{\frac{1}{2}}=.65$ fm in the doublet S-wave channel.  This parameter is split up into pieces depending on the order at which we are working, so that $H_{0}$ doesn't have to be refit at each order.  The value of $H_{2}$ is chosen such that we get the correct triton binding energy of $B_{d}=-8.48$ MeV.  Unlike in the partial-resummation technique the binding energy must be calculated perturbatively here, and we use the technique as outlined by Phillips and Ji \cite{Ji:2012nj}.  Now with all these complications in mind we must solve a set of coupled integral equations represented in Fig. \ref{fig:LO_Integral_Equations_Doublet}.  Note that diagrams with three-body forces will only contribute to the doublet S-wave channel.

\noindent In Fig. \ref{fig:LO_Integral_Equations_Doublet} the double dashed line represents the LO singlet propagator and the thick solid (blue) line represents a sum of two diagrams, one containing a deuteron propagator and the other having a spin singlet dibaryon propagator in place of the solid (blue) line.

Having carried out all necessary projections in the doublet channel as in Refs. \cite{Gabbiani:1999yv,Griesshammer:2004pe}, it is convenient to represent the integral equations in cluster-configuration space \cite{Griesshammer:2004pe}.  The master Eq. (\ref{eq:primary}) in the quartet channel now becomes (Note the subscript $d$ on the amplitude $\mathbf{t}$ refers to the doublet channel.)

\begin{align} 
\label{eq:Doubletprimary}
&\mathbf{t}^{l}_{0,d}(k,p)+\mathbf{t}^{l}_{1,d}(k,p)+\mathbf{t}^{l}_{2,d}(k,p)=\mathbf{B}^{l}_{0}(k,p)+\mathbf{B}^{l}_{1}(k,p)+\mathbf{B}^{l}_{2}(k,p)\\\nonumber
&\hspace{1cm}+(\mathbf{K}^{l}_{0}(q,p,E)+\mathbf{K}^{l}_{1}(q,p,E)+\mathbf{K}^{l}_{2}(q,p,E))\otimes ( \mathbf{t}^{l}_{0,d}(k,q) +\mathbf{t}^{l}_{1,d}(k,q) +\mathbf{t}^{l}_{2,d}(k,q) ),
\end{align}

\noindent where the bold script denotes that this is now a matrix equation in cluster-configuration space.  The vector $\mathbf{t}^{l}_{n,d}(k,q)$ defined by

\begin{equation}
\label{eq:tDef}
\mathbf{t}^{l}_{n,d}(k,q)=\left(
\begin{array}{c}
t^{l}_{n,Nt\to Nt}(k,q)\\
t^{l}_{n,Nt\to Ns}(k,q)
\end{array}\right),
\end{equation}

\noindent contains an amplitude $t^{l}_{n,Nt\to Nt}(k,q)$  ($t^{l}_{n,Nt\to Ns}(k,q)$) that corresponds to $nd$ scattering (a neutron and deuteron scattering to a neutron and spin singlet dibaryon).  The vector $\mathbf{B}^{l}_{n}(k,p)$ is defined by

\begin{align}
\label{eq:inhomDoublet}
&\mathbf{B}^{l}_{0}(k,p)=\left(\begin{array}{c}
\frac{M_{N}}{2}y_{t}^{2}\left[\frac{1}{pk}Q_{l}\left(\frac{p^{2}+k^{2}-M_{N}E-i\epsilon}{pk}\right)+\mathcal{H}_{0}(E,\Lambda)\delta_{l0}\right]\\
-\frac{M_{N}}{2}y_{t}y_{s}\left[\frac{3}{pk}Q_{l}\left(\frac{p^{2}+k^{2}-M_{N}E-i\epsilon}{pk}\right)+\mathcal{H}_{0}(E,\Lambda)\delta_{l0}\right]
\end{array}\right)\\\nonumber
&\mathbf{B}^{l}_{1}(k,p)=\left(\begin{array}{c}
\frac{M_{N}}{2}y_{t}^{2}\mathcal{H}_{1}(E,\Lambda)\delta_{l0}\\
-\frac{M_{N}}{2}y_{t}y_{s}\mathcal{H}_{1}(E,\Lambda)\delta_{l0}
\end{array}\right)
,\mathbf{B}^{l}_{2}(k,p)=\left(\begin{array}{c}
\frac{M_{N}}{2}y_{t}^{2}\mathcal{H}_{2}(E,\Lambda)\delta_{l0}\\
-\frac{M_{N}}{2}y_{t}y_{s}\mathcal{H}_{2}(E,\Lambda)\delta_{l0}
\end{array}\right),
\end{align}

\noindent and the kernel matrix $\mathbf{K}^{l}_{n}(q,p,E)$ defined as

\begin{align}
\label{eq:homDoublet}
\mathbf{K}^{l}_{n}(q,p,E)=&\frac{M_{N}}{8\pi}\frac{1}{qp}Q_{l}\left(\frac{q^{2}+p^{2}-M_{N}E-i\epsilon}{qp}\right)\left(\begin{array}{cc}
-y_{t}^{2} & 3y_{t}y_{s} \\
3y_{s}y_{t} & -y_{s}^{2} 
\end{array}\right)\mathbf{D}^{(n)}\left(E-\frac{\vect{q}^{2}}{2M_{N}},\vect{q}\right)\\\nonumber
&+\frac{M_{N}}{8\pi}\delta_{l0}\sum_{j=0}^{n}\mathcal{H}_{n-j}(E,\Lambda)\left(\begin{array}{cc}
-y_{t}^{2} & y_{t}y_{s} \\
y_{t}y_{s} & -y_{s}^{2}
\end{array}\right)\mathbf{D}^{(j)}\left(E-\frac{\vect{q}^{2}}{2M_{N}},\vect{q}\right),
\end{align}

\noindent where $\mathbf{D}^{(n)}(E,\vect{q})$ is a matrix of dibaryon propagators given by

\begin{equation}
\label{eq:DibMatrix}
\mathbf{D}^{n}(E,\vect{q})=
\left(
\begin{array}{cc}
D^{(n)}_{t}(E,\vect{q}) & 0 \\
0&D^{(n)}_{s}(E,\vect{q})  
\end{array}\right),
\end{equation}

\noindent which multiplies the matrix occurring in the definitions of $\mathbf{K}^{l}_{n}(q,p,E)$ via standard matrix multiplication to get the full form of $\mathbf{K}^{l}_{n}(q,p,E)$.

With the master Eq. (\ref{eq:Doubletprimary}) in hand one can simply pick out all the strictly LO, NLO, and NNLO pieces to obtain an expression for the amplitude at each order, which are given by

\begin{align} 
\label{eq:DoubletLO}
&\mathbf{t}^{l}_{0,d}(k,p)=\mathbf{B}^{l}_{0}(k,p)+\mathbf{K}^{l}_{0}(q,p,E)\otimes\mathbf{t}^{l}_{0,d}(k,q)
\end{align}

\begin{align} 
\label{eq:DoubletNLO}
&\mathbf{t}^{l}_{1,d}(k,p)=\mathbf{B}^{l}_{1}(k,p)+\mathbf{K}^{l}_{1}(q,p,E)\otimes\mathbf{t}^{l}_{0,d}(k,q)+\mathbf{K}^{l}_{0}(q,p,E)\otimes\mathbf{t}^{l}_{1,d}(k,q)
\end{align}

\begin{align} 
\label{eq:DoubletNNLO}
\mathbf{t}^{l}_{2,d}(k,p)=\mathbf{B}^{l}_{2}(k,p)&+\mathbf{K}^{l}_{2}(q,p,E)\otimes\mathbf{t}^{l}_{0,d}(k,q) \\\nonumber
&+\mathbf{K}^{l}_{1}(q,p,E)\otimes\mathbf{t}^{l}_{1,d}(k,q) +\mathbf{K}^{l}_{0}(q,p,E)\otimes\mathbf{t}^{l}_{2,d}(k,q) 
\end{align}

\noindent These equations are analogous to those of the quartet channel.  The primary difference is that they are now defined in cluster-configuartion space.  At each order the lower order amplitudes are put into the inhomogeneous
part of the integral equation, and also the kernel at each order is always the same just as in the quartet case.

Note that all of the amplitudes are unrenormalized in these equations.  In order to find the renormalized amplitude we must multiply $t^{l}_{n,Nt\to Nt}(k,p)$ by the appropriate order of the deuteron wavefunction renormalization $Z_{D}$.  Lastly since the channel $t^{l}_{n,Nt\to Ns}(k,p)$ is of no direct physical interest to us we can renormalize it in any manner we wish.  We choose to renormalize $t^{l}_{n,Nt\to Ns}(k,p)$ such that after renormalizing  $t^{l}_{n,Nt\to Nt}(k,p)$ by the appropriate order of $Z_{D}$, all dependence on $y_{s}$ and $y_{t}$ are removed from our integral equations.

\vspace{.2cm}
\section{SD mixing}

Finally at NNLO there is an SD-mixing term that will cause splittings in partial waves of different $J$ values ($\vect{J}=\vect{L}+\vect{S}$ is orbital plus spin angular momentum) as well as mixing different partial waves with the same $J$ values.  The Lagrangian in the auxiliary field formalism for the NNLO SD-mixing term is given by 

\begin{equation}
\label{eq:SDLagrangian}
{\mathcal{L}}_{Nd}^{SD}=y_{SD}\hat{d}_{i}^{\dagger}\left[\hat{N}^{T}\left((\stackrel{\scriptscriptstyle\rightarrow}{\partial}-\stackrel{\scriptscriptstyle\leftarrow}{\partial})^{i}(\stackrel{\scriptscriptstyle\rightarrow}{\partial}-\stackrel{\scriptscriptstyle\leftarrow}{\partial})^{j}-\frac{1}{3}\delta^{ij}(\stackrel{\scriptscriptstyle\rightarrow}{\partial}-\stackrel{\scriptscriptstyle\leftarrow}{\partial})^{2}\right)P_{j}\hat{N}\right] +h.c., 
\end{equation}

\noindent and in terms of nucleon fields the leading SD-mixing term in \EFT is given by

\begin{equation}
\label{eq:SDLagrangianEFT}
{\mathcal{L}}_{Nd}^{SD}=\frac{1}{4}C^{SD}_{0,-1}(\hat{N}^{T}P_{i}\hat{N})^{\dagger}\left[\hat{N}^{T}\left((\stackrel{\scriptscriptstyle\rightarrow}{\partial}-\stackrel{\scriptscriptstyle\leftarrow}{\partial})^{i}(\stackrel{\scriptscriptstyle\rightarrow}{\partial}-\stackrel{\scriptscriptstyle\leftarrow}{\partial})^{j}-\frac{1}{3}\delta^{ij}(\stackrel{\scriptscriptstyle\rightarrow}{\partial}-\stackrel{\scriptscriptstyle\leftarrow}{\partial})^{2}\right)P_{j}\hat{N}\right] +h.c. .
\end{equation}

\noindent The parameter $C^{SD}_{0,-1}$ is fit by producing the correct asymptotic D/S ratio in the deuteron wavefunction \cite{Chen:1999tn} which yields 

\begin{equation}
\label{eq:CSD}
C^{SD}_{0,-1}=-\eta_{sd}\frac{6\sqrt{2}\pi}{M_{N}\gamma_{t}^{2}(\mu-\gamma_{t})},
\end{equation}

\noindent where $\eta_{sd}=.02543\pm .00007$ \cite{Stoks:1994wp} is the asymptotic D/S mixing ratio of the deuteron wavefunction.  Integrating out the auxiliary fields and performing a field redefinition of the nucleon field one can relate $y_{SD}$ to $C^{SD}_{0,-1}$ via

\begin{equation}
\label{eq:matching}
\frac{y_{SD}}{y_{t}}=\frac{1}{4}\frac{\Delta^{({}^{3}\!S_{1})}_{(-1)}}{y_{t}^{2}}C^{SD}_{0,-1}.
\end{equation}

\noindent Then using Eq (\ref{eq:CSD}) one can obtain a value for the parameter $y_{SD}/y_{t}$ which will ultimately appear in all of our SD-mixing amplitudes as we will show below.

\begin{figure}[hbt]
\includegraphics[width=140mm]{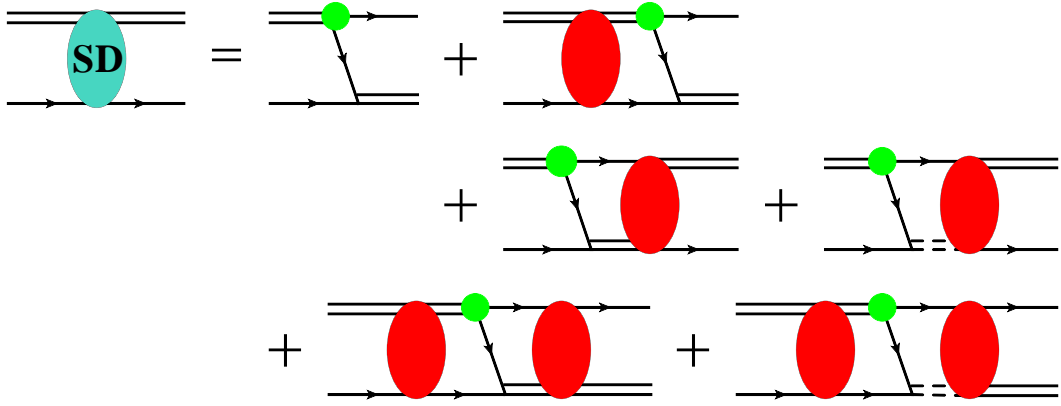}
\caption{\label{fig:NNLO_SDDiagrams}(Color online)SD-mixing diagrams at NNLO (Note diagrams where lower vertices contain SD-mixing terms are not shown)}
\end{figure}

The resulting amplitude from the SD-mixing term is given by the sum of diagrams in Fig. \ref{fig:NNLO_SDDiagrams}.  Note that these diagrams are not projected out in either angular momentum or spin, and upon projection in certain channels specific diagrams will
give zero contribution.

\noindent The sum of all diagrams in Fig. \ref{fig:NNLO_SDDiagrams} in cluster-configuration space gives 

\begin{align}
\label{eq:LO_Amp}
\left(it^{xw}_{SD}\right)^{\beta b}_{\alpha a}(\vect{k},\vect{p})&=\frac{M_{N}}{2}\frac{i}{\vect{k}^{2}+\vect{k}\cdot\vect{p}+\vect{p}^{2}-M_{N}E-i\epsilon}\left(\mathcal{K}^{11}_{SD}{}^{xw}\right)^{\beta b}_{\alpha a}(\vect{p},\vect{k})\\\nonumber
&+\frac{M_{N}}{2}\int\frac{d^{4}q}{(2\pi)^{4}}\mathbf{v}_{p}^{T}(i\tilde{\boldsymbol{\mathcal{K}}}^{xy})^{\beta b}_{\gamma c}(\vect{q},\vect{p},q_{0})i\mathbf{D}^{(0)}\left(\frac{\vect{k}^{2}}{4M_{N}}-\frac{\gamma_{t}^{2}}{M_{N}}+q_{0},\vect{q}\right)\\\nonumber
&\hspace{1.5cm}\times\left(\left(i\mathbf{t}^{yw}\right)^{\gamma c}_{\alpha a}(\vect{k},\vect{q})\right)\propagator{\frac{\vect{k}^{2}}{2M_{N}}-q_{0}}{q}\\\nonumber
&+\frac{M_{N}}{2}\int\frac{d^{4}q}{(2\pi)^{4}}\left(\left(i\mathbf{t}^{xy}\right)^{\beta b}_{\gamma c}(\vect{p},\vect{q})\right)^{T}i\mathbf{D}^{(0)}\left(\frac{\vect{k}^{2}}{4M_{N}}-\frac{\gamma_{t}^{2}}{M_{N}}+q_{0},\vect{q}\right)\\\nonumber
&\hspace{1.5cm}\times(i\tilde{\boldsymbol{\mathcal{K}}}^{yw})^{\gamma c}_{\alpha a}(\vect{k},\vect{q},q_{0})\mathbf{v}_{p}\propagator{\frac{\vect{k}^{2}}{2M_{N}}-q_{0}}{q}\\\nonumber
&+\frac{M_{N}}{2}\int\frac{d^{4}q}{(2\pi)^{4}}\int\frac{d^{4}\ell}{(2\pi)^{4}}\left(\left(i\mathbf{t}^{xz}\right)^{\beta b}_{\delta d}(\vect{p},\vectS{\ell})\right)^{T}i\mathbf{D}^{(0)}\left(\frac{\vect{k}^{2}}{4M_{N}}-\frac{\gamma_{t}^{2}}{M_{N}}+q_{0},\vect{q}\right)\\\nonumber
&\hspace{1.5cm}(i\tilde{\boldsymbol{\mathcal{K}}}^{zy})^{\delta d}_{\gamma c}(\vect{q},\vectS{\ell},q_{0}+\ell_{0})i\mathbf{D}^{(0)}\left(\frac{\vect{k}^{2}}{4M_{N}}-\frac{\gamma_{t}^{2}}{M_{N}}+\ell_{0},\vectS{\ell}\right)\left(\left(i\mathbf{t}^{yw}\right)^{\gamma
c}_{\alpha a}(\vect{k},\vect{q})\right)\\\nonumber
&\hspace{1.5cm}\times\frac{i}{\frac{\vect{k}^{2}}{2M_{N}}-q_{0}-\frac{\vect{q}^{2}}{2M_{N}}+i\epsilon}\frac{i}{\frac{\vect{k}^{2}}{2M_{N}}-\ell_{0}-\frac{\vectS{\ell}^{2}}{2M_{N}}+i\epsilon},\\\nonumber
\end{align}

\noindent where the Greek letters represent nucleon spinor indices, Latin letters $a$,$b$,$c$, and $d$ the nucleon isospinor indices, and finally Latin letters $x$,$w$,$y$, and $z$ the deuteron or ${}^{1}\!S_{0}$ dibaryon polarization depending on the respective matrix element in cluster-configuration space.  As before, the vector $\vect{k}$ ($\vect{p}$) represents the the momentum of the incoming (outgoing) particles in the c.m. frame.  The vector $\mathbf{v}_{p}$ projects out the $nd$ amplitude in cluster-configuration space and is defined by \cite{Griesshammer:2004pe}

\begin{equation}
\mathbf{v}_{p}=\left(
\begin{array}{c}
1\\
0
\end{array}
\right),
\end{equation}

\noindent and the amplitude $\left(\left(i\mathbf{t}^{xw}\right)^{\beta b}_{\alpha a}(\vect{k},\vect{q})\right)$ is a vector defined by

\begin{equation}
\left(\left(i\mathbf{t}^{xw}\right)^{\beta b}_{\alpha a}(\vect{k},\vect{q})\right)=\left(
\begin{array}{c}
\left(it^{xw}_{Nt\to Nt}\right)^{\beta b}_{\alpha a}(\vect{k},\vect{q})\\
\left(it^{xw}_{Nt\to Ns}\right)^{\beta b}_{\alpha a}(\vect{k},\vect{q})
\end{array}\right),
\end{equation}

\noindent where $t_{Nt\to Nt}$ is the LO amplitude for $nd$ scattering and $t_{Nt\to Ns}$ is the LO amplitude for $nd$ going to a nucleon and a spin singlet combination of remaining nucleons.  (Note that we have not yet projected out quartet or doublet channels.)  The matrix $\mathbf{D}^{(0)}(E,\vect{q})$ was given earlier in Eq. (\ref{eq:DibMatrix}), and $(i\tilde{\boldsymbol{\mathcal{K}}}^{xw})^{\beta b}_{\alpha a}(\vect{q},\vectS{\ell},q_{0})$ is a matrix defined by

\begin{align}
\label{eq:K}
(i\tilde{\boldsymbol{\mathcal{K}}}^{xw})^{\beta b}_{\alpha a}(\vect{q},\vectS{\ell},q_{0})=&\frac{i}{\frac{1}{2}\vect{q}^{2}+\vect{q}\cdot\vectS{\ell}+\frac{1}{2}\vectS{\ell}^{2}+\frac{1}{4}\vect{k}^{2}+\gamma_{t}^{2}-M_{N}q_{0}-i\epsilon}\\\nonumber
&\quad\quad\quad\times\left(
\begin{array}{cc}
\left(\mathcal{K}_{SD}^{11}{}^{xw}\right)^{\beta b}_{\alpha a}(\vect{q},\vectS{\ell}) & \left(\mathcal{K}_{SD}^{12}{}^{xw}\right)^{\beta b}_{\alpha a}(\vect{q},\vectS{\ell})\\
\left(\mathcal{K}_{SD}^{21}{}^{xw}\right)^{\beta b}_{\alpha a}(\vect{q},\vectS{\ell}) & \left(\mathcal{K}_{SD}^{22}{}^{xw}\right)^{\beta b}_{\alpha a}(\vect{q},\vectS{\ell})
\end{array}\right),
\end{align}

\noindent where the functions $\left(\mathcal{K}_{SD}^{XY}{}^{xw}\right)^{\beta b}_{\alpha a}(\vect{q},\vectS{\ell})$, which contain all of the angular dependence are defined as

\begin{subequations}
\label{seq:Kequations}
\begin{align}
\label{eq:K11}
\left(\mathcal{K}_{SD}^{11}{}^{xw}\right)^{\beta b}_{\alpha a}(\vect{k},\vect{p})&=y_{t}y_{SD}(\sigma^{y}\sigma^{x})^{\beta}_{\alpha}\delta^{b}_{a}\left[(2\vect{p}+\vect{k})^{w}(2\vect{p}+\vect{k})^{y}-\frac{1}{3}\delta_{yw}(2\vect{p}+\vect{k})^{2}\right]\\\nonumber
&+y_{t}y_{SD}(\sigma^{w}\sigma^{y})^{\beta}_{\alpha}\delta^{b}_{a}\left[(2\vect{k}+\vect{p})^{x}(2\vect{k}+\vect{p})^{y}-\frac{1}{3}\delta_{yx}(2\vect{k}+\vect{p})^{2}\right]
\end{align}
\vspace{-2cm}

\begin{align}
\left(\mathcal{K}_{SD}^{12}{}^{xA}\right)^{\beta b}_{\alpha a}(\vect{k},\vect{p})&=y_{s}y_{SD}(\sigma^{y})^{\beta}_{\alpha}(\tau^{A})^{b}_{a}\left[(2\vect{k}+\vect{p})^{x}(2\vect{k}+\vect{p})^{y}-\frac{1}{3}\delta_{yx}(2\vect{k}+\vect{p})^{2}\right]
\end{align}
\vspace{-2cm}

\begin{align}
\left(\mathcal{K}_{SD}^{21}{}^{Bw}\right)^{\beta b}_{\alpha a}(\vect{k},\vect{p})&=y_{t}y_{SD}(\sigma^{y})^{\beta}_{\alpha}(\tau^{B})^{b}_{a}\left[(2\vect{p}+\vect{k})^{w}(2\vect{p}+\vect{k})^{y}-\frac{1}{3}\delta_{yw}(2\vect{p}+\vect{k})^{2}\right]
\end{align}
\vspace{-2cm}

\begin{align}
\left(\mathcal{K}_{SD}^{22}{}^{BA}\right)^{\beta b}_{\alpha a}(\vect{k},\vect{p})&=0\quad.
\end{align}
\end{subequations}

\noindent (Note that the capital letters $A$,$B$, and $C$ are used for the singlet auxiliary field polarization and the lowercase letters $w$,$x$, and $y$ are used for the deuteron auxiliary field polarization.)  Integrating over the energy and picking up the poles from the nucleon propagators in our diagrams, Eq. (\ref{eq:LO_Amp}) becomes.

\begin{align}
\label{eq:LO_Amp_partial}
\left(t^{xw}_{SD}\right)^{\beta b}_{\alpha a}(\vect{k},\vect{p})&=\frac{M_{N}}{2}\mathbf{v}_{p}^{T}\left(\boldsymbol{\mathcal{K}}^{xw}\right)^{\beta b}_{\alpha a}(\vect{k},\vect{p})\mathbf{v}_{p}\\\nonumber
&-\frac{M_{N}}{2}\int\frac{d^{3}q}{(2\pi)^{3}}\mathbf{v}_{p}^{T}(\boldsymbol{\mathcal{K}}^{xy})^{\beta b}_{\gamma c}(\vect{q},\vect{p})\mathbf{D}^{(0)}\left(E-\frac{\vect{q}^{2}}{2M_{N}},\vect{q}\right)\left(\left(\mathbf{t}^{yw}\right)^{\gamma c}_{\alpha a}(\vect{k},\vect{q})\right)\\\nonumber
&-\frac{M_{N}}{2}\int\frac{d^{3}q}{(2\pi)^{3}}\left(\left(\mathbf{t}^{xy}\right)^{\beta b}_{\gamma c}(\vect{q},\vect{p})\right)^{T}\mathbf{D}^{(0)}\left(E-\frac{\vect{q}^{2}}{2M_{N}},\vect{q}\right)(\boldsymbol{\mathcal{K}}^{yw})^{\gamma c}_{\alpha a}(\vect{k},\vect{q})\mathbf{v}_{p}\\\nonumber
&+\frac{M_{N}}{2}\int\frac{d^{3}q}{(2\pi)^{3}}\int\frac{d^{3}\ell}{(2\pi)^{3}}\left(\left(\mathbf{t}^{xz}\right)^{\beta b}_{\delta d}(\vectS{\ell},\vect{p},)\right)^{T}\mathbf{D}^{(0)}\left(E-\frac{\vect{q}^{2}}{2M_{N}},\vect{q}\right)\\\nonumber
&\hspace{1.5cm}(\boldsymbol{\mathcal{K}}^{zy})^{\delta d}_{\gamma c}(\vect{q},\vectS{\ell})\mathbf{D}^{(0)}\left(E-\frac{\vectS{\ell}^{2}}{2M_{N}},\vectS{\ell}\right)\left(\left(\mathbf{t}^{yw}\right)^{\gamma c}_{\alpha a}(\vect{k},\vect{q})\right),
\end{align}

\noindent where 

\begin{align}
\label{eq:Krelation}
(\boldsymbol{\mathcal{K}}^{xw})^{\beta b}_{\alpha a}(\vect{q},\vectS{\ell})=&\frac{1}{\vect{q}^{2}+\vect{q}\cdot\vectS{\ell}+\vectS{\ell}^{2}-M_{N}E-i\epsilon}\\\nonumber
&\quad\quad\quad\times\left(
\begin{array}{cc}
\left(\mathcal{K}_{SD}^{11}{}^{xw}\right)^{\beta b}_{\alpha a}(\vect{q},\vectS{\ell}) & \left(\mathcal{K}_{SD}^{12}{}^{xw}\right)^{\beta b}_{\alpha a}(\vect{q},\vectS{\ell})\\
\left(\mathcal{K}_{SD}^{21}{}^{xw}\right)^{\beta b}_{\alpha a}(\vect{q},\vectS{\ell}) & \left(\mathcal{K}_{SD}^{22}{}^{xw}\right)^{\beta b}_{\alpha a}(\vect{q},\vectS{\ell})
\end{array}\right).
\end{align}

Now since the SD-mixing term mixes spin and orbital angular momentum we will project out in total angular momentum $\vect{J}$, and the amplitude for the SD-mixing term can be written in a partial wave basis as

\begin{equation}
t_{SD}(\vect{k},\vect{p})=\sum_{J=\nicefrac{1}{2}}^{\infty}\sum_{M=-J}^{M=J}\sum_{L=|J-S|}^{J+S}\sum_{L'=|J-S'|}^{J+S'}\sum_{S,S'}4\pi {t_{SD}}^{JM}_{L'S',LS}(k,p)\mathscr{Y}^{M}_{J,L'S'}(\hat{\mathbf{p}})\left(\mathscr{Y}^{M}_{J,LS}(\hat{\mathbf{k}})\right)^{*}
\end{equation}

\noindent where the spin angle functions are 

\begin{equation}
\mathscr{Y}^{M}_{J,LS}(\hat{\mathbf{k}})=\sum_{m_{L},m_{S}}C_{L,S;J}^{m_{L},m_{S},M}Y_{L}^{m_{L}}(\hat{\mathbf{k}})\chi_{S}^{m_{S}}.
\end{equation}

\noindent Here, $\chi_{S}^{m_{S}}$ is the spinor part of the spin-angle functions, $C_{L,S,J}^{m_{L},m_{S},M}$ the appropriate Clebsch-Gordan coefficient, and $Y_{L}^{m_{L}}(\hat{\mathbf{k}})$ the appropriate spherical harmonic.  Since the spin-angle functions are orthogonal, we can project out the amplitudes in our angular momentum basis 

\begin{equation}
 {t_{SD}}^{JM}_{L'S',LS}(k,p)=\frac{1}{4\pi}\int d\Omega_{k}\int d\Omega_{p}\left(\mathscr{Y}^{M}_{J,L'S'}(\hat{\mathbf{p}})\right)^{*} t_{SD}(\vect{k},\vect{p})\mathscr{Y}^{M}_{J,LS}(\hat{\mathbf{k}})
\end{equation}

Finally projecting out the isospin and carrying out the spin angle projection on Eq. (\ref{eq:LO_Amp_partial}) we obtain 

\begin{align}
\label{eq:LO_Amp_Projected}
{t_{SD}}^{JM}_{L'S',LS}(k,p)&=\frac{M_{N}}{8\pi}\mathbf{v}_{p}^{T}\boldsymbol{\mathcal{K}}(k,p)^{J}_{L'S',LS}\mathbf{v}_{p}\\\nonumber
&-\frac{M_{N}}{16\pi^{3}}\int_{0}^{\infty}dqq^{2}\mathbf{v}_{p}^{T}\boldsymbol{\mathcal{K}}(q,p)^{J}_{L'S',LS}\mathbf{D}^{(0)}\left(E-\frac{\vect{q}^{2}}{2M_{N}},\vect{q}\right)\left(\mathbf{t}^{JM}_{LS,LS}(k,q)\right)\\\nonumber
&-\frac{M_{N}}{16\pi^{3}}\int_{0}^{\infty}dqq^{2}\left(\mathbf{t}^{JM}_{L'S',L'S'}(q,p)\right)^{T}\mathbf{D}^{(0)}\left(E-\frac{\vect{q}^{2}}{2M_{N}},\vect{q}\right)\boldsymbol{\mathcal{K}}(k,q)^{J}_{L'S',LS}\mathbf{v}_{p}\\\nonumber
&+\frac{M_{N}}{32\pi^{5}}\int_{0}^{\infty}dq q^{2}\int_{0}^{\infty}d\ell\ell^{2}\left(\mathbf{t}^{JM}_{L'S',L'S'}(p,\ell)\right)^{T}\mathbf{D}^{(0)}\left(E-\frac{\vect{q}^{2}}{2M_{N}},\vect{q}\right)\\\nonumber
&\quad\quad\times\boldsymbol{\mathcal{K}}(q,\ell)^{JM}_{L'S',LS}\mathbf{D}^{(0)}\left(E-\frac{\vectS{\ell^{2}}}{2M_{N}},\vectS{\ell}\right)\left(\mathbf{t}^{JM}_{LS,LS}(k,q)\right),
\end{align}

\noindent where $\mathbf{t}^{JM}_{LS,LS}(k,q)$ are the LO scattering amplitudes projected out in a partial wave basis and have already been calculated in the preceding sections.  The matrix $\boldsymbol{\mathcal{K}}(k,p)^{J}_{L'S',LS}$ is the kernel function of Eq. (\ref{eq:Krelation}) fully projected out in isospin and total angular momentum.  The kernel matrix $(\boldsymbol{\mathcal{K}}^{xw})^{\beta b}_{\alpha a}(\vect{k},\vect{p})$ can be fully projected out in isospin and total angular momentum by use of Racah algebra obtaining general expressions for any values of angular momentum in terms of 3n-j symbols and has been done in detail in Ref. \cite{Vanasse:2012th}.  The matrix element $\left[\boldsymbol{\mathcal{K}}(k,p)^{J}_{LS,L'S'}\right]_{22}=0$ (the subscript outside the square brackets represents the specific matrix element) since there is no two-body SD-mixing term coupling to the ${}^{1}\!S_{0}$ dibaryon.  The matrix elements $\left[\boldsymbol{\mathcal{K}}(k,p)^{J}_{LS,L'S'}\right]_{12}$ and $\left[\boldsymbol{\mathcal{K}}(k,p)^{J}_{LS,L'S'}\right]_{21}$ are related by

\begin{equation}
\label{eq:K12K21}
\left[\boldsymbol{\mathcal{K}}(k,p)^{J}_{LS,L'S'}\right]_{12}=\left[\boldsymbol{\mathcal{K}}(p,k)^{J}_{L'S',LS}\right]_{21},
\end{equation}

\noindent because of time reversal invariance.  Finally $\left[\boldsymbol{\mathcal{K}}(k,p)^{J}_{LS,L'S'}\right]_{11}$ and $\left[\boldsymbol{\mathcal{K}}(k,p)^{J}_{LS,L'S'}\right]_{12}$ are given by  

\begin{align}
\label{eq:ProjectedSDDtoD}
&\left[\boldsymbol{\mathcal{K}}(k,p)^{J}_{L'S',LS}\right]_{11}=4\pi \sqrt{\bar{S}\bar{S}'\bar{L}}\sqrt{\frac{10}{3}}\left(\delta_{S',\nicefrac{1}{2}}+2\delta_{S',\nicefrac{3}{2}}\right)C_{L,2,L'}^{0,0,0}(-1)^{2S'+S+L-J}\\\nonumber
&\times\SixJ{2}{1}{1}{\nicefrac{1}{2}}{S}{S'}\SixJ{S'}{2}{S}{L}{J}{L'}\frac{1}{kp}(k^{2}Q_{L'}(a)+4p^{2}Q_{L}(a))\\\nonumber
&+8\pi\sqrt{\bar{S}\bar{S'}\bar{L}}\left(\delta_{S',\nicefrac{1}{2}}+2\delta_{S',\nicefrac{3}{2}}\right)\sum_{L''}C_{L,1,L''}^{0,0,0}C_{L'',1,L'}^{0,0,0}\\\nonumber
&\hspace{1cm}\times(-1)^{\nicefrac{3}{2}-S'-L-L''}\sqrt{\bar{L''}}
\left\{\begin{array}{ccc}
\nicefrac{1}{2} & 1 & S\\
1 & L'' & L\\
S' & L' & J
\end{array}\right\}Q_{L''}(a)\\\nonumber
&+8\pi\sqrt{\bar{S}\bar{S}'\bar{L}}\left(\delta_{S',\nicefrac{1}{2}}+2\delta_{S',\nicefrac{3}{2}}\right)\sum_{L''}C_{L,1,L''}^{0,0,0}C_{L'',1,L'}^{0,0,0}(-1)^{\nicefrac{1}{2}+S'+L+L''}\\\nonumber
&\hspace{1cm}\times\sqrt{\bar{L''}}\SixJ{\nicefrac{1}{2}}{1}{S'}{L'}{J}{L''}\SixJ{L}{1}{L''}{\nicefrac{1}{2}}{J}{S}Q_{L''}(a)\\\nonumber
&-\frac{16\pi}{3}\frac{1}{\sqrt{\bar{L}}}\left(\delta_{S',\nicefrac{1}{2}}+2\delta_{S',\nicefrac{3}{2}}\right)(-1)^{\nicefrac{1}{2}-S'}\delta_{L,L'}\delta_{S,S'}\sum_{L''}C_{L,1,L''}^{0,0,0}C_{1,L'',L}^{0,0,0}\sqrt{\bar{L}''}Q_{L''}(a)\\\nonumber
&+(S\longleftrightarrow S')(L\longleftrightarrow L')(k\longleftrightarrow p),
\end{align}

\noindent and \vspace{-1cm}

\begin{align}
\label{eq:ProjectedSDStoD}
&\left[\boldsymbol{\mathcal{K}}(k,p)^{J}_{L'S',LS}\right]_{21}=8\pi\sqrt{5}\sqrt{\bar{S}\bar{L}}\delta_{S'\nicefrac{1}{2}}C_{L,2,L'}^{0,0,0}(-1)^{1+S+L-J}\\\nonumber
&\SixJ{2}{1}{1}{\nicefrac{1}{2}}{S}{S'}\SixJ{S'}{2}{S}{L}{J}{L'}\frac{1}{kp}(k^{2}Q_{L'}(a)+4p^{2}Q_{L}(a))\\\nonumber
&+8\pi\sqrt{6}\sqrt{\bar{L}\bar{S}}\delta_{S'\nicefrac{1}{2}}\sum_{L''}\sqrt{\bar{L}''}\CG{L}{0}{1}{0}{L''}{0}\CG{L''}{0}{1}{0}{L'}{0}
\left\{\begin{array}{ccc}
\nicefrac{1}{2} & 1 & S\\
1 & L'' & L \\
S' & L' & J
\end{array}\right\}Q_{L''}(a)\\\nonumber
&+8\pi\sqrt{6}\sqrt{\bar{L}\bar{S}}\delta_{S'\nicefrac{1}{2}}\sum_{L''}(-1)^{1+L+L''}\sqrt{\bar{L}''}\CG{L}{0}{1}{0}{L''}{0}\CG{L''}{0}{1}{0}{L'}{0}\\\nonumber
&\hspace{1cm}\times\SixJ{L'}{1}{L''}{\nicefrac{1}{2}}{J}{S'}\SixJ{L}{1}{L''}{\nicefrac{1}{2}}{J}{S}Q_{L''}(a)\\\nonumber
&+\frac{16\pi}{\sqrt{3}}(-1)^{L''-L}\sqrt{\frac{1}{\bar{L}'}}\sum_{L''}\sqrt{\bar{L}''} C_{L,1,L''}^{0,0,0}C_{L'',1,L'}^{0,0,0}\delta_{S'\nicefrac{1}{2}}\delta_{S'S}\delta_{L'L}Q_{L''}(a),\\\nonumber
\end{align}

\noindent where $Q_{L}(a)$ are defined by Eq. (\ref{eq:LegnedgreQ}), $a=(k^{2}+p^{2}-M_{N}E-i\epsilon)/pk$, and the bar notation is defined as $\bar{x}=2x+1$.

In order to evaluate Eq. (\ref{eq:LO_Amp_Projected}) we multiply both sides by $Z_{LO}$, and use the renormalized scattering amplitudes in the integrals. Doing this this will remove all factors of $y_{s}$ and all terms will be multiplied by factors $y_{SD}/y_{t}$, which has already been fit to physical data.  Then numerically integrating the renormalized scattering amplitudes with their matching kernel we obtain the renormalized SD-mixing amplitude.  Note that the renormalization of the $t^{l}_{n,Nt\to Ns}(k,p)$ was arbitrary and chosen to remove any dependence on $y_{t}$ and $y_{s}$ in the doublet integral equations.  As long as we are consistent with our normalization choice all factors of $y_{s}$ should also cancel out in the SD-mixing terms leaving only $y_{SD}/y_{t}$.


\section{Phase Shift Analysis}

After calculating the quartet, doublet, and SD-mixing amplitudes one can calculate the scattering amplitude $T^{J}_{L'S',LS}$ up to NNLO, where $L$ ($S$) is the initial orbital (spin) angular momentum, $L'$ ($S'$) the final orbital (spin) angular momentum, and $J$ the total angular momentum.  At LO the scattering amplitude is given by 

\begin{align}
\label{eq:LOScattAmp}
T^{J(0)}_{L'S',LS}(p)=&\delta_{LL'}\delta_{SS'}\delta_{S'\nicefrac{1}{2}}Z_{LO}t^{L}_{0,Nt\to Nt}(p)+\delta_{LL'}\delta_{SS'}\delta_{S'\nicefrac{3}{2}}Z_{LO}t^{L}_{0,q}(p),
\end{align}

\noindent where the quartet and doublet scattering amplitudes are multiplied by the LO deuteron wavefunction renormalization. (Note the superscript (n) on $T^{J(n)}_{L'S',LS}$ refers to the order of the amplitude.)  The NLO scattering amplitude is given by

\begin{align}
\label{eq:NLOScattAmp}
T^{J(1)}_{L'S',LS}(p)=&\delta_{LL'}\delta_{SS'}\delta_{S'\nicefrac{1}{2}}(Z_{NLO}t^{L}_{0,Nt\to Nt}(p)+Z_{LO}t^{L}_{1,Nt\to Nt}(p))\\\nonumber
&+\delta_{LL'}\delta_{SS'}\delta_{S'\nicefrac{3}{2}}(Z_{NLO}t^{L}_{0,q}(p)+Z_{LO}t^{L}_{1,q}(p)).
\end{align}

\noindent For both LO and NLO the scattering amplitude has no $J$ dependence and there is also no mixing between different spin and partial waves.  Finally the NNLO scattering amplitude is given by

\begin{align}
\label{eq:NNLOScattAmp}
T^{J(2)}_{L'S',LS}(p)=&\delta_{LL'}\delta_{SS'}\delta_{S'\nicefrac{1}{2}}(Z_{NNLO}t^{L}_{0,Nt\to Nt}(p)+Z_{NLO}t^{L}_{1,Nt\to Nt}(p)+Z_{LO}t^{L}_{2,Nt\to Nt}(p))\\\nonumber
&+\delta_{LL'}\delta_{SS'}\delta_{S'\nicefrac{3}{2}}(Z_{NNLO}t^{L}_{0,q}(p)+Z_{NLO}t^{L}_{1,q}(p)+Z_{LO}t^{L}_{2,q}(p))+{t_{SD}}^{J}_{L'S',LS}(p),
\end{align}

\noindent and at this order the SD-mixing amplitudes are necessary.  Note the SD-mixing amplitudes in Eq. (\ref{eq:NNLOScattAmp}) are already renormalized by the LO deuteron wavefunction renormalization and thus do not have a factor of $Z_{LO}$.  The SD-mixing amplitude at this order will cause splittings for different $J$ values in the quartet channel.  However, the SD-mixing amplitude is zero for the doublet channel and no splitting occurs.  Also the SD-mixing amplitude allows all possible (up to conservation of parity and total angular momentum) mixings of different initial and final partial waves and spin.

Since the total angular momentum is a good quantum number the S-matrix can be decomposed into a direct sum of irreducible representations of $\vect{J}$.  For each $J$ value, states of different parity can be further factorized since parity is conserved.  Starting at $J=\frac{1}{2}$ the S-matrix is decomposed into a 2$\times$2 matrix for positive and negative parity giving

\begin{equation}
\label{eq:SOneHalf}
\mathbf{S}^{\frac{1}{2}+}=\left(\begin{array}{cc}
S^{\frac{1}{2}}_{2\frac{3}{2},2\frac{3}{2}} & S^{\frac{1}{2}}_{2\frac{3}{2},0\frac{1}{2}}\\
S^{\frac{1}{2}}_{0\frac{1}{2},2\frac{3}{2}} & S^{\frac{1}{2}}_{0\frac{1}{2},0\frac{1}{2}}
\end{array}\right),
\mathbf{S}^{\frac{1}{2}-}=\left(\begin{array}{cc}
S^{\frac{1}{2}}_{1\frac{1}{2},1\frac{1}{2}} & S^{\frac{1}{2}}_{1\frac{1}{2},1\frac{3}{2}}\\
S^{\frac{1}{2}}_{1\frac{3}{2},1\frac{1}{2}} & S^{\frac{1}{2}}_{1\frac{3}{2},1\frac{3}{2}}
\end{array}\right).
\end{equation}

\noindent The 2$\times$2 S-matrix in the Blatt and Biedenharn parametrization \cite{Blatt:1952zz} is given by

\begin{equation}
\label{eq:Sdecomp}
\mathbf{S}^{J\pi}=\left(\mathbf{u}^{J\pi}\right)^{T}e^{2i\boldsymbol{\delta}^{J\pi}}\mathbf{u}^{J\pi}
\end{equation}

\noindent where the $\boldsymbol{\delta}^{J\pi}$ and $\mathbf{u}^{J\pi}$ (note $\pi$ represents parity) are matrices defined as

\begin{equation}
\boldsymbol{\delta}^{\frac{1}{2}+}=\left(\begin{array}{cc}
\delta^{\frac{1}{2}}_{2\frac{3}{2}} & 0 \\
0 & \delta^{\frac{1}{2}}_{0\frac{1}{2}} 
\end{array}\right),
\boldsymbol{\delta}^{\frac{1}{2}-}=\left(\begin{array}{cc}
\delta^{\frac{1}{2}}_{1\frac{1}{2}} & 0 \\
0 & \delta^{\frac{1}{2}}_{1\frac{3}{2}} 
\end{array}\right),
\end{equation}

\begin{equation}
\mathbf{u}^{\frac{1}{2}+}=\left(\begin{array}{cc}
\cos\eta^{\frac{1}{2}+} & \sin\eta^{\frac{1}{2}+}\\
-\sin\eta^{\frac{1}{2}+} & \cos\eta^{\frac{1}{2}+}
\end{array}\right),
\mathbf{u}^{\frac{1}{2}-}=\left(\begin{array}{cc}
\cos\epsilon^{\frac{1}{2}-} & \sin\epsilon^{\frac{1}{2}-}\\
-\sin\epsilon^{\frac{1}{2}-} & \cos\epsilon^{\frac{1}{2}-}
\end{array}\right)
\end{equation}

In the limit where the mixing parameters $\eta^{\frac{1}{2}+}$ and $\epsilon^{\frac{1}{2}-}$ are zero the matrix $\mathbf{u}^{J\pi}$ becomes the identity matrix and the S-matrix becomes diagonal.   Therefore, the $\delta^{J\pi}_{LS}$ eigenphase parameters become the standard phase-shifts in the limit of zero mixing angles. To obtain fits for the phase-shift parameters one must relate the S-matrix to the T-matrix and this can be done by projecting out the operator equation $\hat{\mathbf{S}}=\boldsymbol{1}+i\hat{\mathbf{T}}$ in total angular momentum yielding

\begin{equation}
\label{eq:S-Trelation}
S^{J}_{L'S',LS}=\delta_{LL'}\delta_{SS'}+i\frac{2M_{N}p}{3\pi}T^{J}_{L'S',LS}(p).
\end{equation}

\noindent Now one expands the eigenphase parameters $\delta^{J\pi}_{LS}$ and the mixing parameters perturbatively up to NNLO giving the S-matrix in a perturbative expansion.  Then using Eq. (\ref{eq:S-Trelation}) one expands the S-matrix perturbatively by expanding the T-matrix perturbatively and making $\delta_{LL'}\delta_{SS'}$ LO.  Finally matching these two perturbative expansions onto each other one obtains

\begin{subequations}
\label{eq:PhaseShift}

\begin{equation}
\label{eq:PhaseShiftLO}
\delta_{LS}^{J(0)}=\frac{1}{2i}\log\left(1+i\frac{2M_{N}p}{3\pi}T^{J(0)}_{LS,LS}(p)\right)
\end{equation}

\begin{equation}
\delta_{LS}^{J(1)}=\frac{1}{2i}\frac{i\frac{2M_{N}p}{3\pi}T^{J(1)}_{LS,LS}(p)}{1+i\frac{2M_{N}p}{3\pi}T^{J(0)}_{LS,LS}(p)}
\end{equation}

\begin{equation}
\label{eq:PhaseShiftNNLO}
\delta_{LS}^{J(2)}=\frac{1}{2i}\frac{i\frac{2M_{N}p}{3\pi}T^{J(2)}_{LS,LS}(p)}{1+i\frac{2M_{N}p}{3\pi}T^{J(0)}_{LS,LS}(p)}-\frac{1}{4i}\frac{\left(i\frac{2M_{N}p}{3\pi}T^{J(1)}_{LS,LS}(p)\right)^{2}}{\left(1+i\frac{2M_{N}p}{3\pi}T^{J(0)}_{LS,LS}(p)\right)^{2}},
\end{equation}

\end{subequations}

\noindent for the perturbative expansion of the eigenphase parameters and

\begin{equation}
\label{eq:mixing}
\eta^{\frac{1}{2}+}=\frac{T^{\frac{1}{2}(2)}_{2\frac{3}{2},0\frac{1}{2}}}{T^{\frac{1}{2}(0)}_{2\frac{3}{2},2\frac{3}{2}}-T^{\frac{1}{2}(0)}_{0\frac{1}{2},0\frac{1}{2}}},
\epsilon^{\frac{1}{2}-}=\frac{T^{\frac{1}{2}(2)}_{1\frac{1}{2},1\frac{3}{2}}}{T^{\frac{1}{2}(0)}_{1\frac{1}{2},1\frac{1}{2}}-T^{\frac{1}{2}(0)}_{1\frac{3}{2},1\frac{3}{2}}},
\end{equation}

\noindent for the mixing parameters, in terms of the T-matrix.  Note that the mixing parameters first occur at order NNLO.

For $J\geq\frac{3}{2}$ the S-matrix is given by a 3$\times$3 matrix and we choose to parametrize the matrix using the conventions of Seyler \cite{Seyler:1968}, in which case the S-matrix takes the form

\begin{equation}
\label{eq:Smatrix3by3}
\mathbf{S}^{J\pi}=\left(\begin{array}{ccc}
S^{J}_{J\mp\frac{3}{2}\frac{3}{2},J\mp\frac{3}{2}\frac{3}{2}} & S^{J}_{J\mp\frac{3}{2}\frac{3}{2},J\pm\frac{1}{2}\frac{1}{2}} & S^{J}_{J\mp\frac{3}{2}\frac{3}{2},J\pm\frac{1}{2}\frac{3}{2}} \\
S^{J}_{J\pm\frac{1}{2}\frac{1}{2},J\mp\frac{3}{2}\frac{3}{2}} & S^{J}_{J\pm\frac{1}{2}\frac{1}{2},J\pm\frac{1}{2}\frac{1}{2}} & S^{J}_{J\pm\frac{1}{2}\frac{1}{2},J\pm\frac{1}{2}\frac{3}{2}} \\
S^{J}_{J\pm\frac{1}{2}\frac{3}{2},J\mp\frac{3}{2}\frac{3}{2}} & S^{J}_{J\pm\frac{1}{2}\frac{3}{2},J\pm\frac{1}{2}\frac{1}{2}} & S^{J}_{J\pm\frac{1}{2}\frac{3}{2},J\pm\frac{1}{2}\frac{3}{2}}
\end{array}\right).
\end{equation}

\noindent To parametrize the matrix we again use Eq. (\ref{eq:Sdecomp}).  However, the matrix $\boldsymbol{\delta}^{J\pi}$ is now a diagonal 3$\times$3 matrix defined by 

\begin{equation}
\label{eq:delta3by3}
\boldsymbol{\delta}^{J\pi}=\left(\begin{array}{ccc}
\delta^{J}_{J\mp\frac{3}{2}\frac{3}{2}} & 0 & 0 \\
0 & \delta^{J}_{J\pm\frac{1}{2}\frac{1}{2}} & 0 \\
0 & 0 & \delta^{J}_{J\pm\frac{1}{2}\frac{3}{2}} 
\end{array}\right),
\end{equation}

\noindent and we now have three mixing parameters $\epsilon^{J\pi}$,$\zeta^{J\pi}$, and $\eta^{J\pi}$.  The matrix $\mathbf{u}^{J\pi}$ is now defined  by three successive Blatt and Biedenharn ``rotations" 

\begin{equation}
\label{eq:BlattRotations}
\begin{scriptsize}\mathbf{v}^{J\pi}=\left(\begin{array}{ccc}
1 & 0 & 0\\
0 & \cos\epsilon^{J\pi} &\sin\epsilon^{J\pi}\\
0 & -\sin\epsilon^{J\pi} & \cos\epsilon^{J\pi}
\end{array}\right),
\mathbf{w}^{J\pi}=\left(\begin{array}{ccc}
\cos\zeta^{J\pi} & 0 & \sin\zeta^{J\pi}\\
0 & 1 & 0 \\
-\sin\zeta^{J\pi} & 0 & \cos\zeta^{J\pi}
\end{array}\right),
\mathbf{x}^{J\pi}=\left(\begin{array}{ccc}
\cos\eta^{J\pi} & \sin\eta^{J\pi} & 0\\
-\sin\eta^{J\pi} & \cos\eta^{J\pi} & 0 \\
0 & 0 & 1
\end{array}\right)
\end{scriptsize}
\end{equation}
\begin{equation}
\label{eq:u3by3}
\mathbf{u}^{J\pi}=\mathbf{v}^{J\pi}\mathbf{w}^{J\pi}\mathbf{x}^{J\pi}
\end{equation}

Finally, expanding the 3$\times$3 S-matrix perturbatively in terms of the phase-shift parameters and the T-matrix we can again obtain an expression for the perturbative expansion of the phase-shift parameters in terms of a perturbative expansion of the T-matrix.  The equations for the eigenphase parameters are the same as before.  However, the mixing parameters are given in terms of the T-matrix as follows

\begin{align}
\label{eq:mixing3by3}
\eta^{J\pi}=\frac{T^{J(2)}_{J\mp\frac{3}{2}\frac{3}{2},J\pm\frac{1}{2}\frac{1}{2}}}{T^{J(0)}_{J\mp\frac{3}{2}\frac{3}{2},J\mp\frac{3}{2}\frac{3}{2}}-T^{J(0)}_{J\pm\frac{1}{2}\frac{1}{2},J\pm\frac{1}{2}\frac{1}{2}}},&
\epsilon^{J\pi}=\frac{T^{J(2)}_{J\pm\frac{1}{2}\frac{1}{2},J\pm\frac{1}{2}\frac{3}{2}}}{T^{J(0)}_{J\pm\frac{1}{2}\frac{1}{2},J\pm\frac{1}{2}\frac{1}{2}}-T^{J(0)}_{J\pm\frac{1}{2}\frac{3}{2},J\pm\frac{1}{2}\frac{3}{2}}},\\\nonumber
&\hspace{-3cm}\zeta^{J\pi}=\frac{T^{J(2)}_{J\mp\frac{3}{2}\frac{3}{2},J\pm\frac{1}{2}\frac{3}{2}}}{T^{J(0)}_{J\mp\frac{3}{2}\frac{3}{2},J\mp\frac{3}{2}\frac{3}{2}}-T^{J(0)}_{J\pm\frac{1}{2}\frac{3}{2},J\pm\frac{1}{2}\frac{3}{2}}}.
\end{align}


\section{Results}
\begin{figure}[hbt]

	\begin{center}

	\subfloat{\includegraphics[angle=-90,width=88mm]{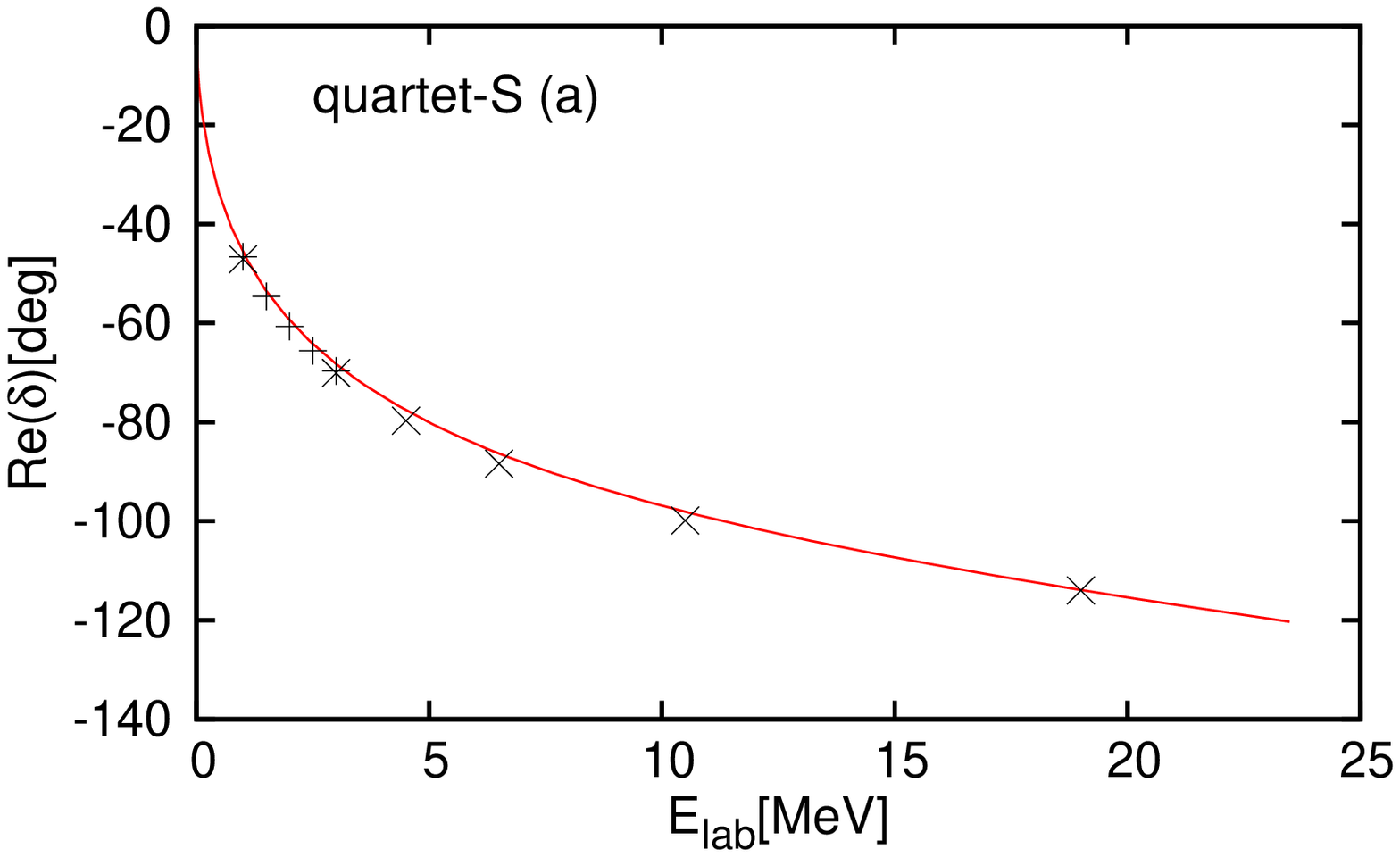}}
	\subfloat{\includegraphics[angle=-90,width=88mm]{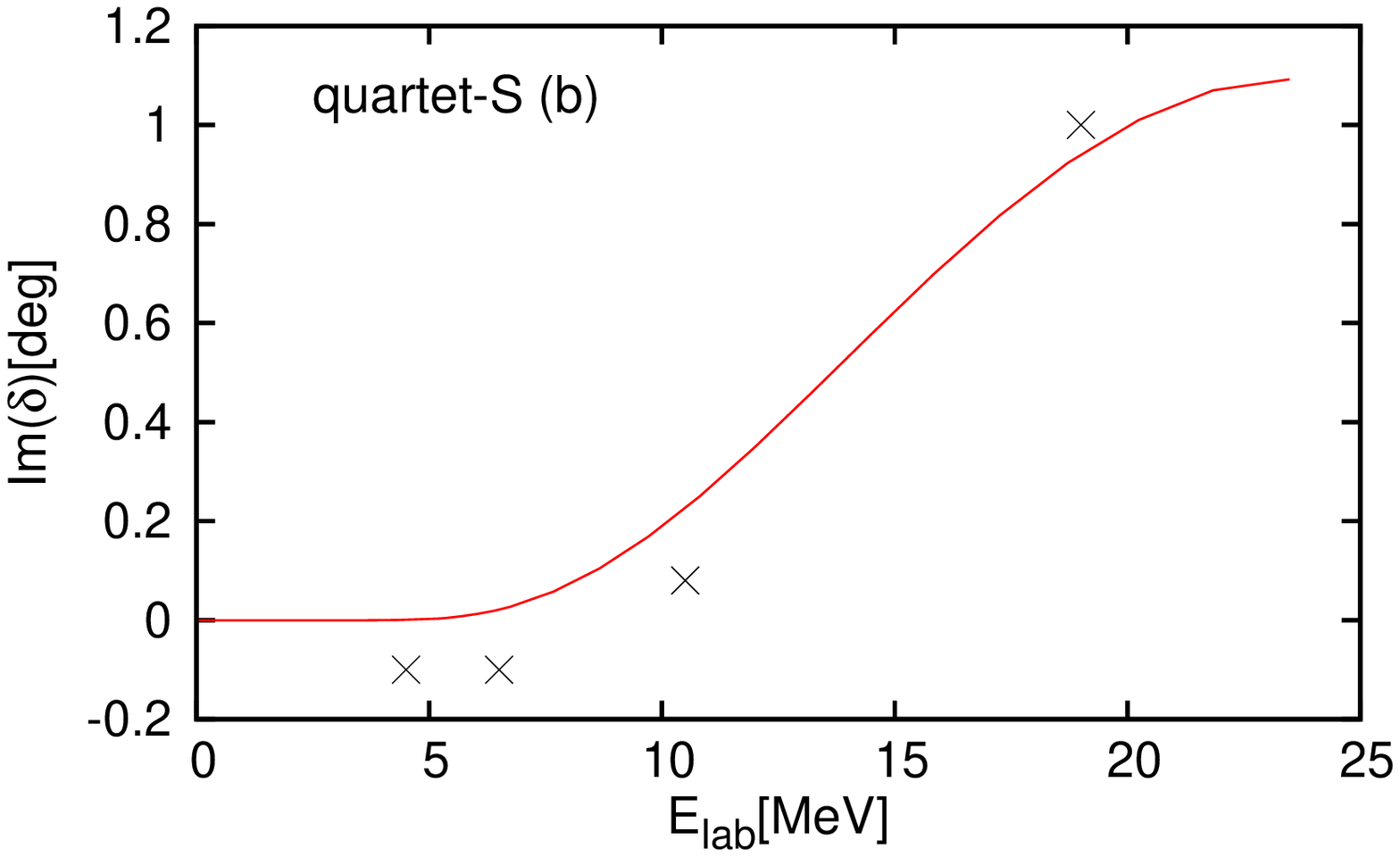}}

	\subfloat{\includegraphics[angle=-90,width=88mm]{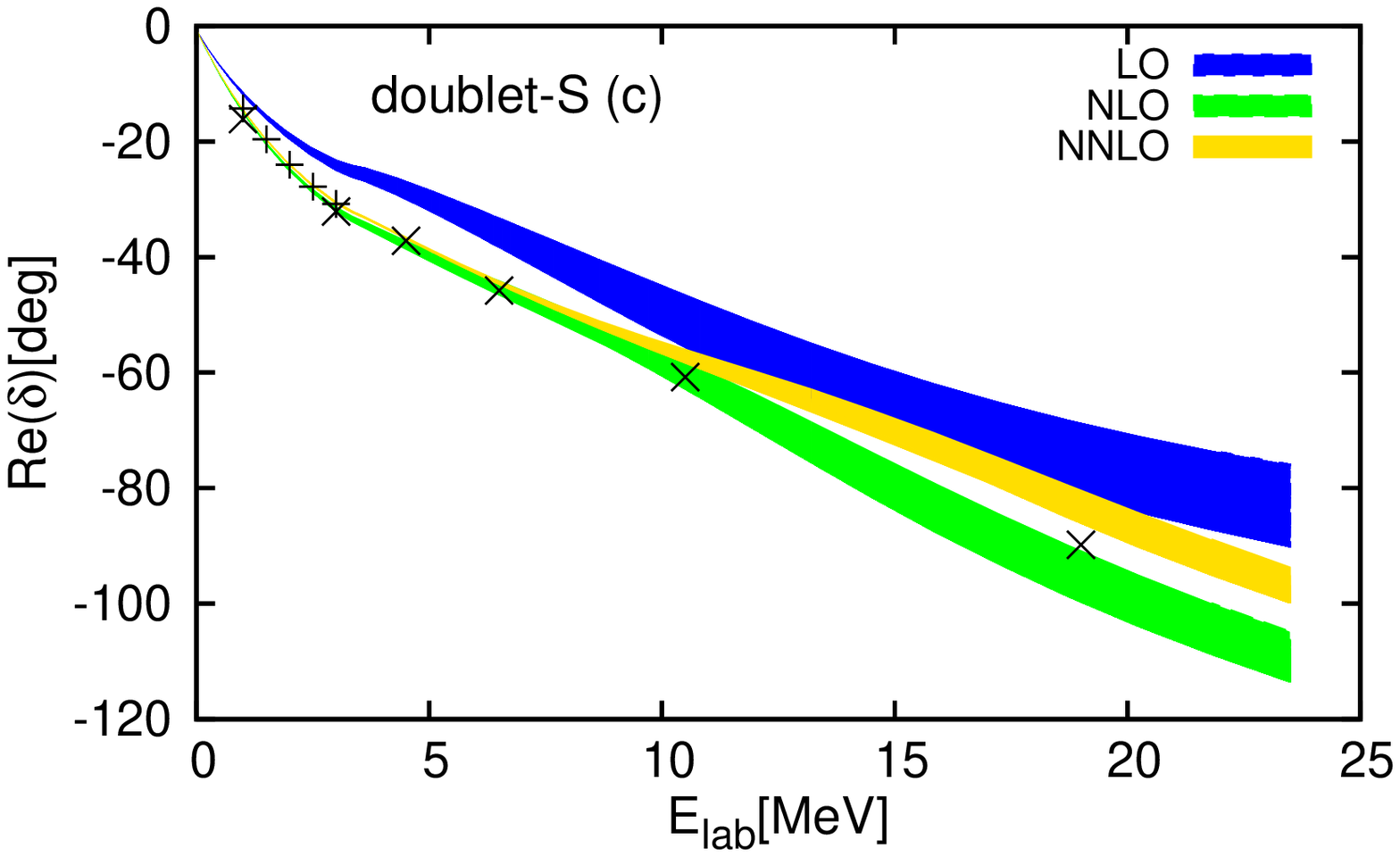}}
	\subfloat{\includegraphics[angle=-90,width=88mm]{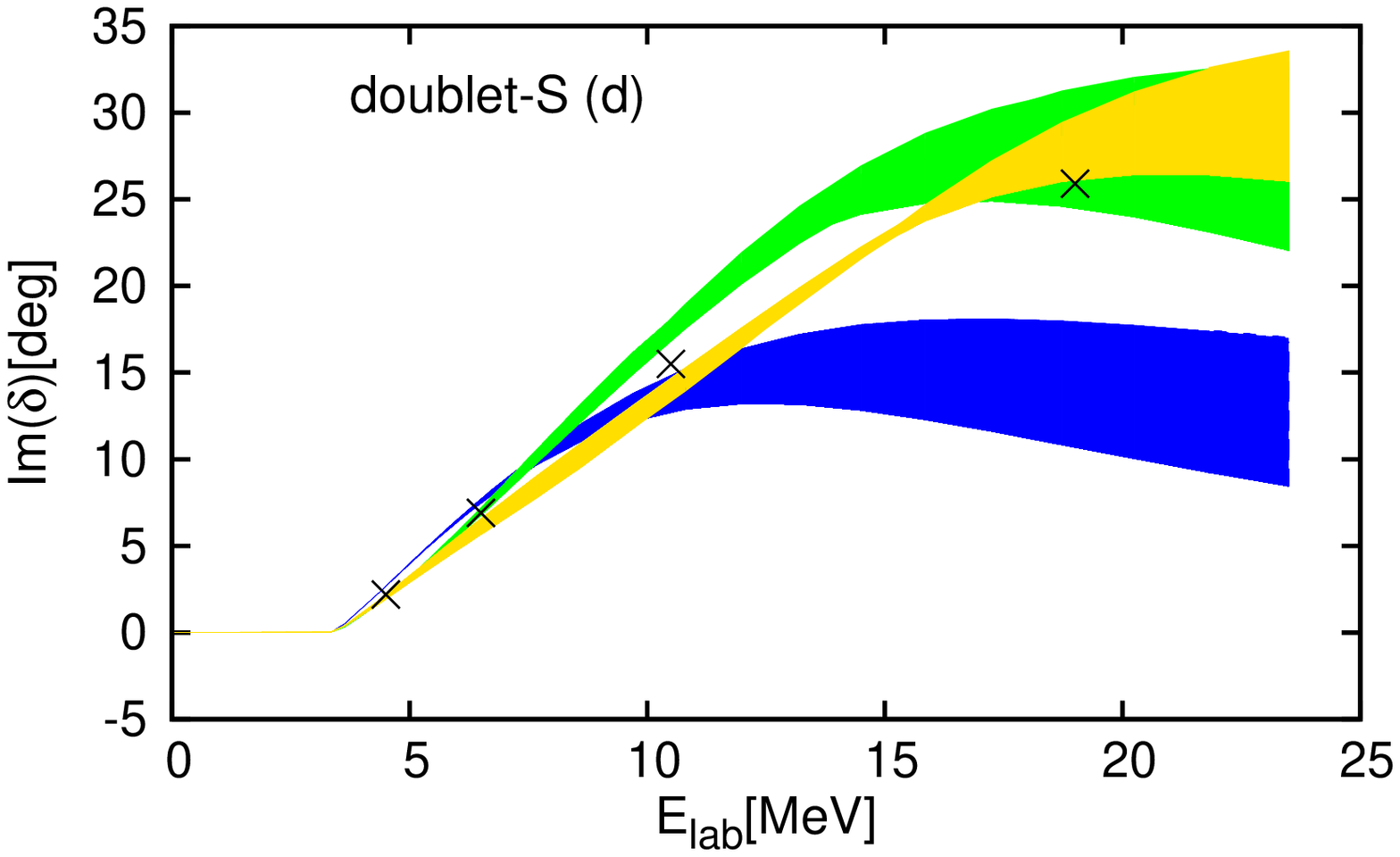}}

	\end{center}

	\caption{\label{fig:SwavePhaseShifts}(Color online) quartet (doublet) S-wave phase shift above (below), real part (imaginary part) left (right) in degrees as function of lab energy.  quartet S-wave given for $\Lambda=1600$ MeV and doublet S-wave cutoff varied from $\Lambda=200-1600$ MeV.  The crosses below breakup are AV18+UIX data \cite{Kievsky:1996ca}, the stars Faddeev equations with Bonn-B \cite{Glockle:1995fb} }

\end{figure}

The only channel for which three-body forces are necessary up to and including NNLO is the doublet S-wave.  After fixing the three-body forces as described above the resulting doublet S-wave phase-shift  is given in Fig \ref{fig:SwavePhaseShifts}.  The data below
 the deuteron breakup threshold DBT is from the Argonne $v_{18}$ and Urbana IX potentials (AV18+UIX) \cite{Kievsky:1996ca} with the Pair Correlated Hyperspherical Harmonic method (crosses), and above and below breakup from the Bonn-B potential with Faddeev equations (stars) \cite{Glockle:1995fb}.  Agreement between the potential model calculations (PMC) and the calculation here is within the expected error of \EFT in the Z-parametrization at NNLO, $((Z_{t}-1)/2)^{3}\approx3\%$.

\noindent 
At each order we vary the cutoff $\Lambda$ between 200 and 1600 MeV.  The resulting range of predictions is shown as a solid band in the figure.  One should also note that each order the cutoff variation diminishes.  Also shown in Fig \ref{fig:SwavePhaseShifts} is the quartet S-wave phase shift which agrees well with the same PMC.  The cutoff variation in the quartet S-wave channel is negligible as no three-body forces are necessary, thus the phase shift is plotted for a cutoff of $\Lambda=1600$ MeV.  Note at roughly $\Lambda=900$ MeV all results begin to converge even in the doublet S-wave and higher partial waves, which is in agreement with previous findings \cite{Griesshammer:2004pe}. The scattering length, ${}^{4}a$, in the quartet S-wave channel at NLO and NNLO is found to be $6.74$ fm and $6.19\pm.030$ fm respectively.  The NNLO scattering length compared to the experimental value of $6.35\pm.02$ fm \cite{Dilg:1971pl} is within the expected $3\%$ error.  In the partial-resummation technique the NNLO scattering length of $6.354\pm.002$ fm \cite{Griesshammer:2004pe} is much closer to the experimental value and is not surprising since the partial-resummation technique contains certain higher order contributions.

\begin{figure}

	\begin{center}

	\subfloat{\includegraphics[angle=-90,width=88mm]{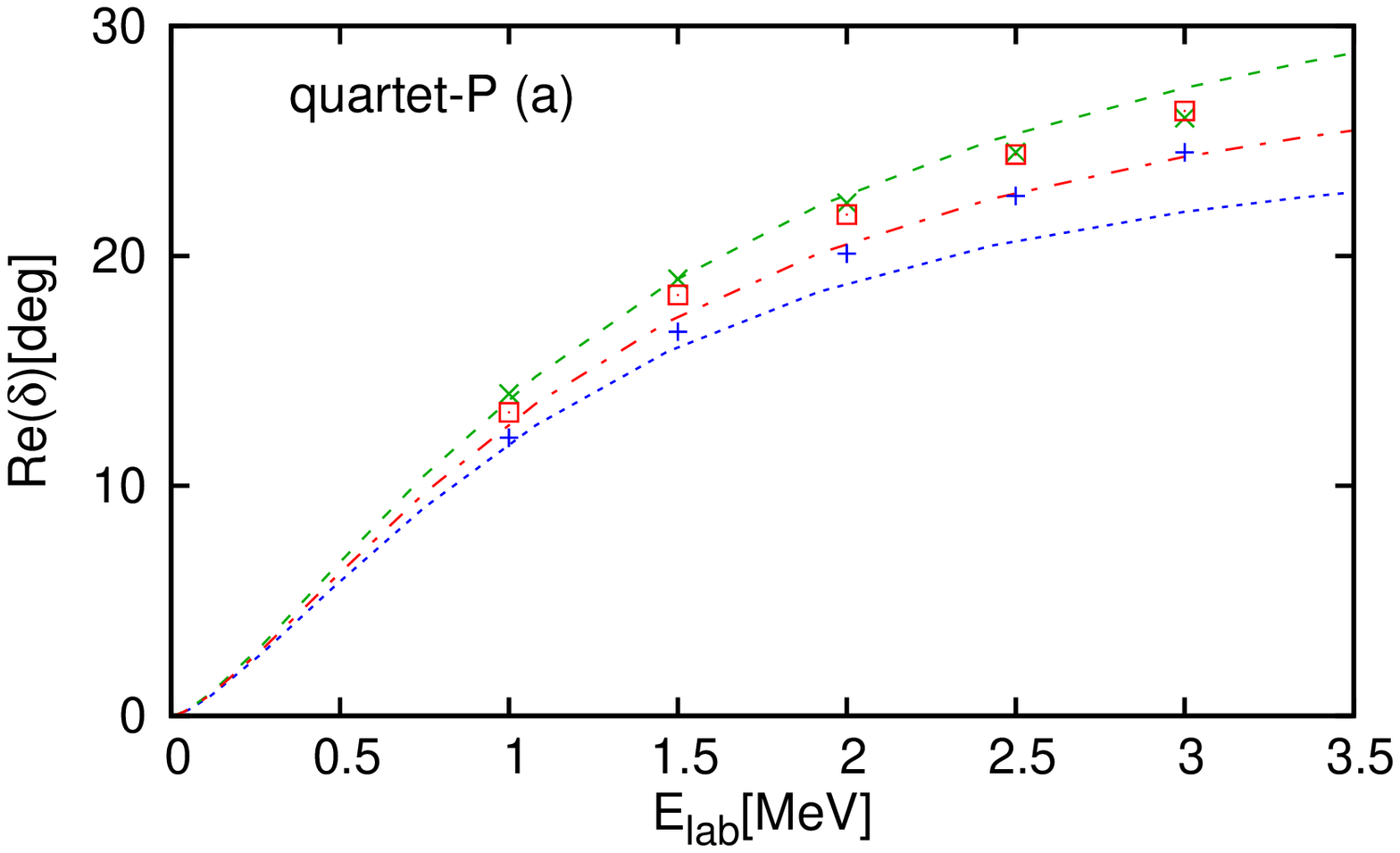}}
	\subfloat{\includegraphics[angle=-90,width=88mm]{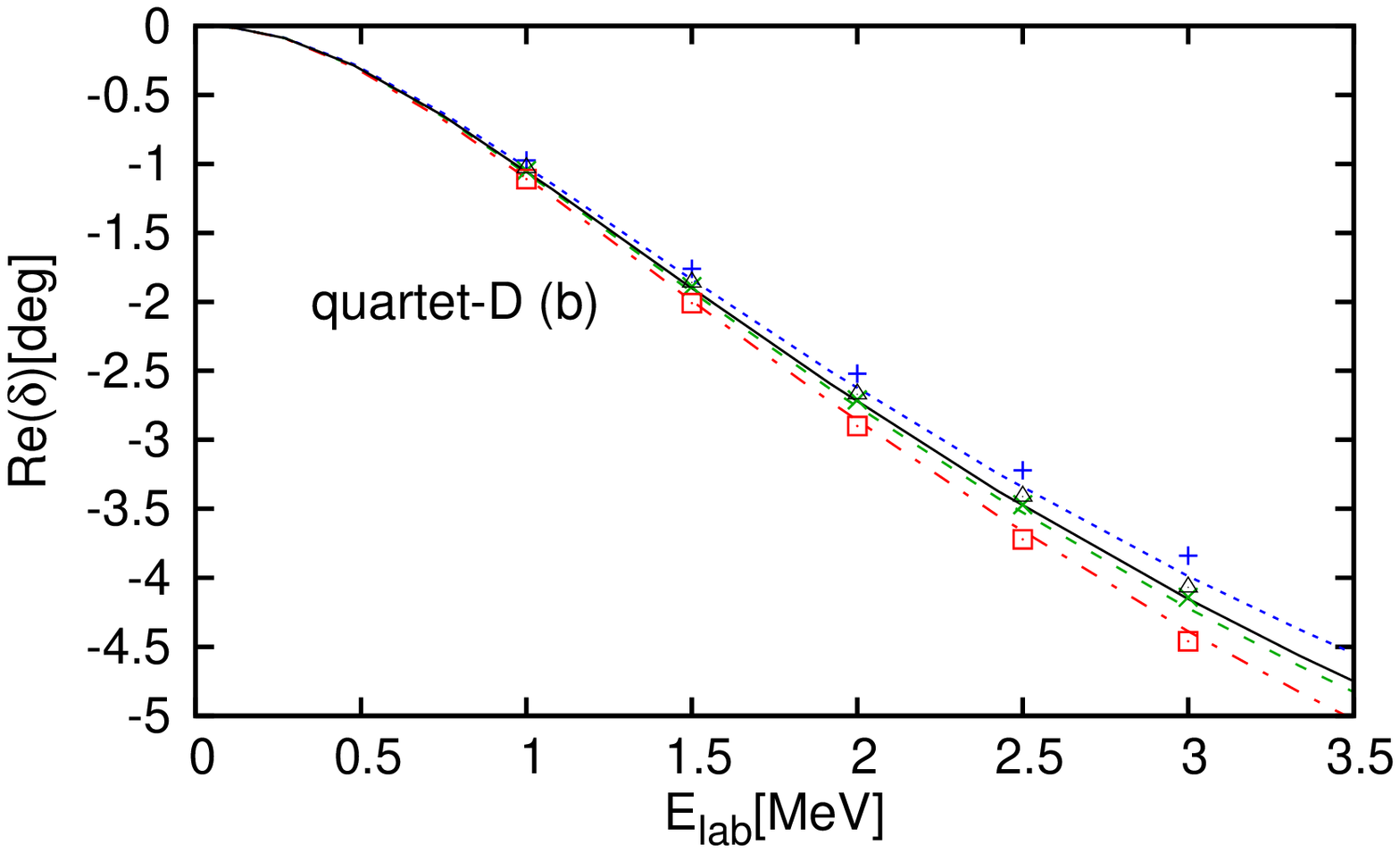}}

	\subfloat{\includegraphics[angle=-90,width=88mm]{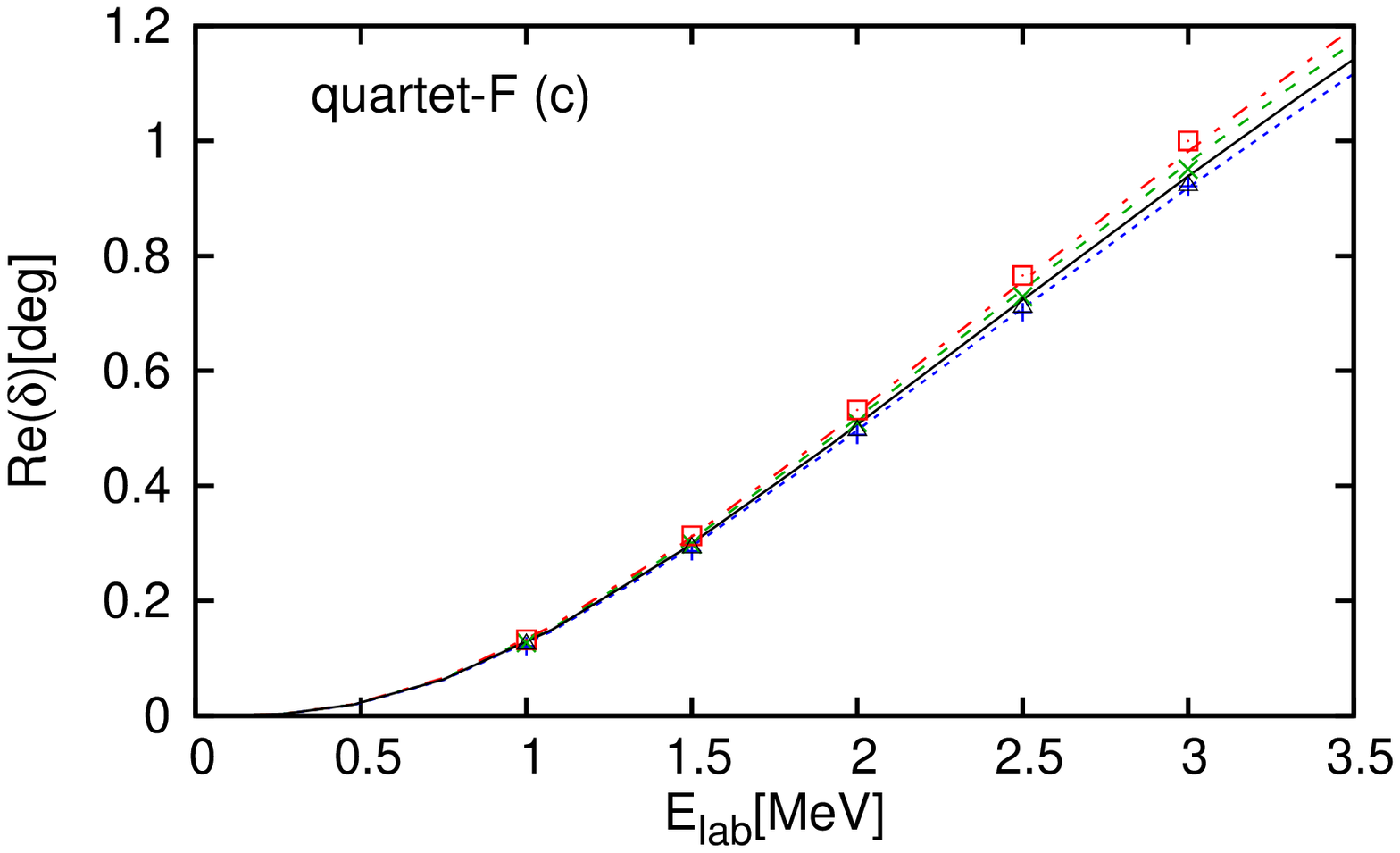}}
	\subfloat{\includegraphics[angle=-90,width=88mm]{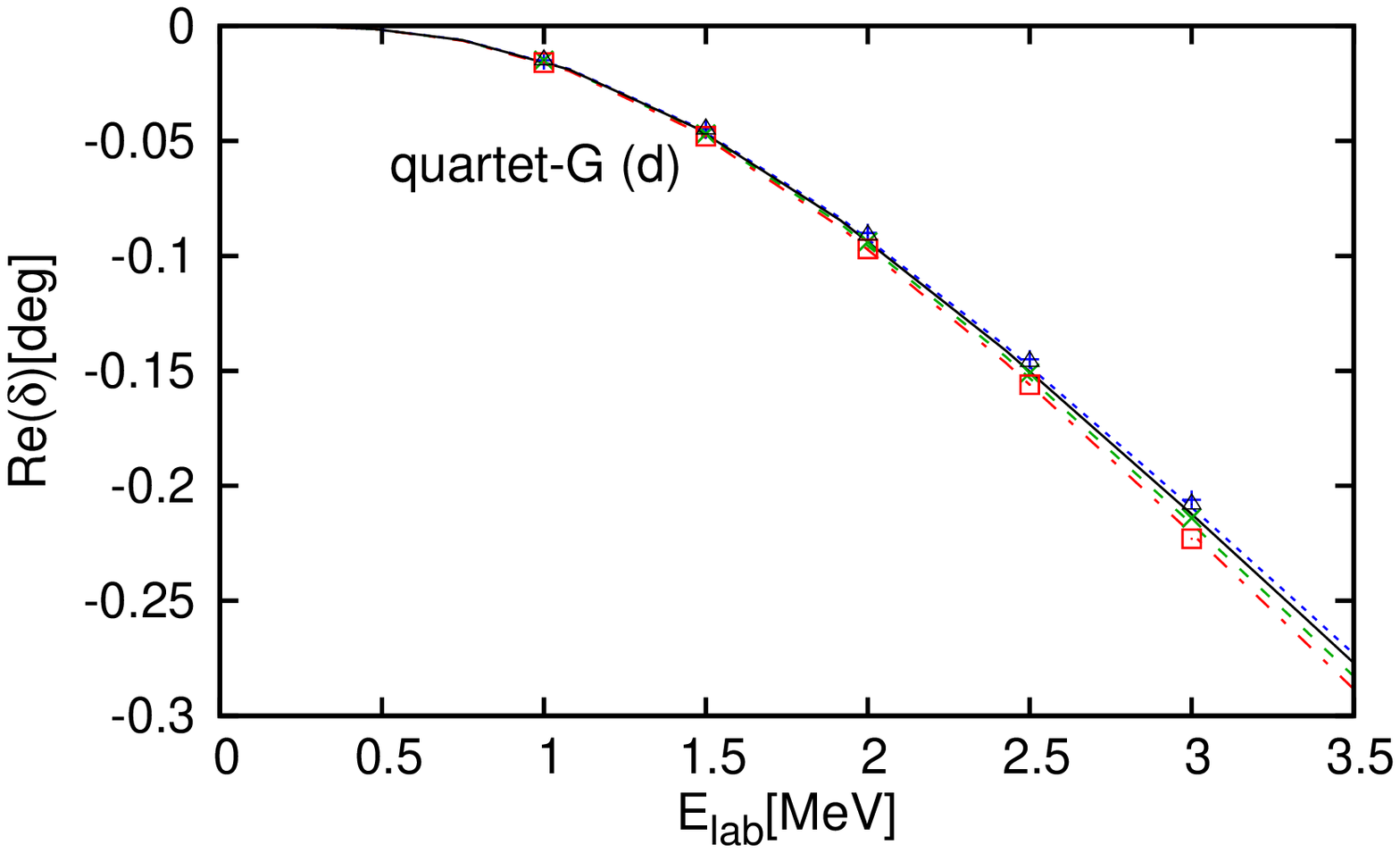}}

	\end{center}

	\caption{\label{fig:QuaretPhaseShiftsDBT}(Color online)$J$ splittings in higher partial waves below DBT. The dotted line is $l-\frac{3}{2}$, dashed line $l-\frac{1}{2}$, dot-dashed line $l+\frac{1}{2}$, and solid line $l+\frac{3}{2}$.  The points from AV18+UIX \cite{Kievsky:1996ca} are cross is $l-\frac{3}{2}$, star $l-\frac{1}{2}$, open square $l+\frac{1}{2}$, and open triangle $l+\frac{3}{2}$. $\Lambda=1600$ MeV }

\end{figure}

\begin{figure}[hbt]

	\begin{center}

	\subfloat{\includegraphics[angle=-90,width=88mm]{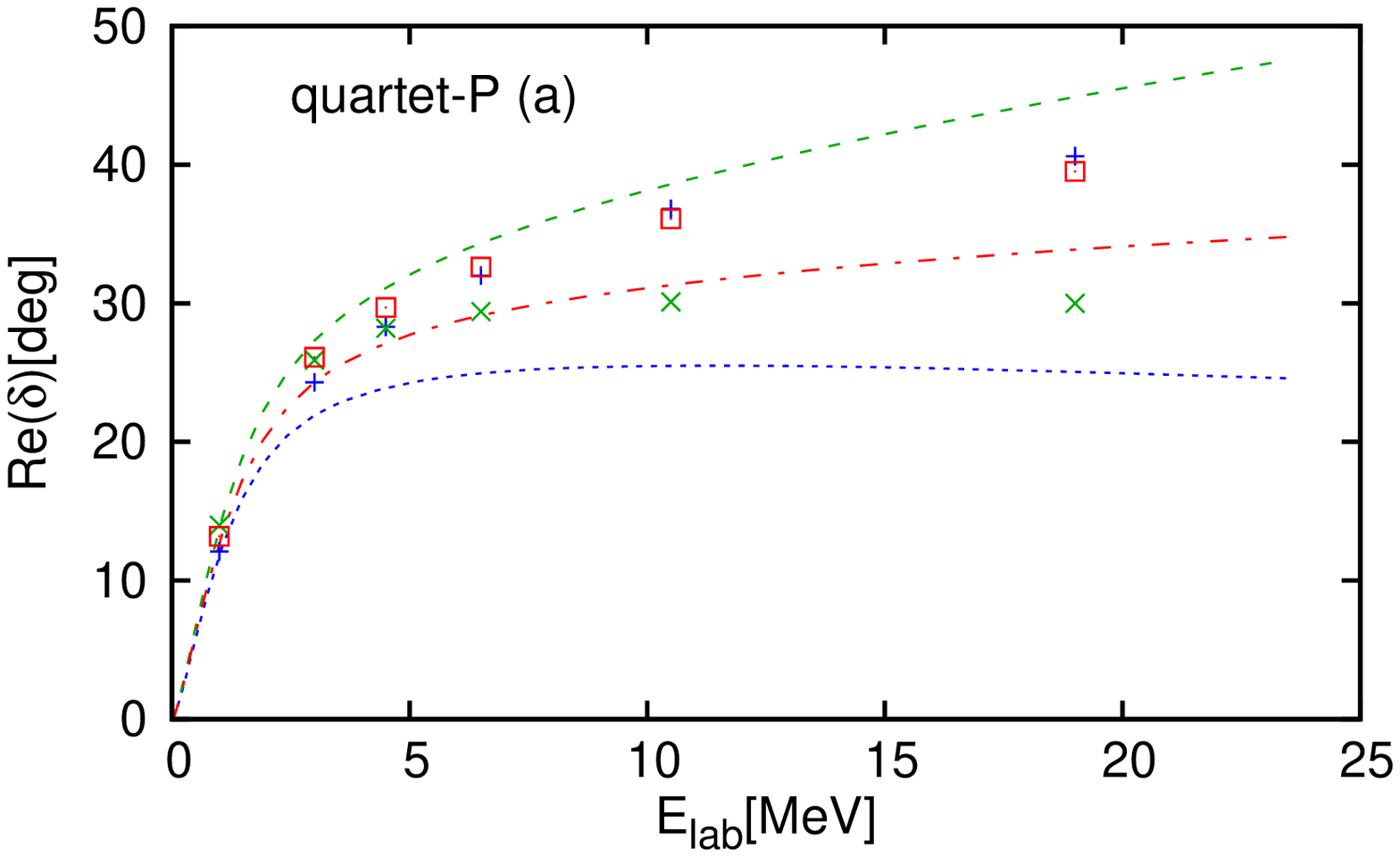}}
	\subfloat{\includegraphics[angle=-90,width=88mm]{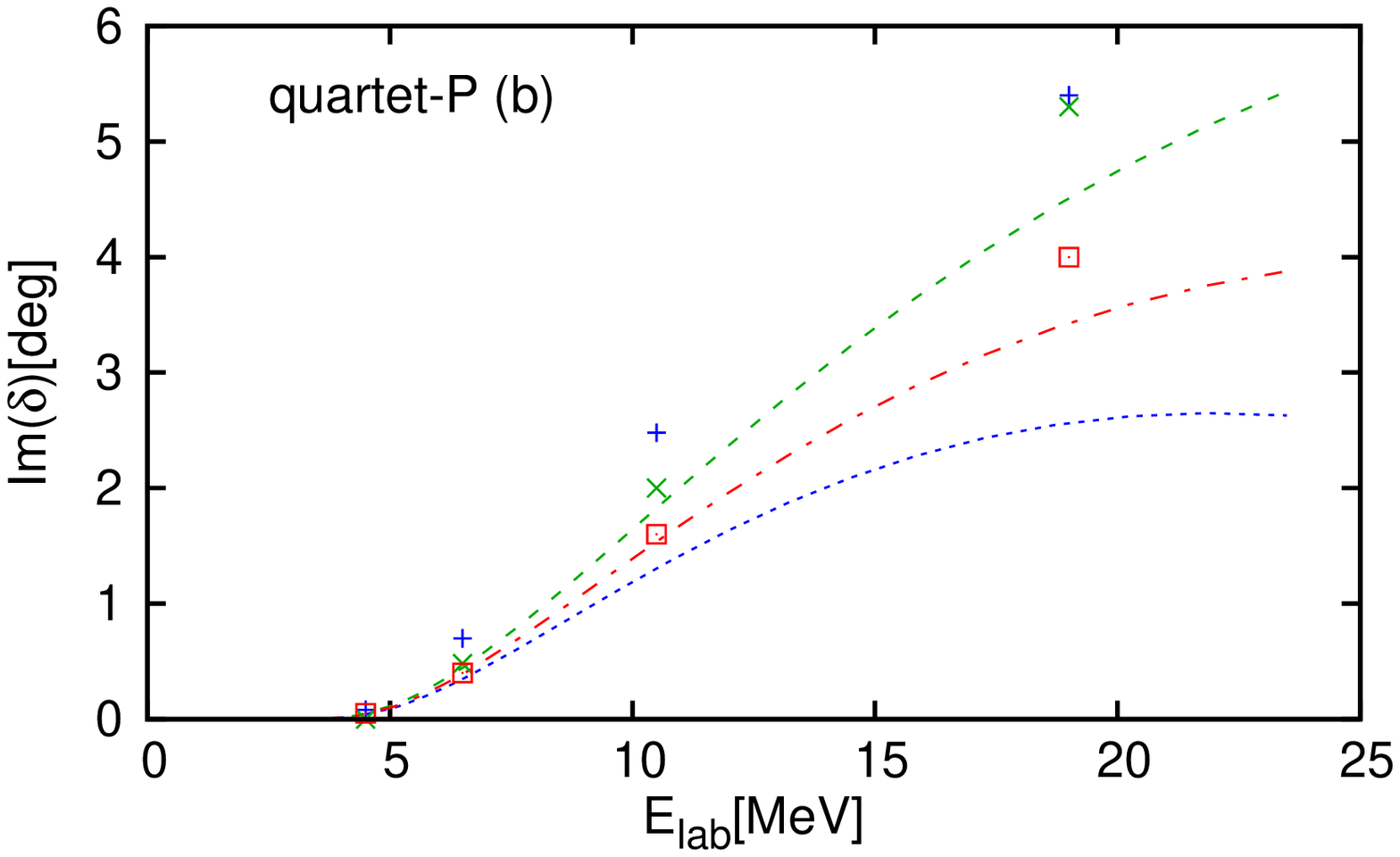}}
	\vspace{-.9cm}
      
	\subfloat{\includegraphics[angle=-90,width=88mm]{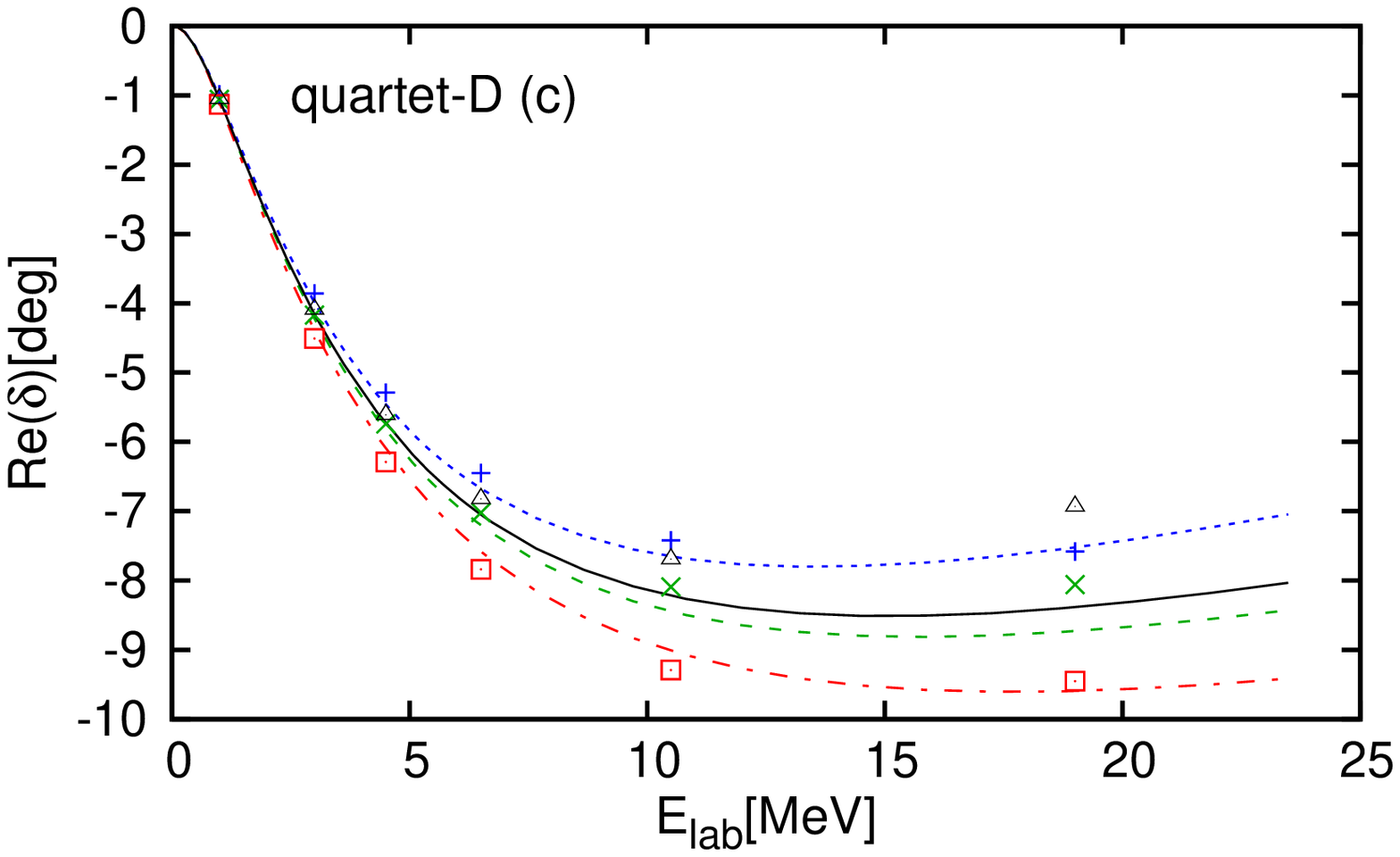}}
	\subfloat{\includegraphics[angle=-90,width=88mm]{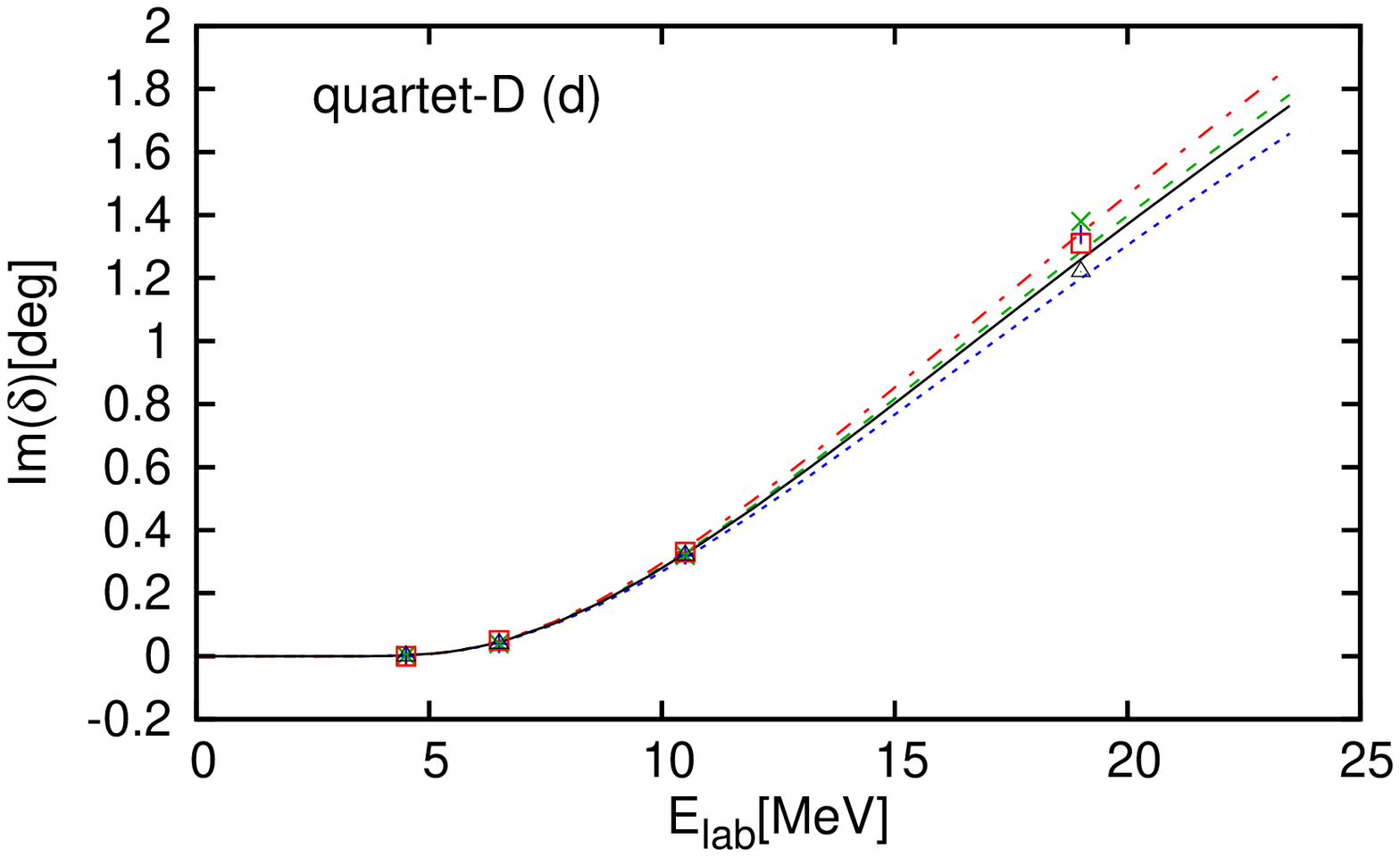}}
	\vspace{-.9cm}

	\subfloat{\includegraphics[angle=-90,width=88mm]{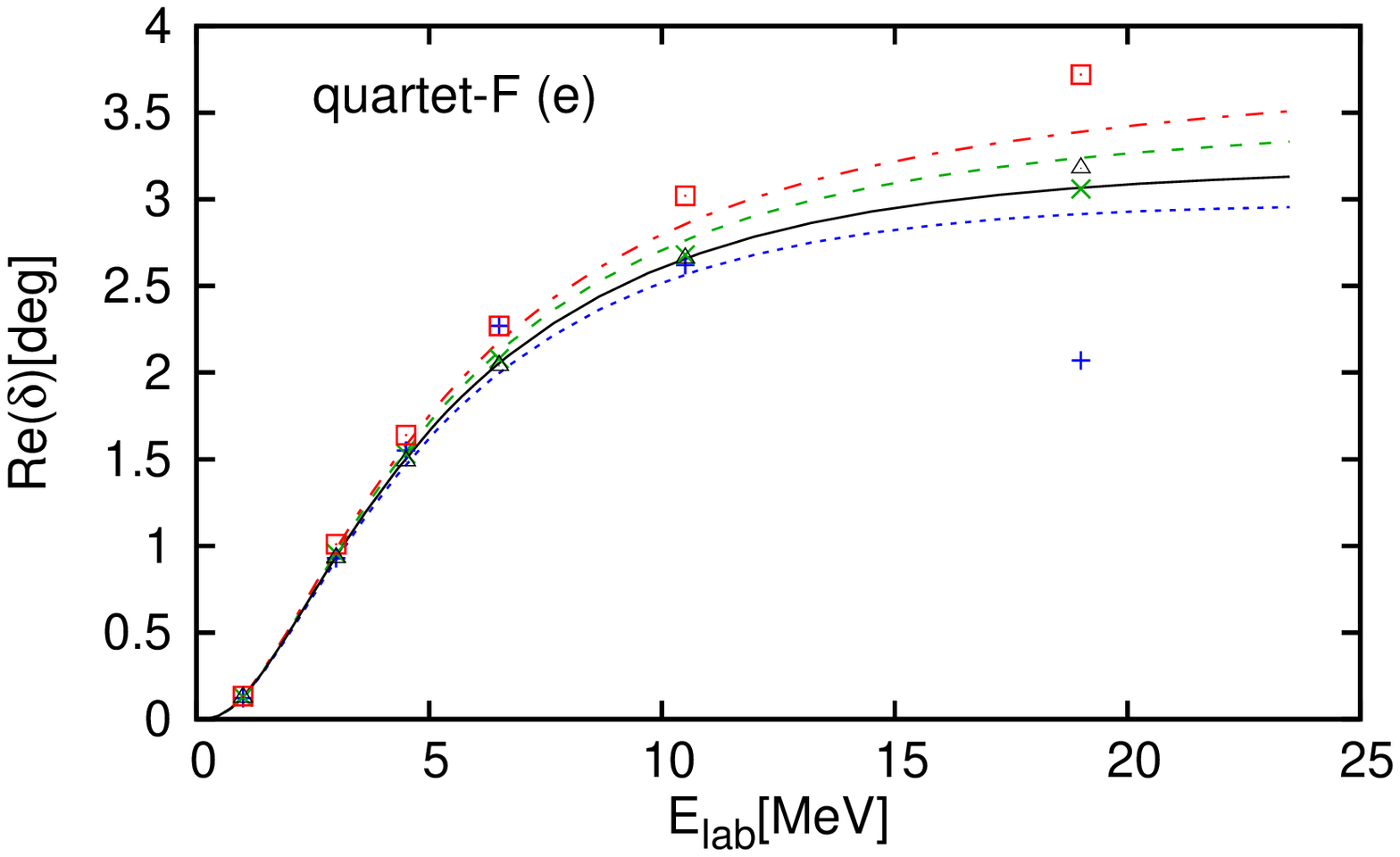}}
	\subfloat{\includegraphics[angle=-90,width=88mm]{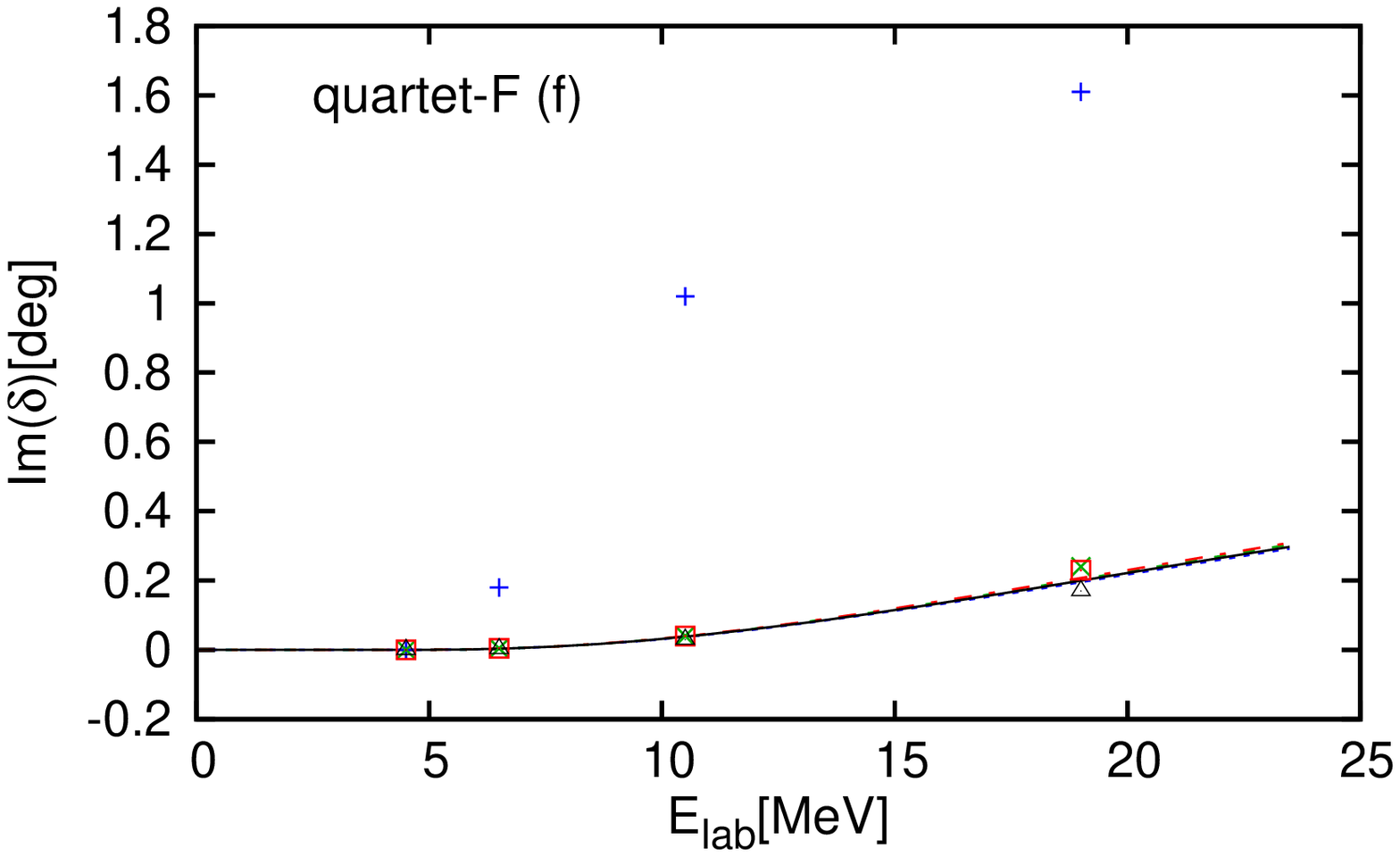}}
	\vspace{-.9cm}

	\subfloat{\includegraphics[angle=-90,width=88mm]{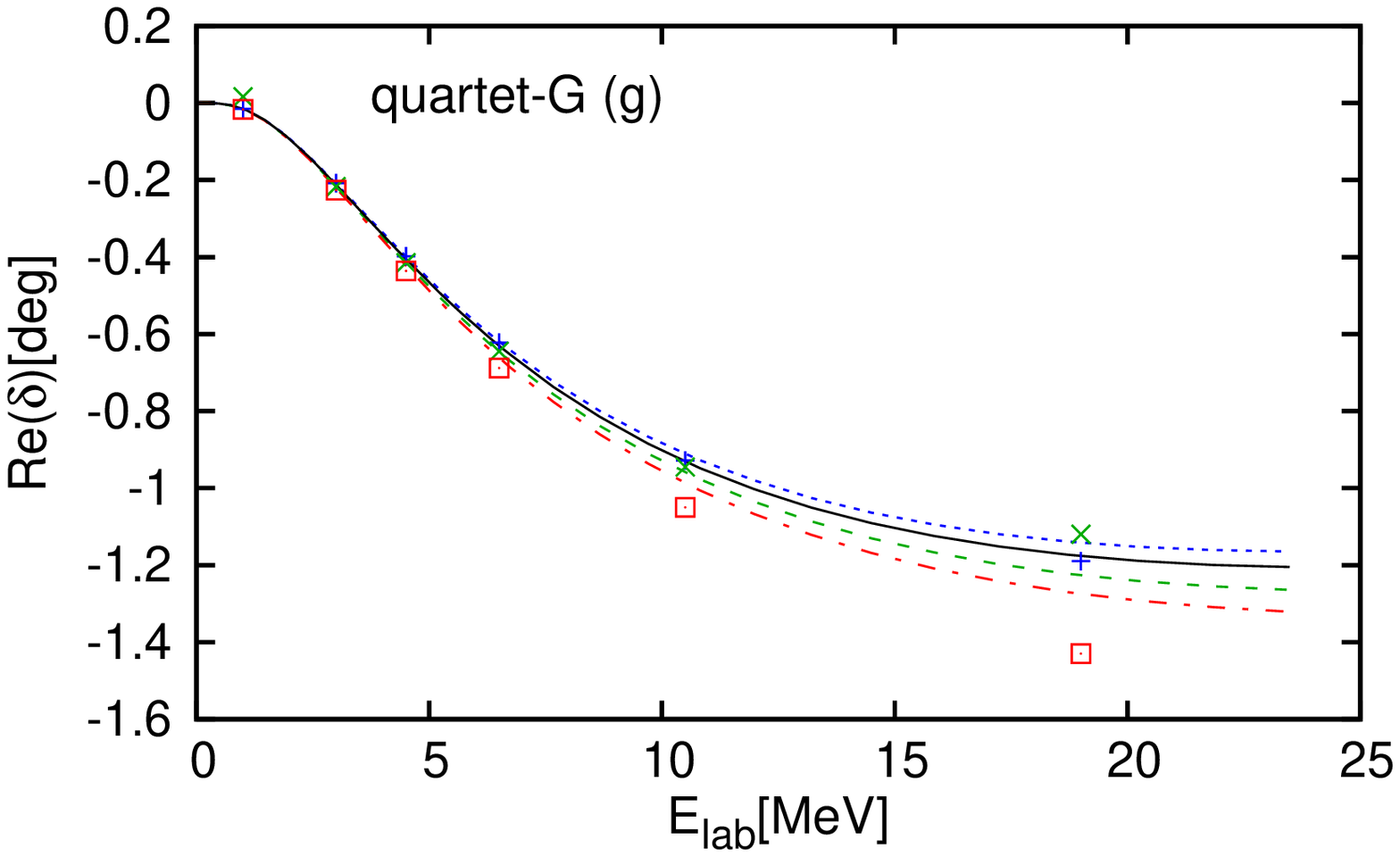}}
	\subfloat{\includegraphics[angle=-90,width=88mm]{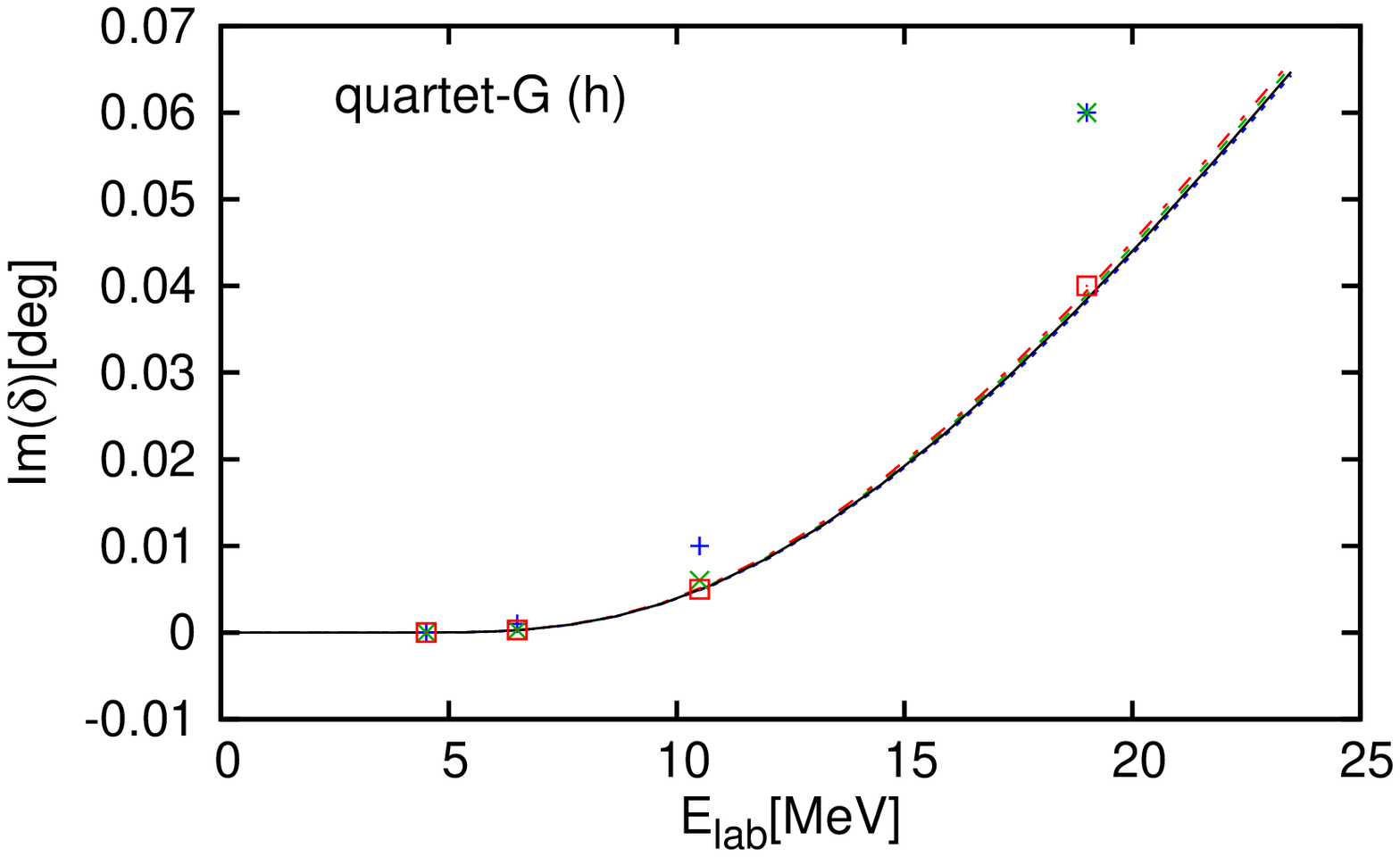}}
	
	\end{center}

	\caption{\label{fig:QuaretPhaseShifts}(Color online)$J$ splittings in higher partial waves below DBT. The dotted line is $l-\frac{3}{2}$, dashed line $l-\frac{1}{2}$, dot-dashed line $l+\frac{1}{2}$, and solid line $l+\frac{3}{2}$.  The points from Bonn-B \cite{Glockle:1995fb} are cross is $l-\frac{3}{2}$, star $l-\frac{1}{2}$, open square $l+\frac{1}{2}$, and open triangle $l+\frac{3}{2}$. $\Lambda=1600$ MeV }

\end{figure}

For higher partial waves in the quartet channel there will be splittings for different $J$ values due to the inclusion of the SD-mixing term at NNLO.  In Fig. \ref{fig:QuaretPhaseShiftsDBT} the eigenphases for P through G-waves are shown for each possible $J$ value and are compared to data from AV18+UIX below the DBT \cite{Kievsky:1996ca}. Agreement between the PMC is apparent for partial waves D through G both qualitatively and quantitatively.  The AV18+UIX data agree within the predicted $3\%$ error and the direction of the splittings is also consistent with our calculation.  However, for the P-wave there are notable discrepancies, namely the AV18+UIX data does not agree within the predicted $3\%$ for larger energies below the DBT or put simply our splitting appears too large.  Despite the large splittings there is still qualitative agreement in the direction of the splittings for different $J$ values.  For Fig \ref{fig:QuaretPhaseShifts} the eigenphases for P through G-waves are shown for energies above the DBT with data from the Bonn-B potential \cite{Glockle:1995fb}.  

The qualitative agreement between the PMC and our calculation at these higher energies is worse since the splittings in the PMC cross each other and we find no such effect.  At the higher energies the quantitative agreement with the Bonn-B data gets poorer (not within $3\%$) as one approaches the breakdown scale of \EFT.  One curious feature of the Bonn-B data is  the $J=\frac{3}{2}$  imaginary part of the F-wave eigenphase is much larger than all of the other $J$ values for the imaginary part of the F-wave eigenphases.  This is in clear contradiction with our results that show the splittings to all be of roughly the same size at this order in \EFT.  Note that the issues in the P-wave noted below the DBT are only further exacerbated above the DBT.

\begin{figure}[hbt]

	\begin{center}

	\subfloat{\includegraphics[angle=-90,width=88mm]{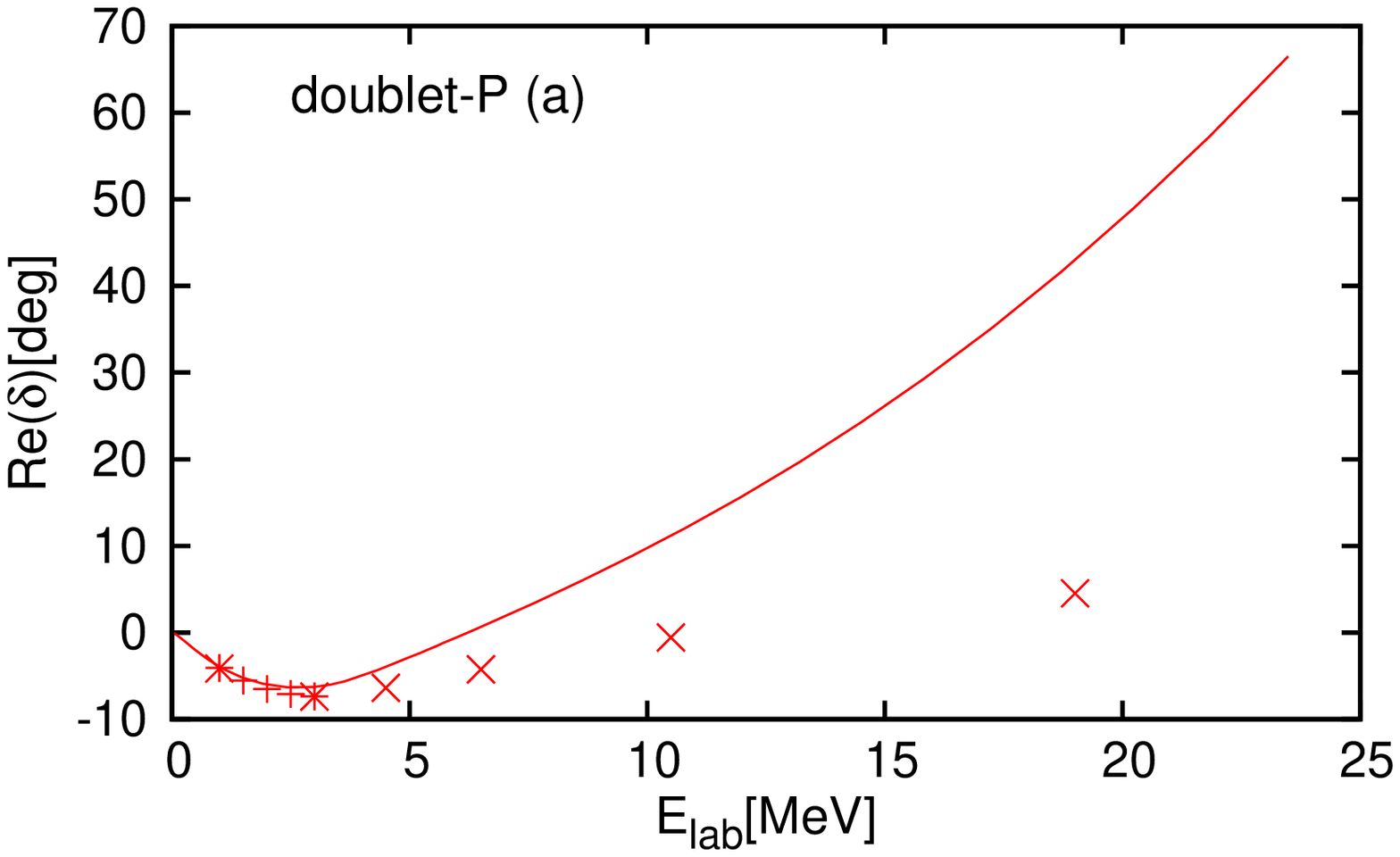}}
	\subfloat{\includegraphics[angle=-90,width=88mm]{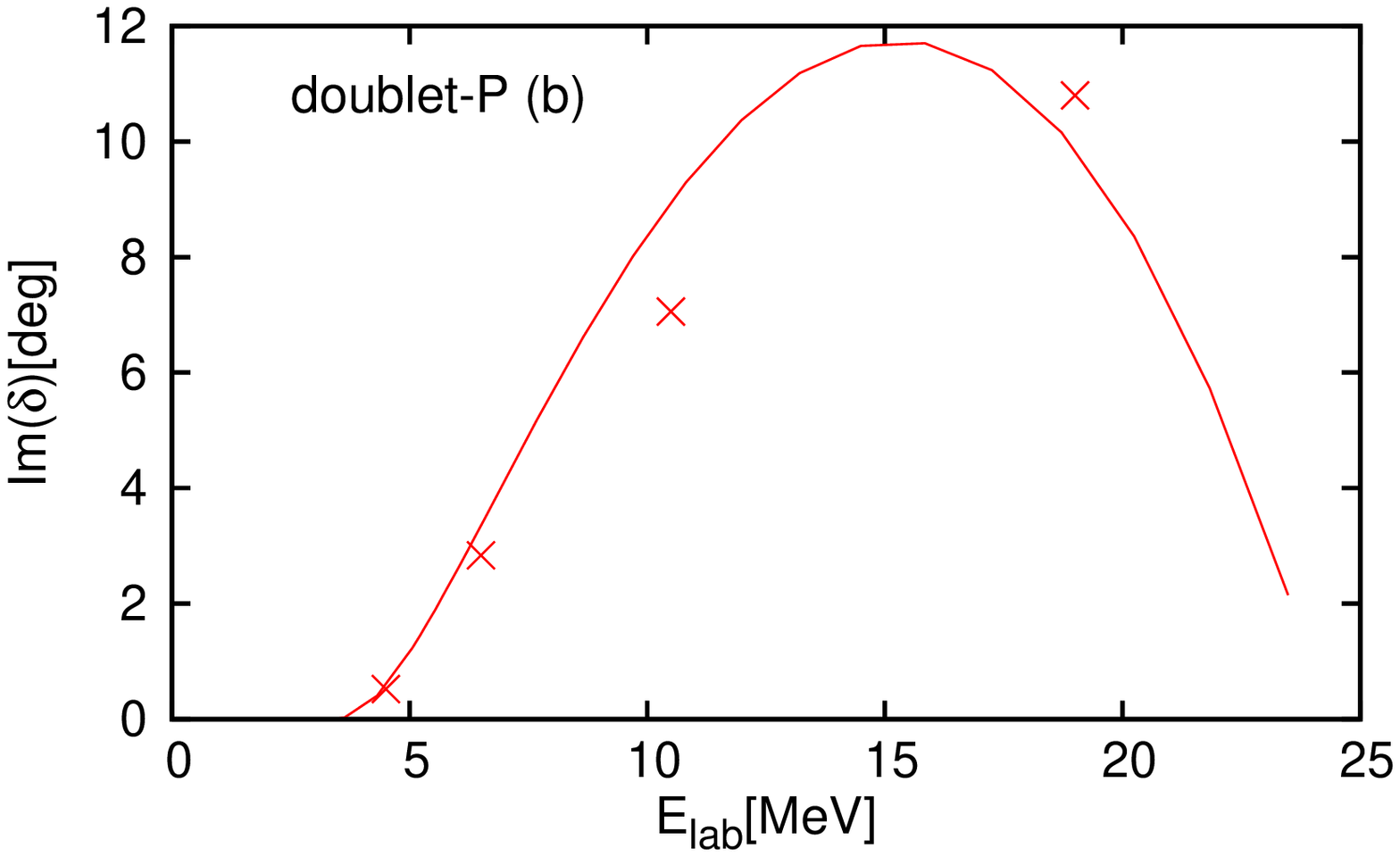}}
	\vspace{-.9cm}
      
	\subfloat{\includegraphics[angle=-90,width=88mm]{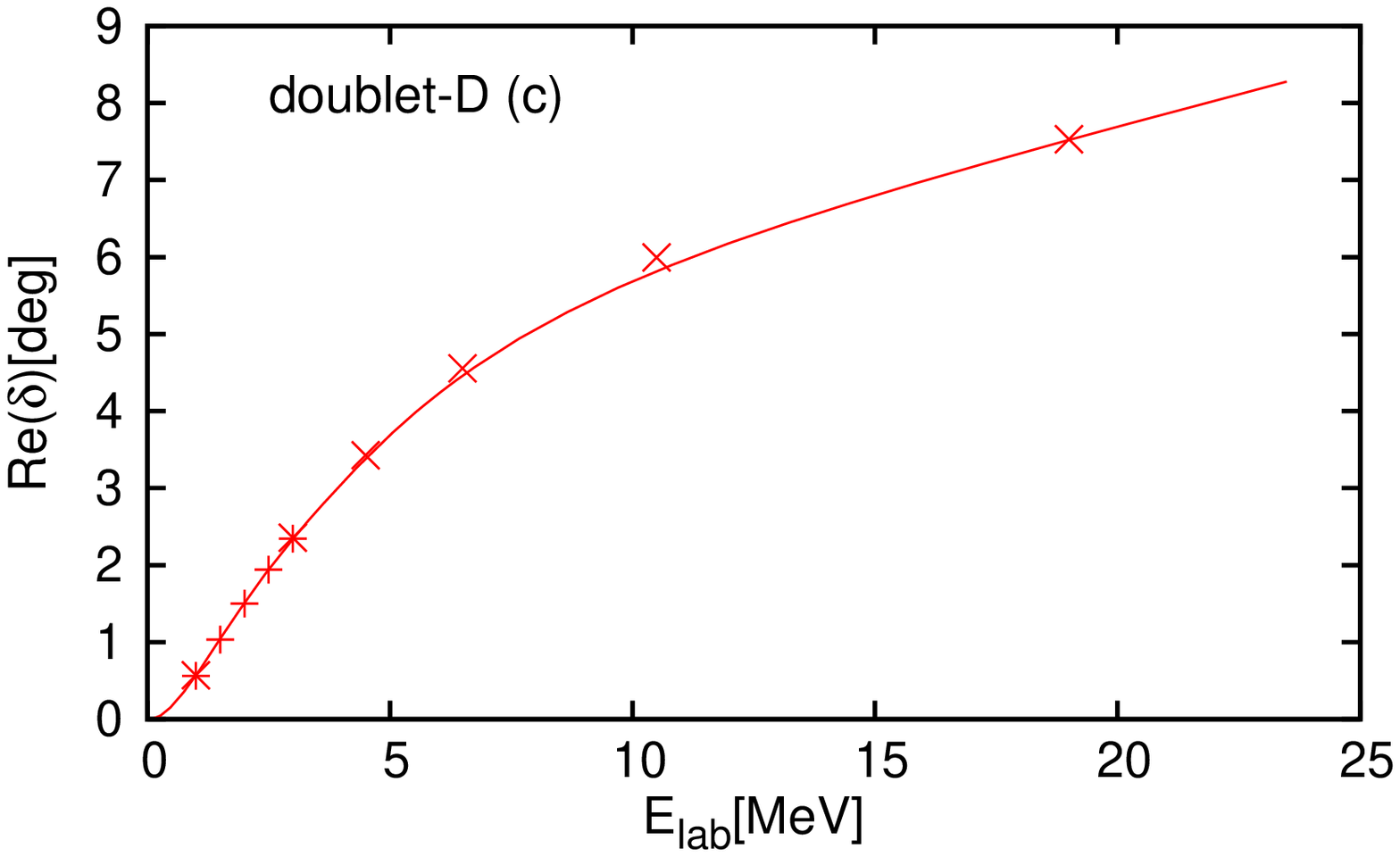}}
	\subfloat{\includegraphics[angle=-90,width=88mm]{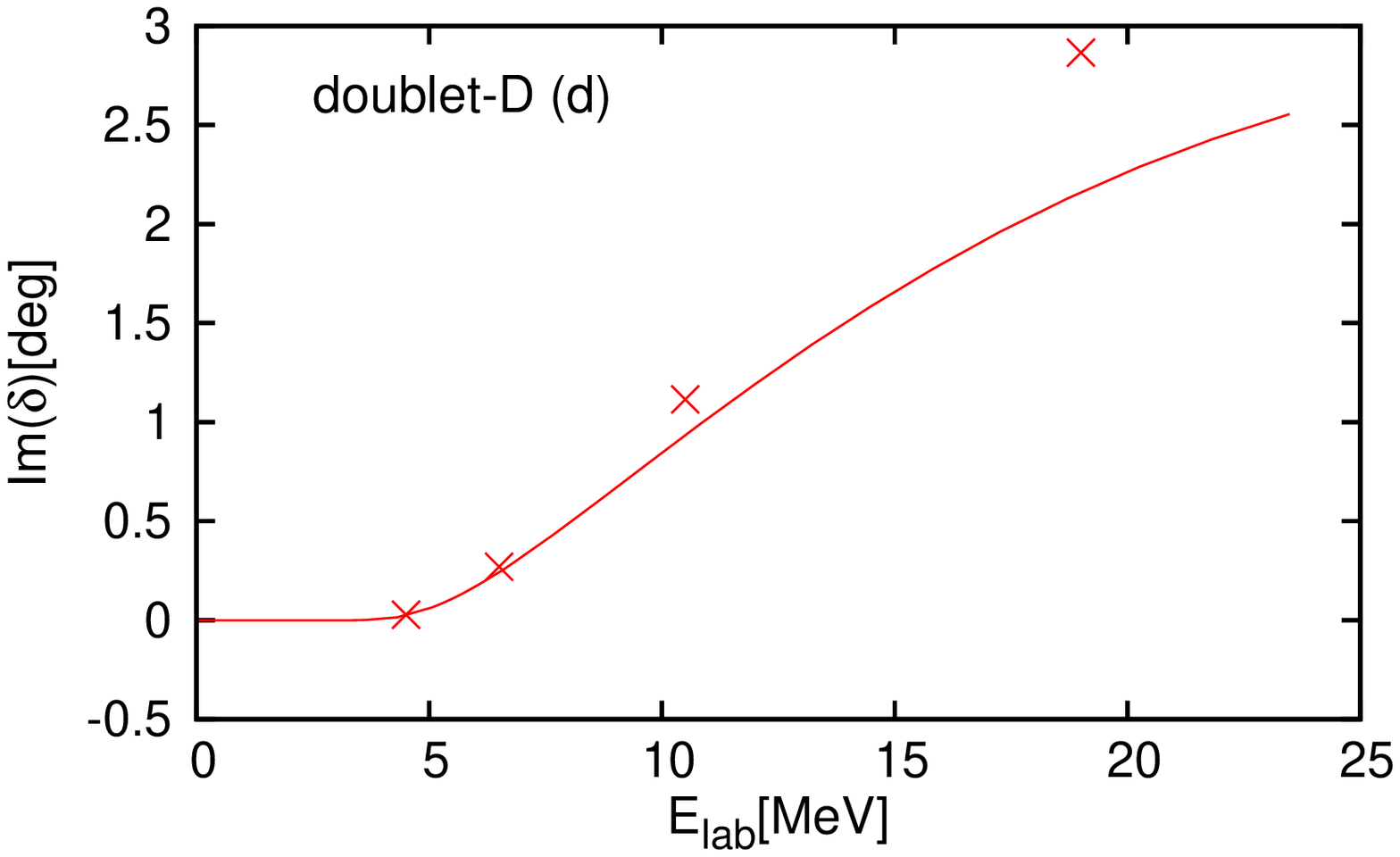}}
	\vspace{-.9cm}

	\subfloat{\includegraphics[angle=-90,width=88mm]{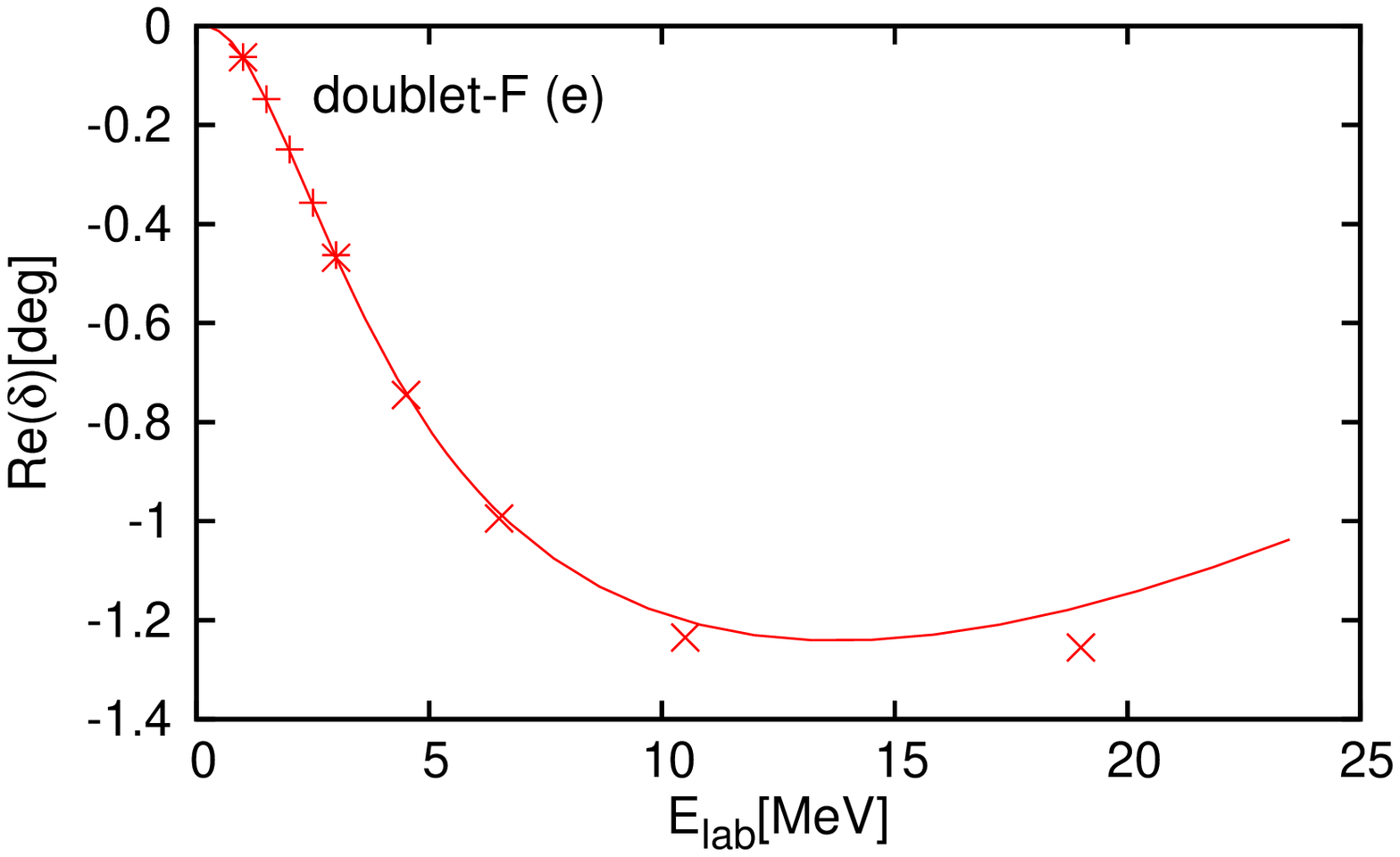}}
	\subfloat{\includegraphics[angle=-90,width=88mm]{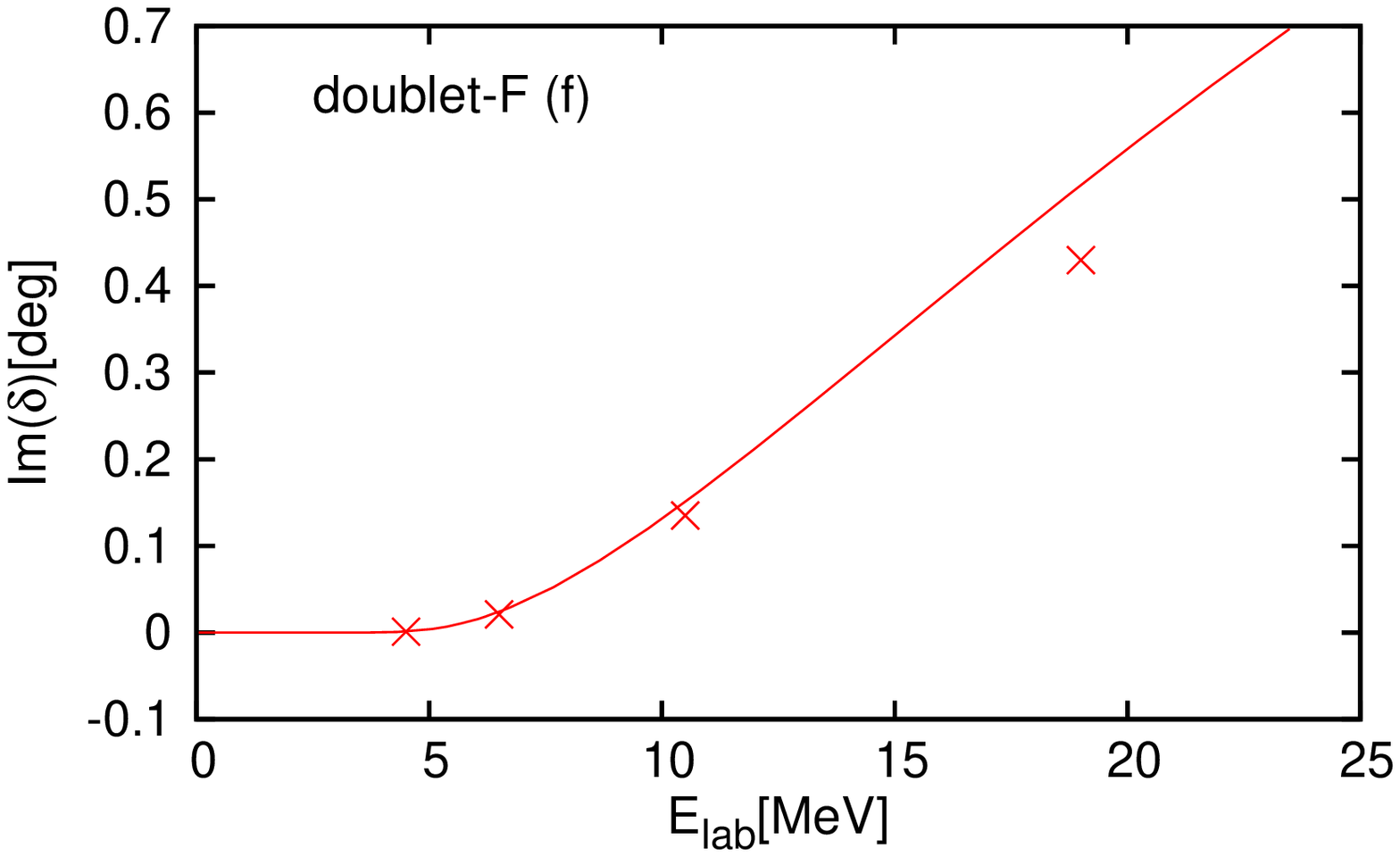}}
	\vspace{-.9cm}

	\subfloat{\includegraphics[angle=-90,width=88mm]{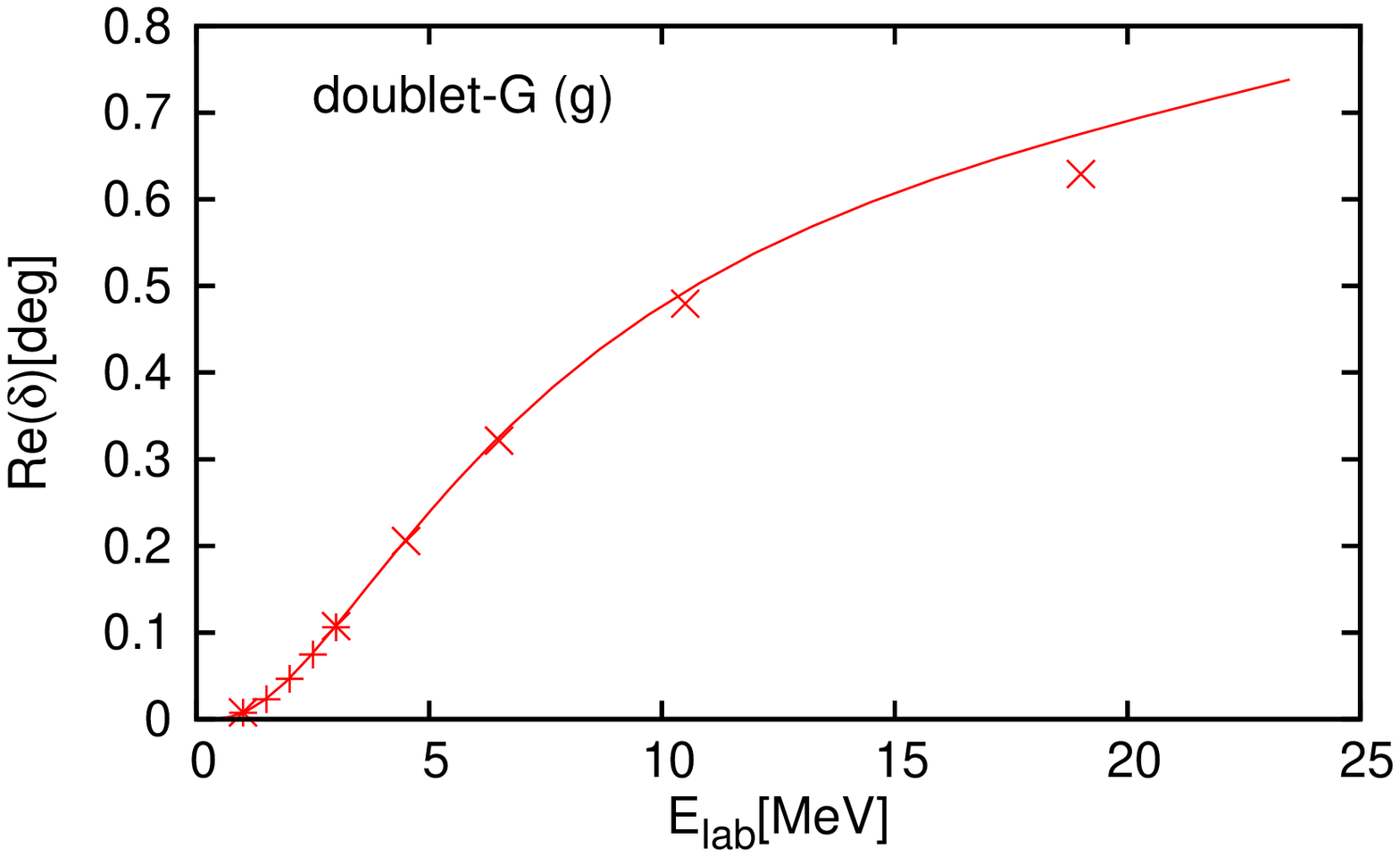}}
	\subfloat{\label{fig:AD_1}\includegraphics[angle=-90,width=88mm]{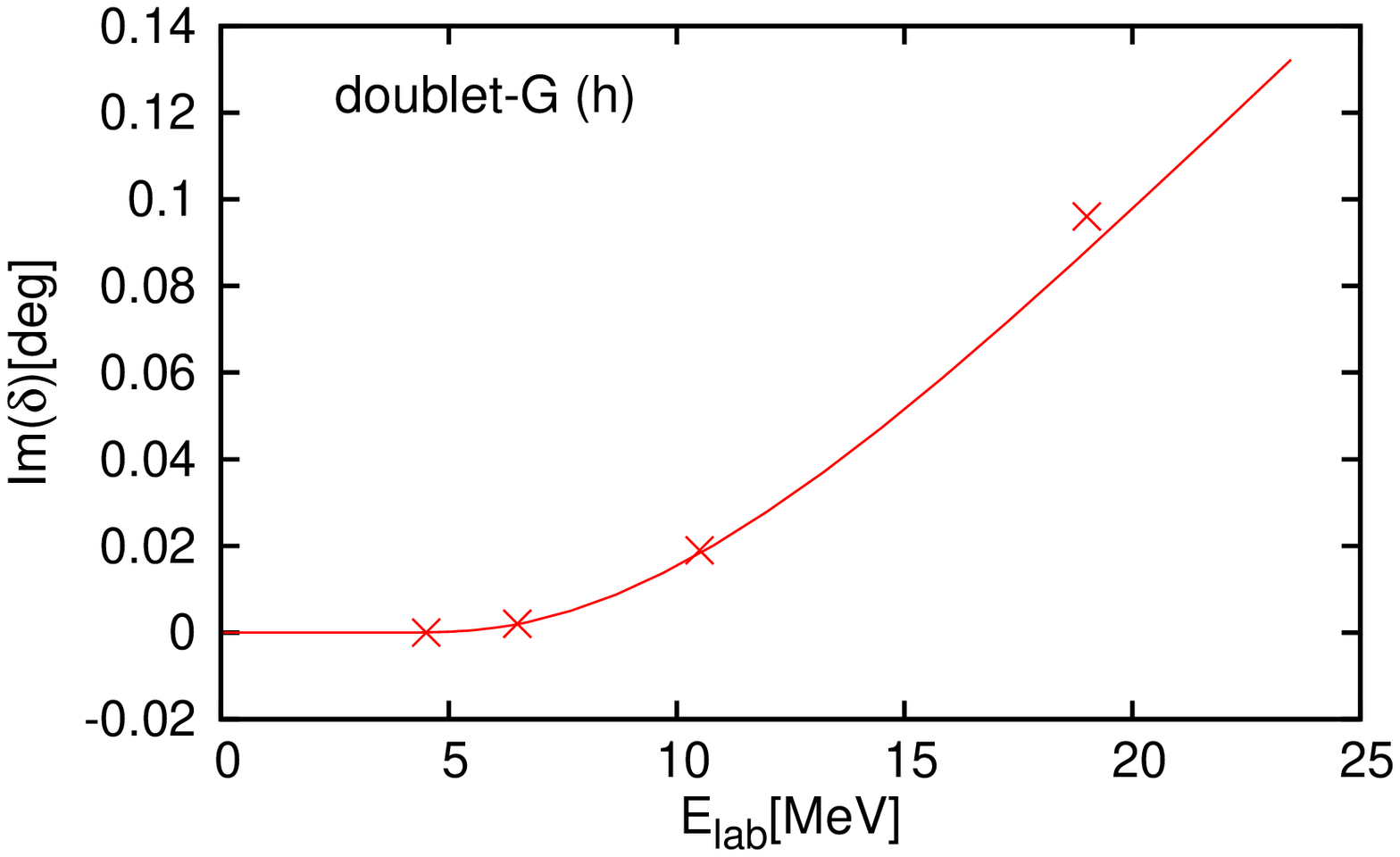}}
	\vspace{-.9cm}
	
	\end{center}

      	\caption{\label{fig:DoubletPhaseShifts}(Color online)Higher partial wave doublet phase shifts as function of lab energy real part (imaginary part) left (right) cutoff of $\Lambda=1600$ MeV.  Data shown below DBT is AV18+UIX (crosses) \cite{Kievsky:1996ca}, above and below DBT is Bonn-B (stars) \cite{Glockle:1995fb}.  Note average $((l+J)+(l-J))/2$ of data is taken}

\end{figure}

Higher partial waves for the doublet channel are shown in Fig. \ref{fig:DoubletPhaseShifts}.  At NNLO the SD-mixing terms do not cause splittings in the doublet channel thus the data from AV18+UIX  \cite{Kievsky:1996ca} below DBT and Bonn-B \cite{Glockle:1995fb} above and below DBT are averaged for all $J$ values to compare to our results.  For the D through G partial  waves good agreement is seen both for the real and imaginary parts within the expected $3\%$ accuracy.  However for the doublet P-wave there are apparent discrepancies above the DBT in agreement with previous results \cite{Griesshammer:2004pe}.  This may be due to the fact that  the rather strong two-body P-wave contributions occur at $\mathrm{N}^{3}\mathrm{LO}$ and their inclusion may resolve this discrepancy.  In addition this could explain the discrepancies observed in the splittings of the quartet P-waves.  Thus further study at higher orders is warranted.

\begin{figure}[hbt]

	\begin{center}

	\subfloat{\includegraphics[angle=-90,width=88mm]{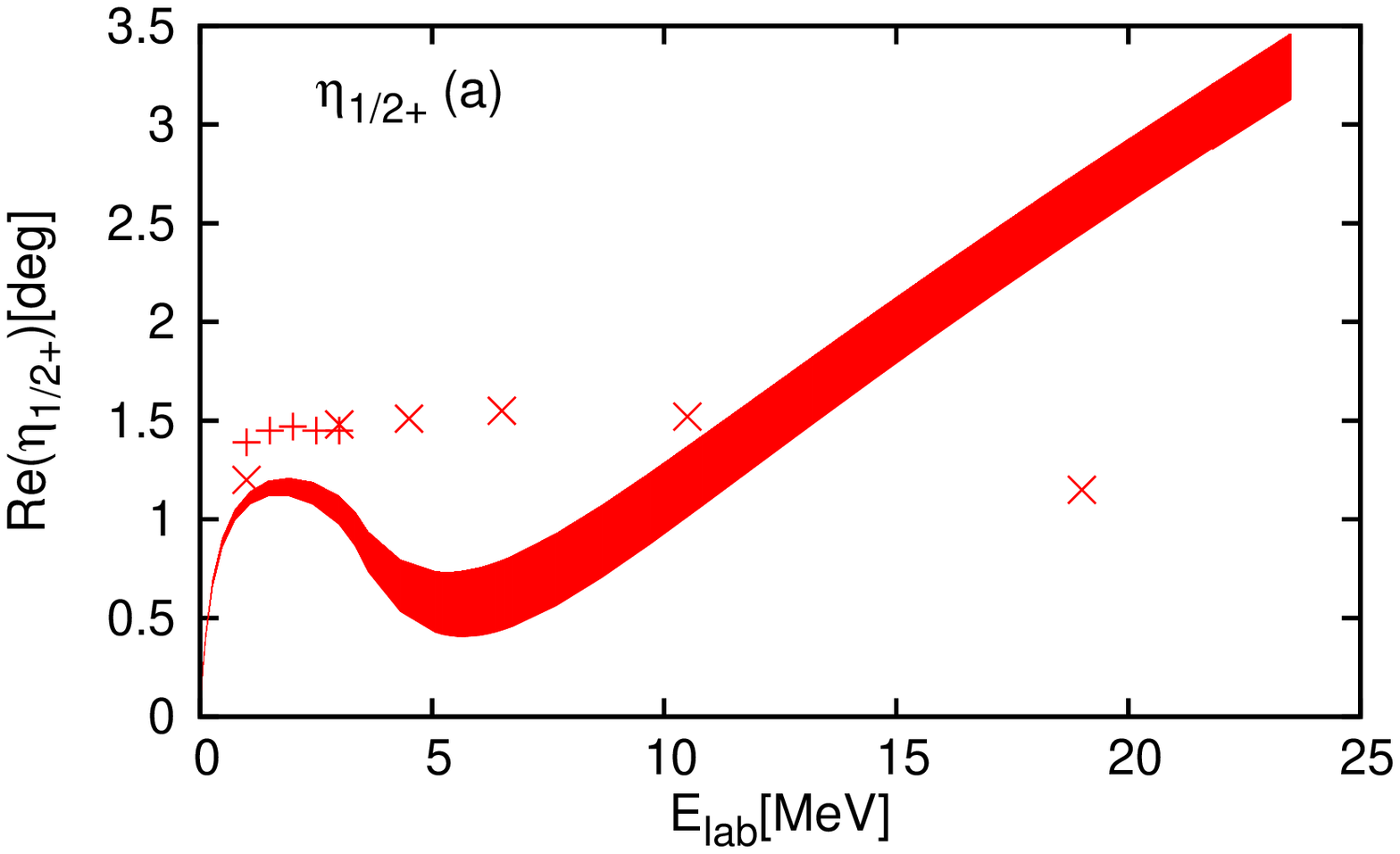}}
	\subfloat{\includegraphics[angle=-90,width=88mm]{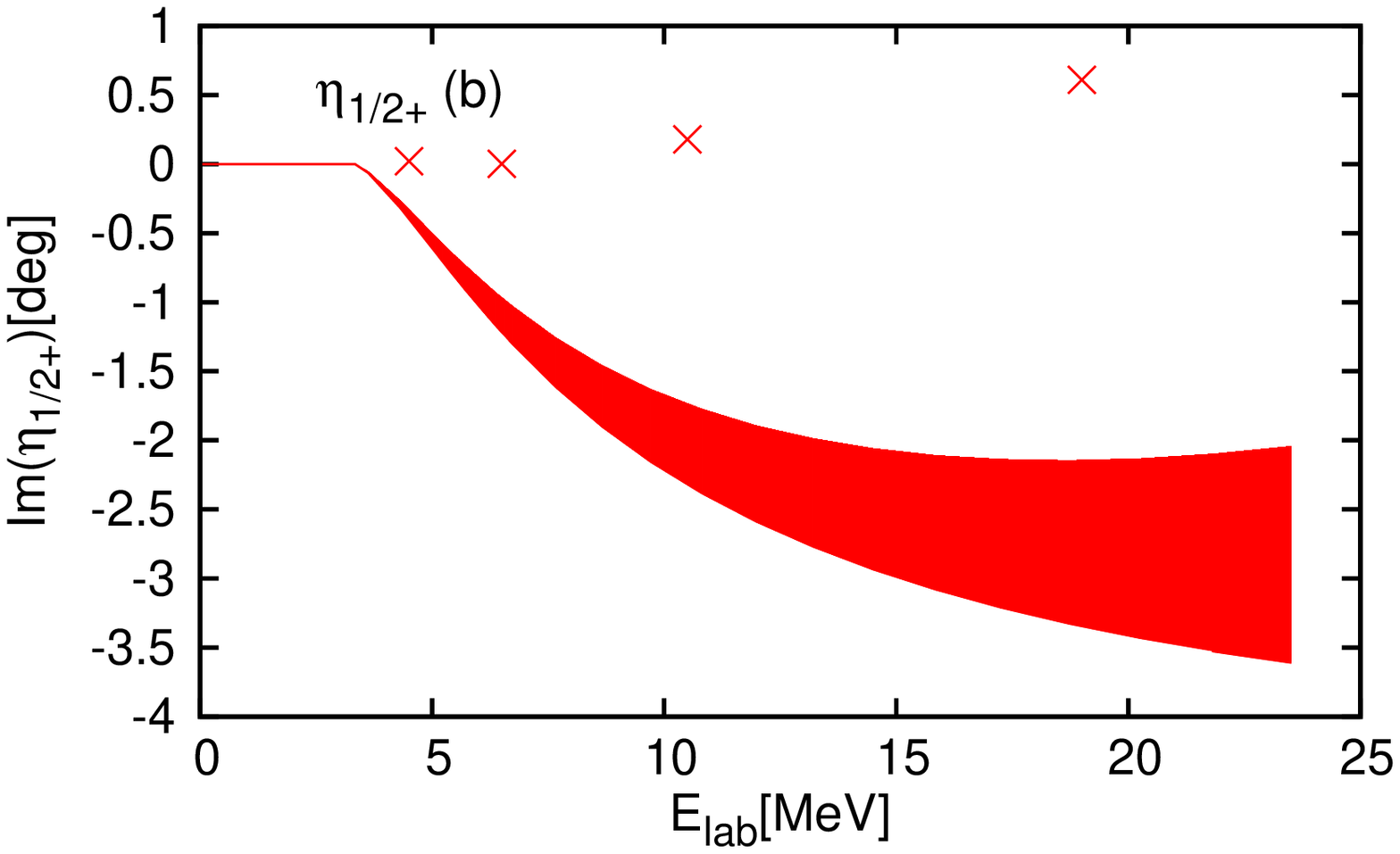}}
	\vspace{-.9cm}

	\subfloat{\includegraphics[angle=-90,width=88mm]{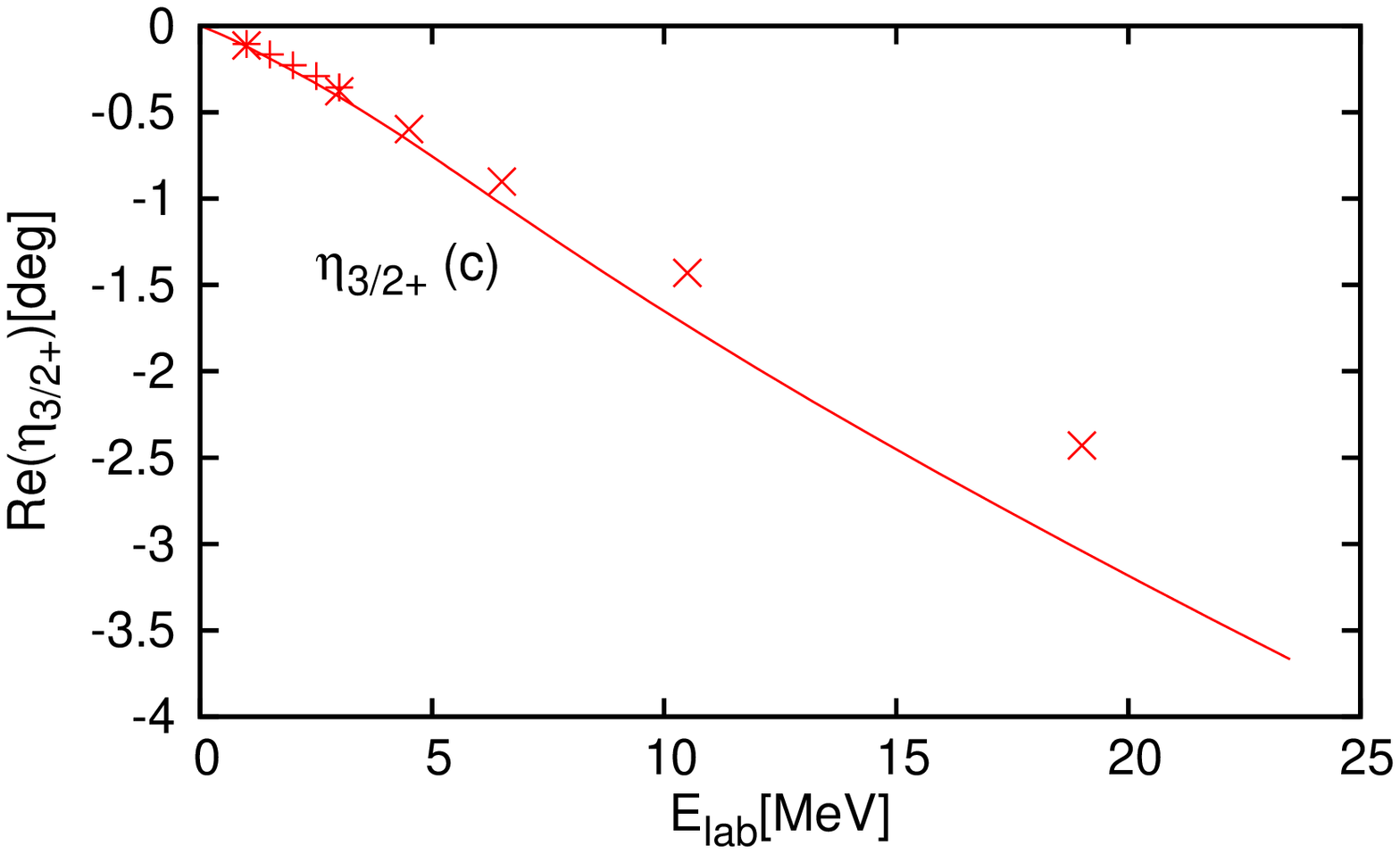}}
	\subfloat{\includegraphics[angle=-90,width=88mm]{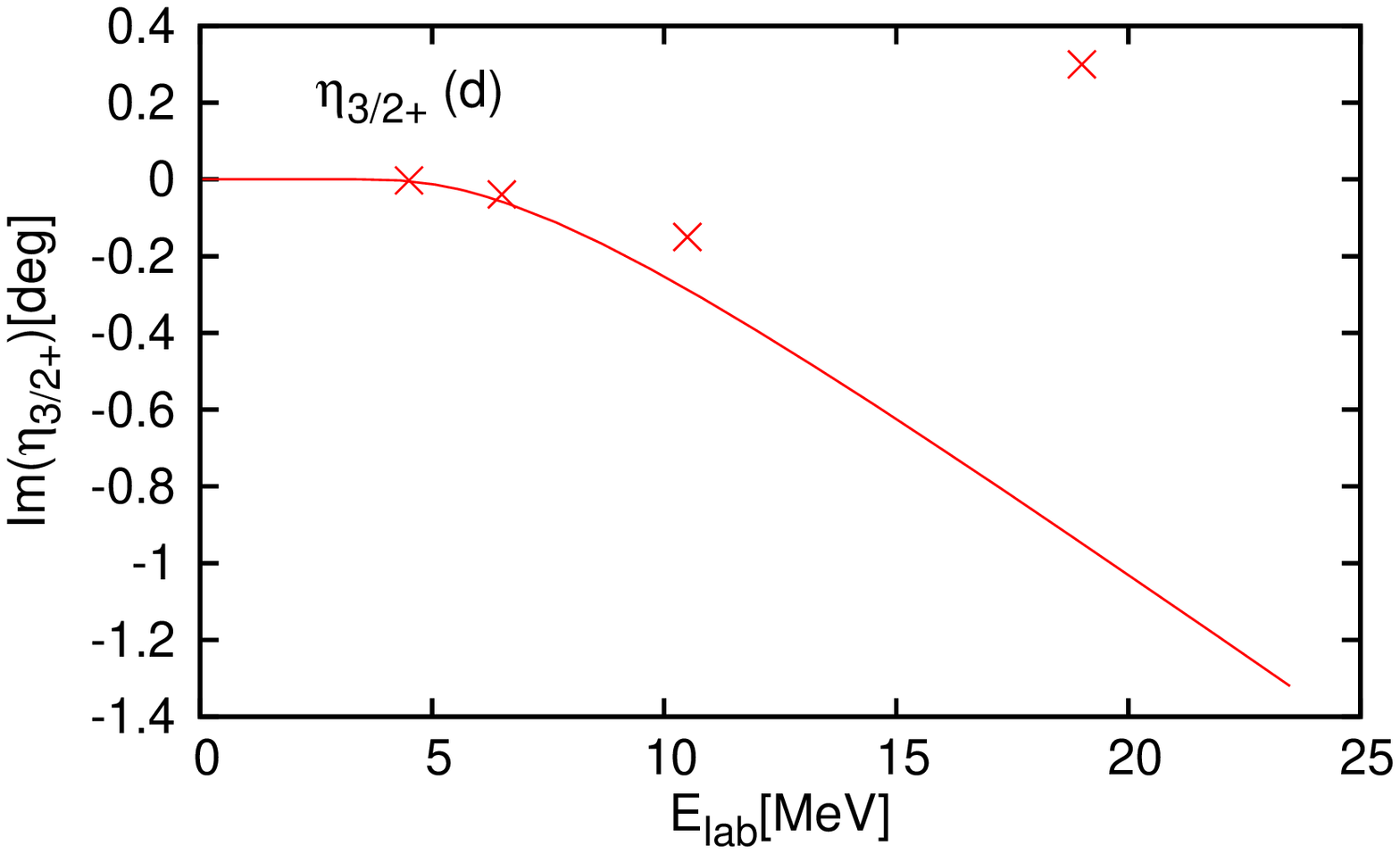}}
	\vspace{-.9cm}

	\subfloat{\includegraphics[angle=-90,width=88mm]{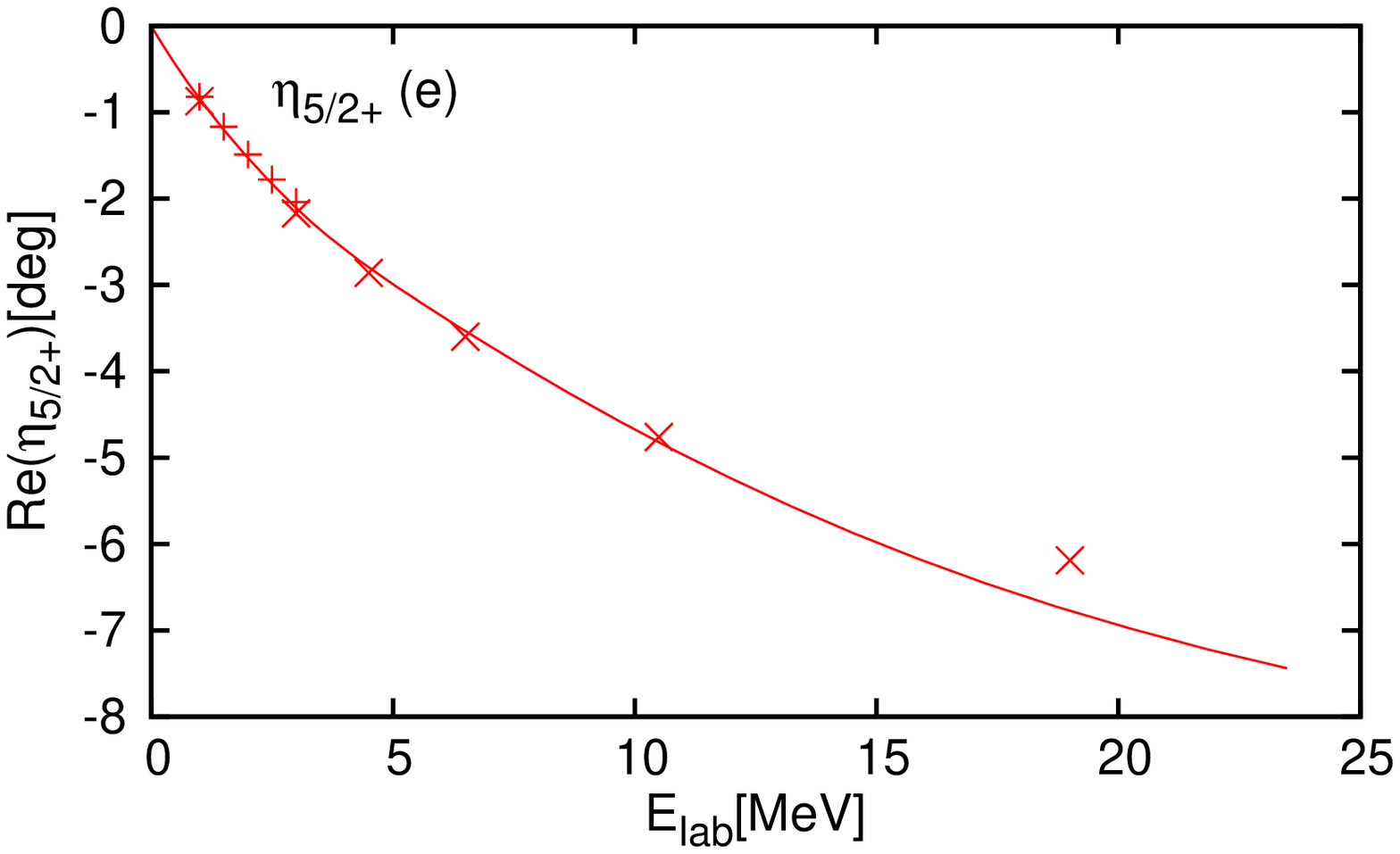}}
	\subfloat{\includegraphics[angle=-90,width=88mm]{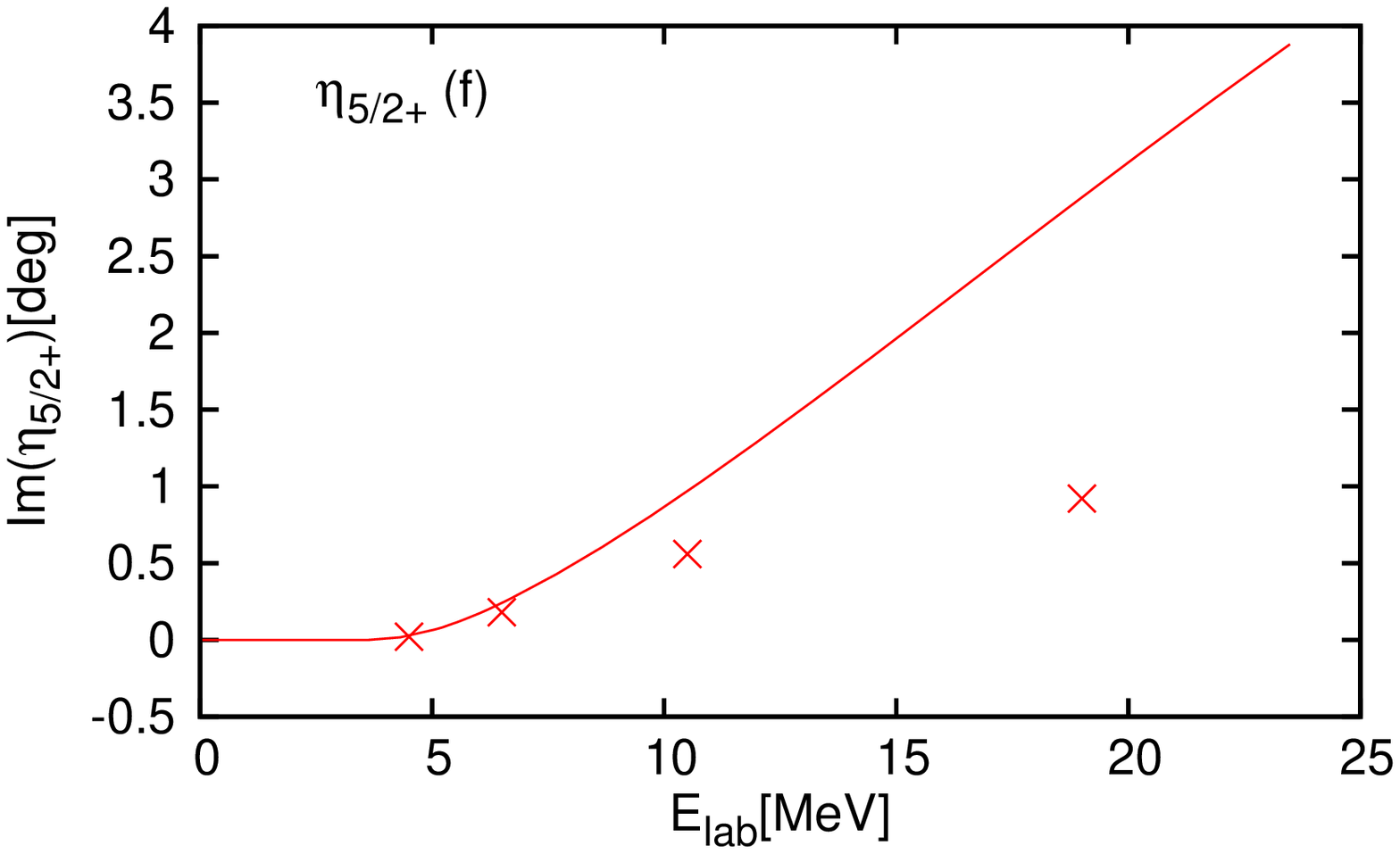}}
	\vspace{-.9cm}

	\subfloat{\includegraphics[angle=-90,width=88mm]{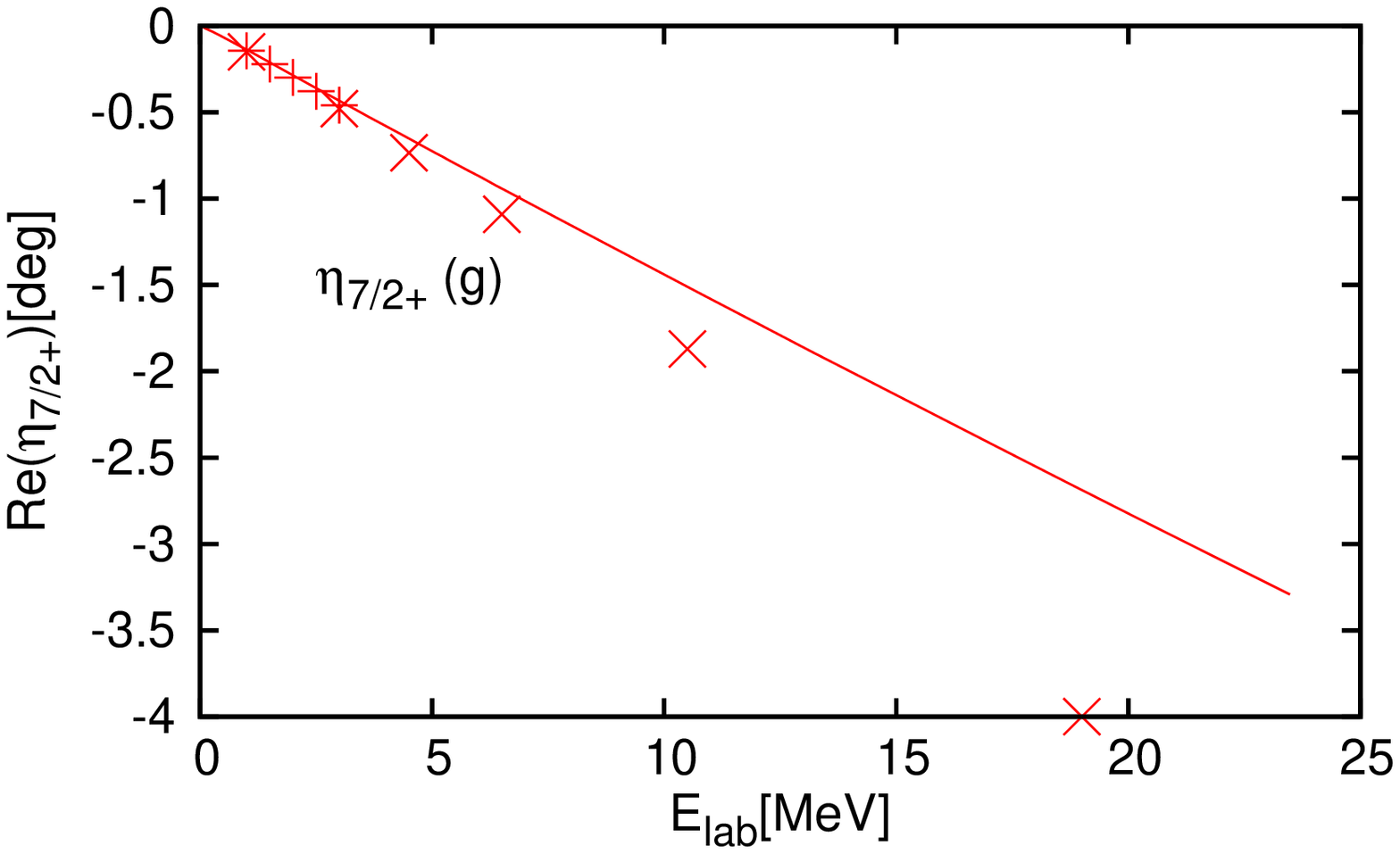}}
	\subfloat{\includegraphics[angle=-90,width=88mm]{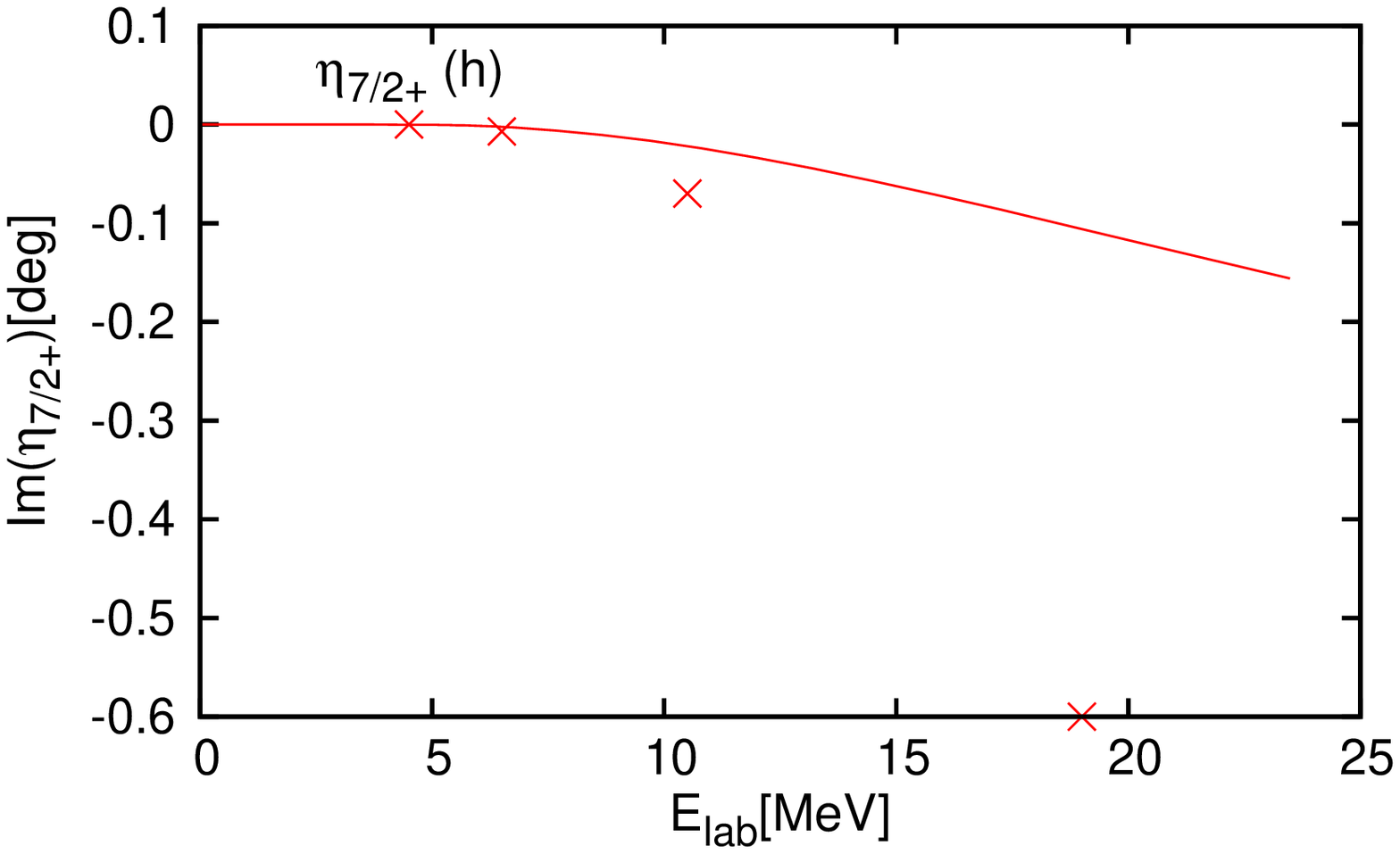}}
	
	\end{center}

	\caption{\label{fig:etap}(Color online)All $\eta^{J+}$ for $\Lambda=1600$ MeV left (right) real part (imaginary part).  $\eta^{\frac{1}{2}+}$ has cutoff variation from $\Lambda=200-1600$ MeV. Data below DBT (crosses) is AV18+UIX \cite{Kievsky:1996ca}, above and below DBT (stars) is Bonn-B \cite{Glockle:1995fb}}

\end{figure}

\begin{figure}[hbt]

	\begin{center}

	\subfloat{\includegraphics[angle=-90,width=88mm]{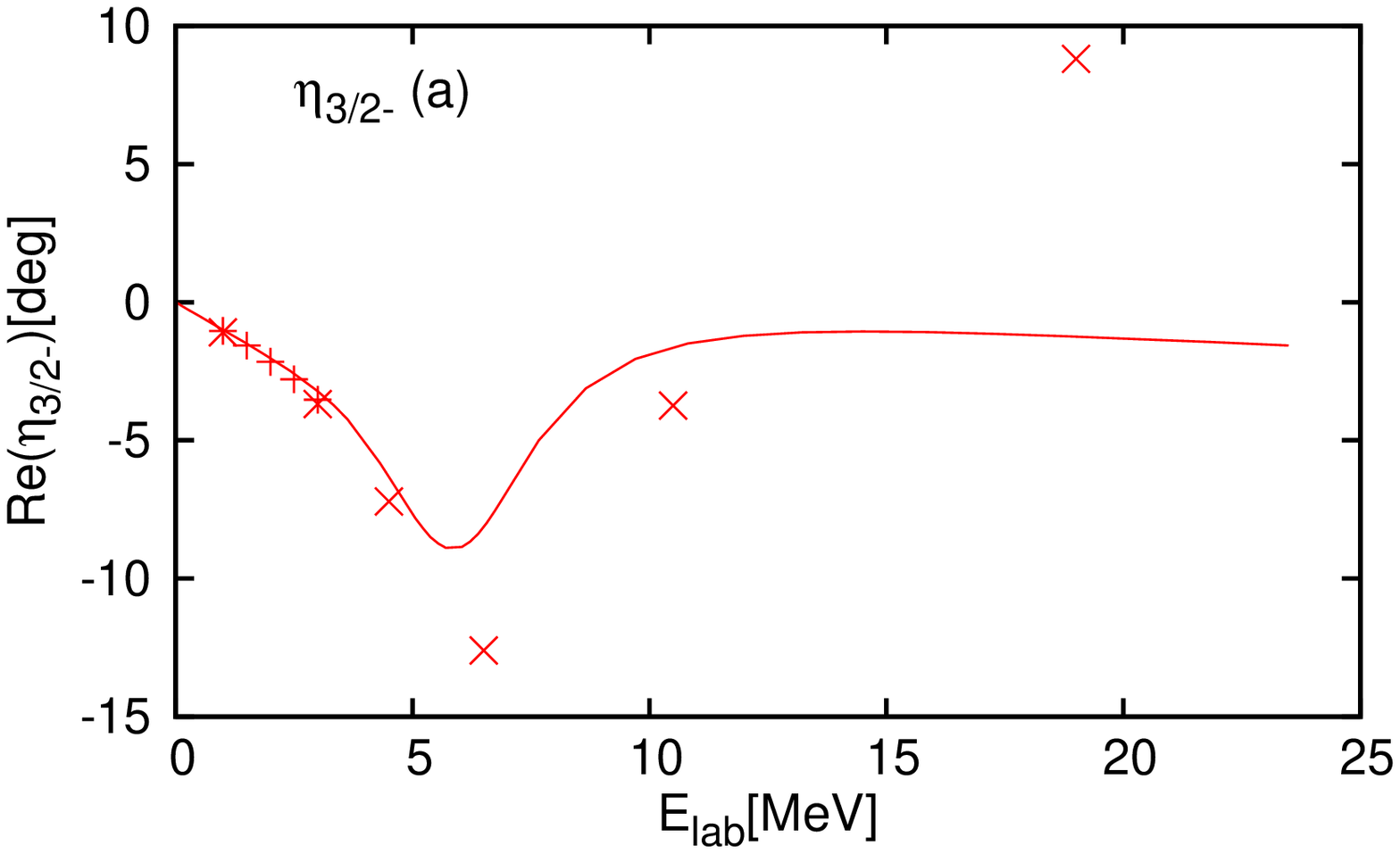}}
	\subfloat{\includegraphics[angle=-90,width=88mm]{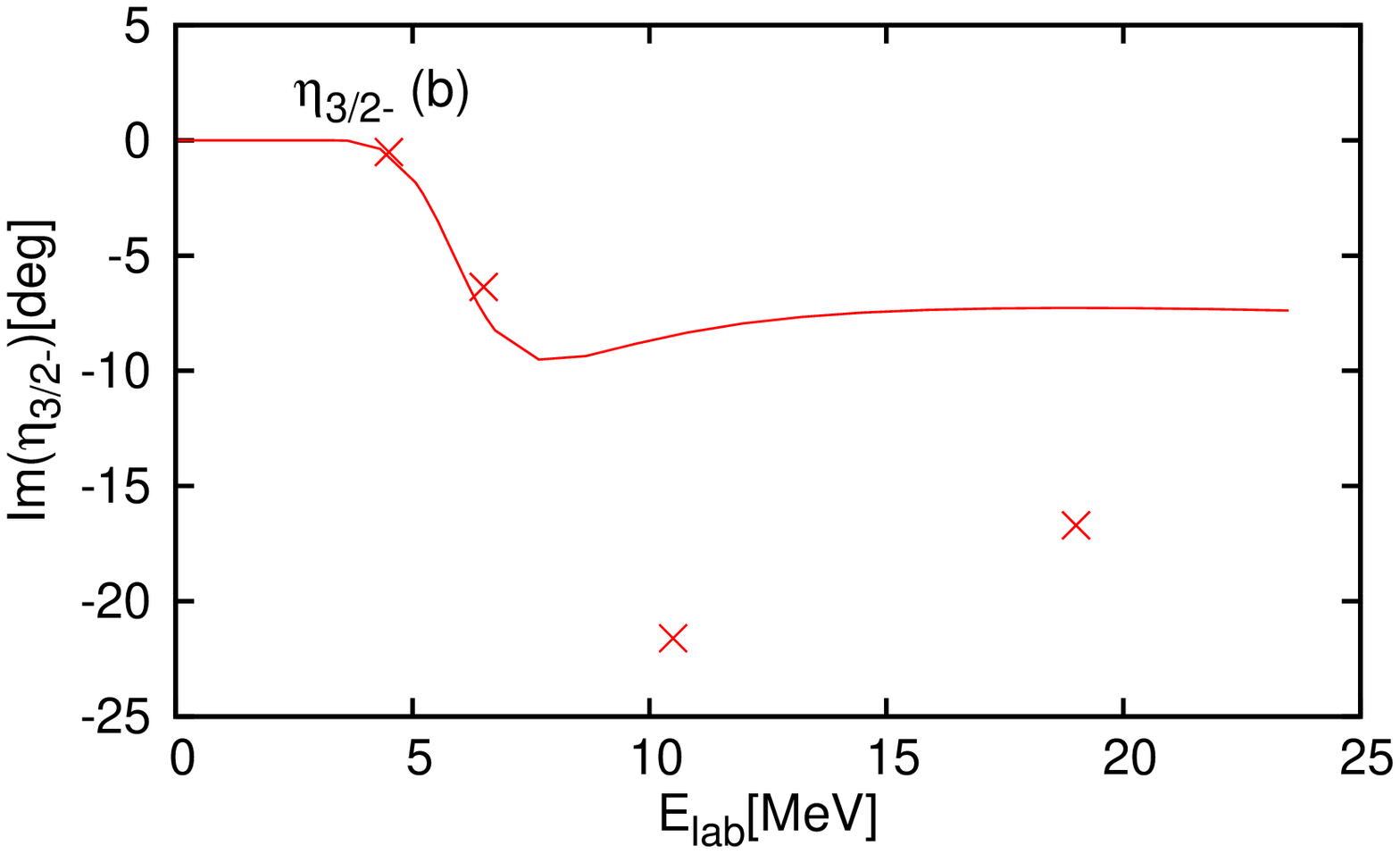}}
	\vspace{-.9cm}

	\subfloat{\includegraphics[angle=-90,width=88mm]{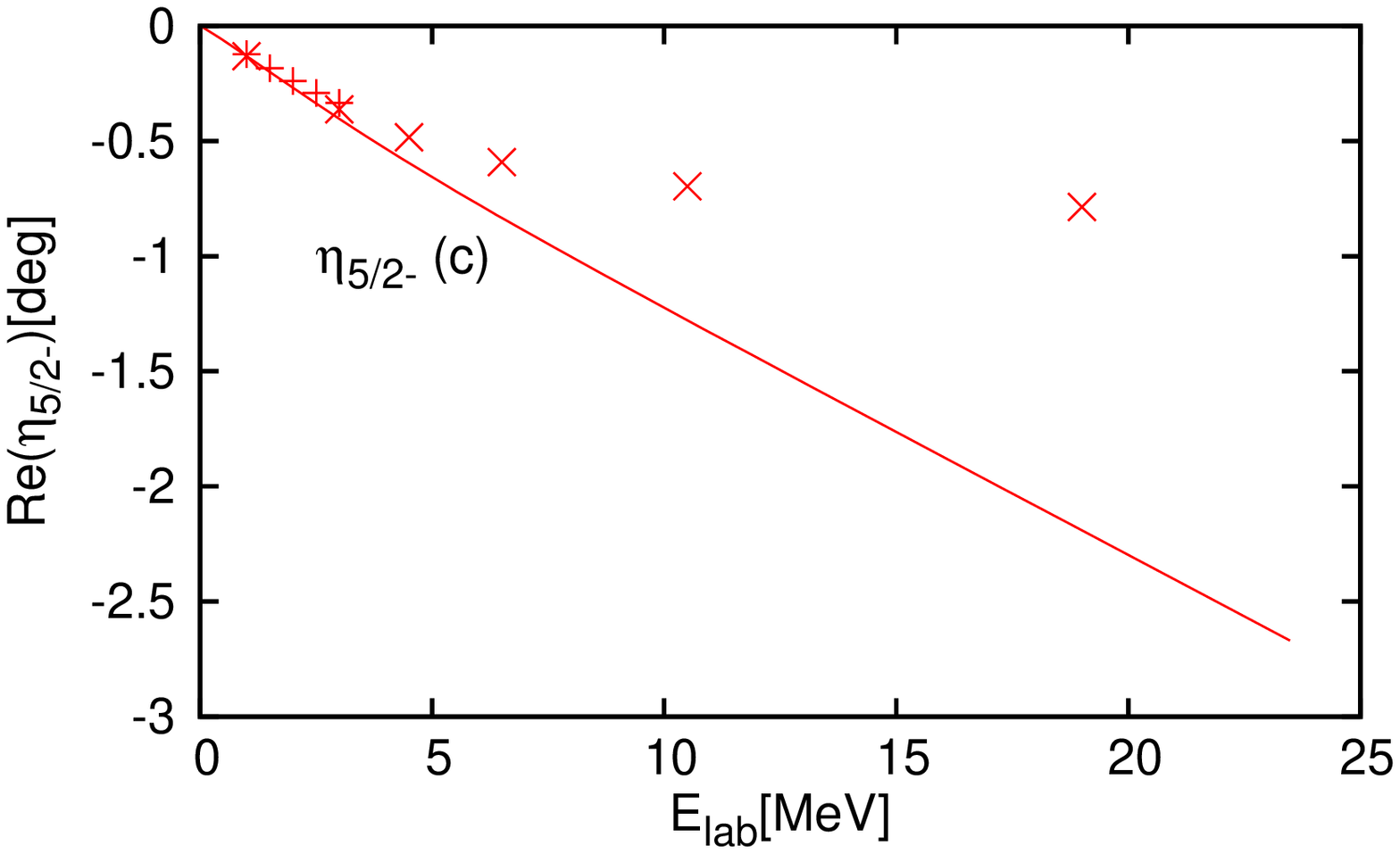}}
	\subfloat{\includegraphics[angle=-90,width=88mm]{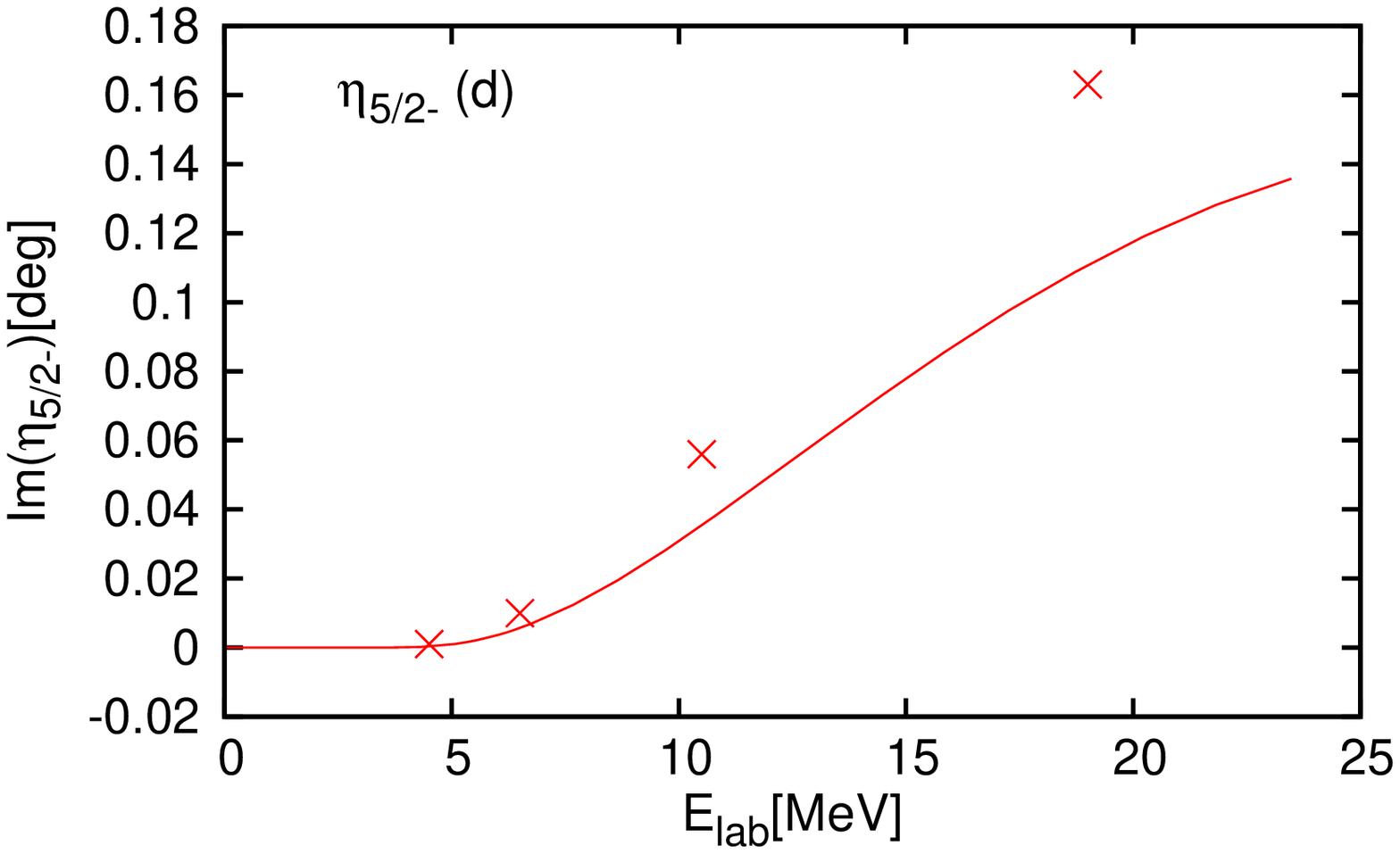}}
	
	\end{center}

	\caption{\label{fig:etan}(Color online)All $\eta^{J-}$ for $\Lambda=1600$ MeV left (right) real part (imaginary part).  Data below DBT (crosses) is AV18+UIX \cite{Kievsky:1996ca}, above and below DBT (stars) is Bonn-B \cite{Glockle:1995fb}  }

\end{figure}

We now consider the mixing angles, starting with the $\eta^{J+}$ shown in Fig. \ref{fig:etap}, which mixes states with the same partial waves but different total spin.  The cutoff variation (from $\Lambda=200$ MeV to $\Lambda=1600$ MeV) for the $\eta_{\frac{1}{2}+}$ mixing angle is shown explicitly since it is the only mixing angle that depends on the doublet S-wave.  With the exception of the $\eta_{\frac{1}{2}+}$ mixing angle good qualitative agreement with PMC can be seen for the real part of the $\eta^{J+}$ mixing angle and at low energies the agreement is within $3\%$.  However, for the imaginary part of $\eta^{J+}$ at higher energies agreement with PMC quickly breaks down.  In Fig. \ref{fig:etan} the mixing angles $\eta^{J-}$ are shown and again good qualitative agreement is seen with the PMC, and at low energies the agreement with PMC is within $3\%$.  However, at higher energies the discrepancies seem to be more pronounced than in the $\eta^{J+}$ mixing angles.  This is likely due to the fact that these mixing angles depend on the P-waves and as already shown there are issues with P-waves that will hopefully be resolved with the inclusion of two-body P-wave terms at $\mathrm{N}^{3}\mathrm{LO}$ in \EFT.

\begin{figure}[hbt]

	\begin{center}

	\subfloat{\includegraphics[angle=-90,width=88mm]{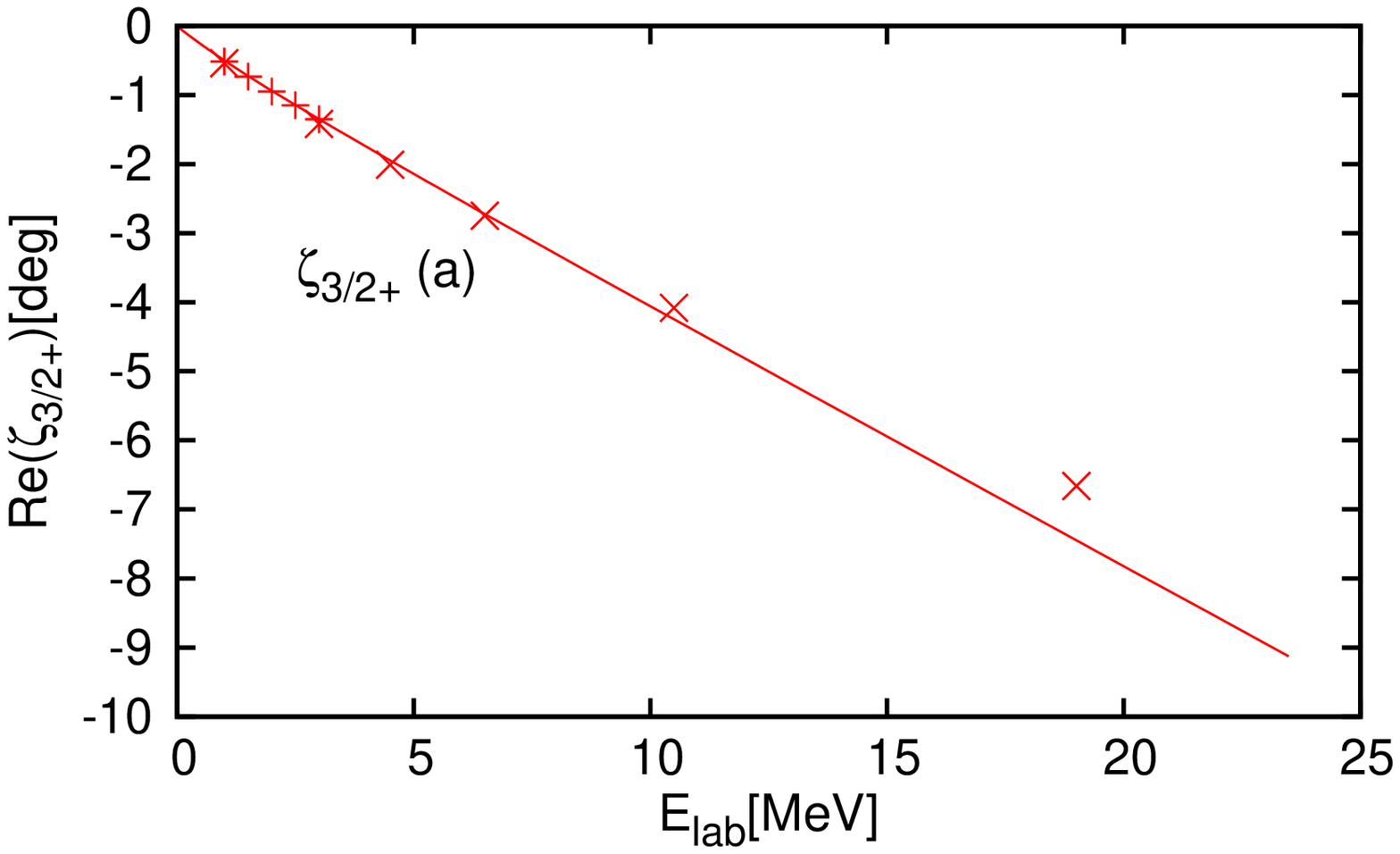}}
	\subfloat{\includegraphics[angle=-90,width=88mm]{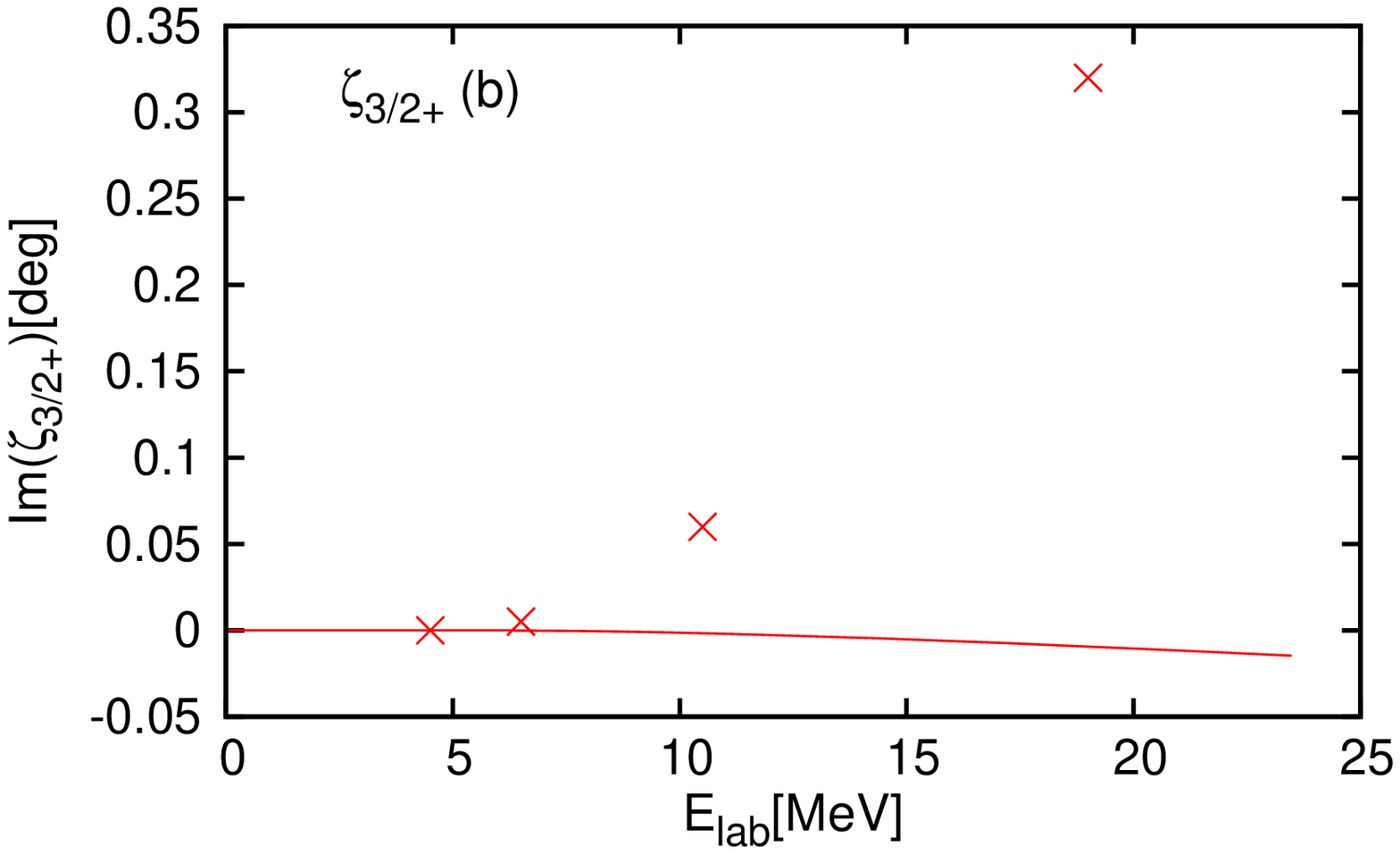}}
	\vspace{-.9cm}

	\subfloat{\includegraphics[angle=-90,width=88mm]{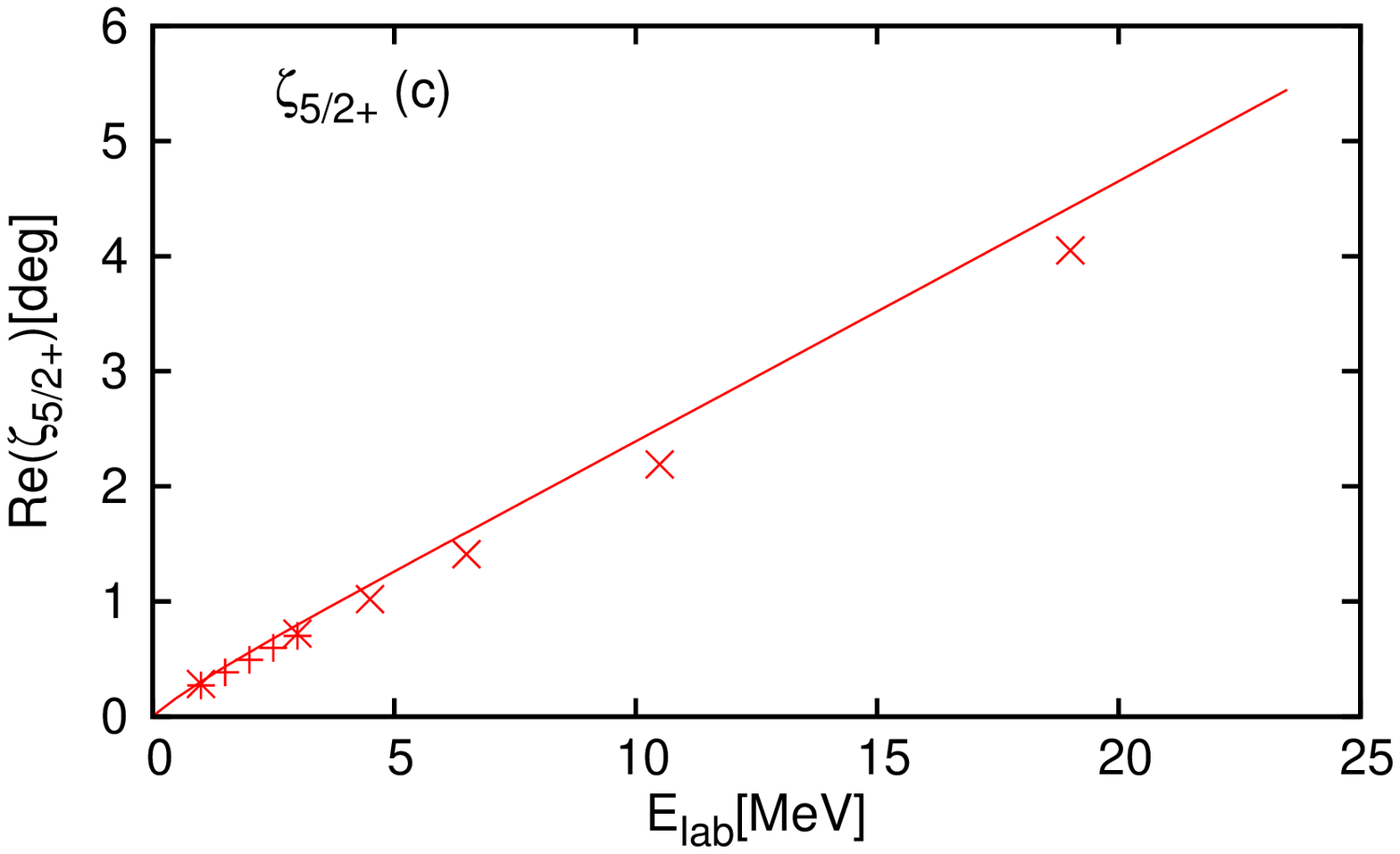}}
	\subfloat{\includegraphics[angle=-90,width=88mm]{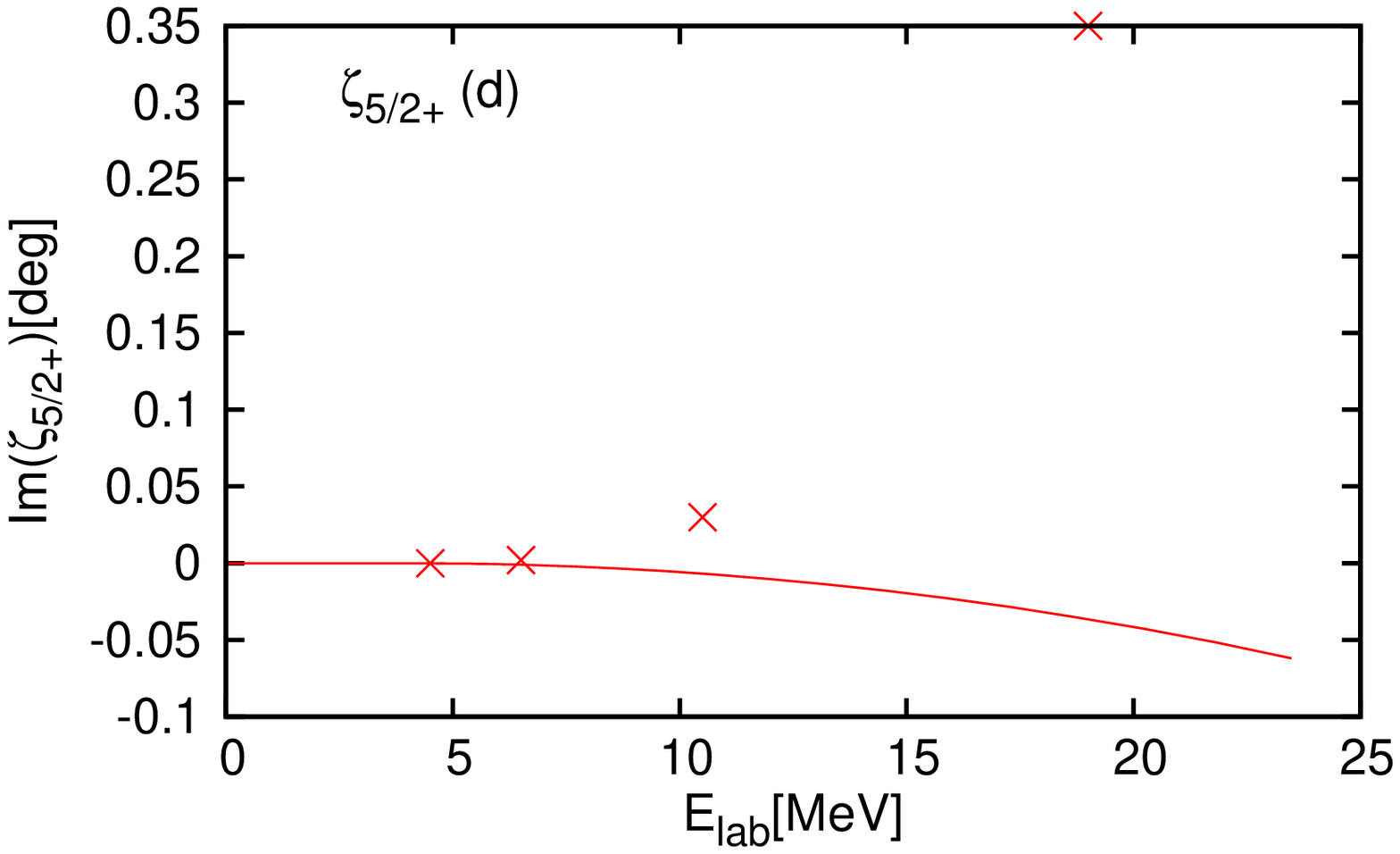}}
	\vspace{-.9cm}

	\subfloat{\includegraphics[angle=-90,width=88mm]{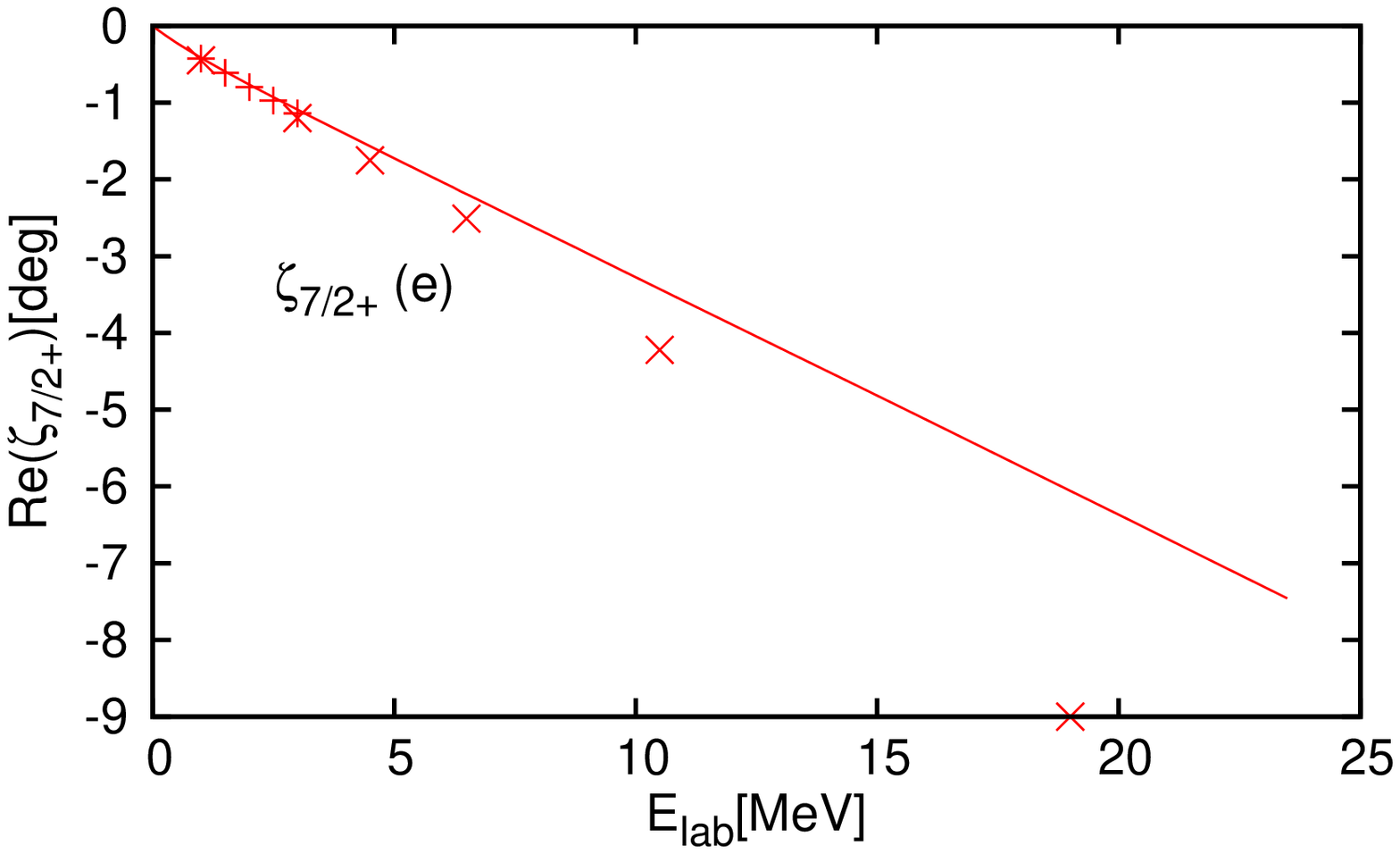}}
	\subfloat{\includegraphics[angle=-90,width=88mm]{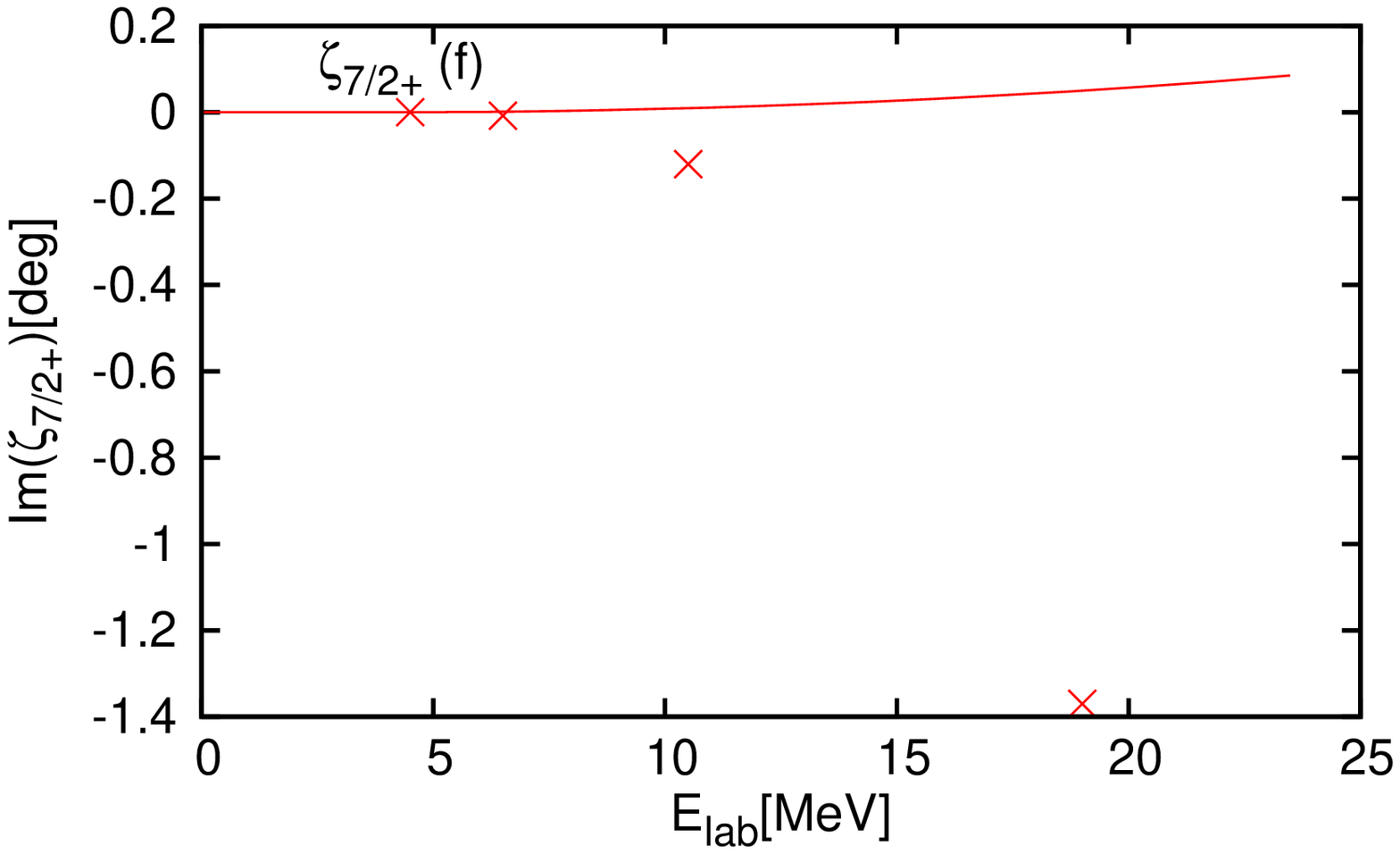}}
	
	\end{center}

	\caption{\label{fig:zetap}(Color online)All $\zeta^{J+}$ for $\Lambda=1600$ MeV left (right) real part (imaginary part).  Data below DBT (crosses) is AV18+UIX \cite{Kievsky:1996ca}, above and below DBT (stars) is Bonn-B \cite{Glockle:1995fb}}

\end{figure}

\begin{figure}[hbt]

	\begin{center}

	\subfloat{\includegraphics[angle=-90,width=88mm]{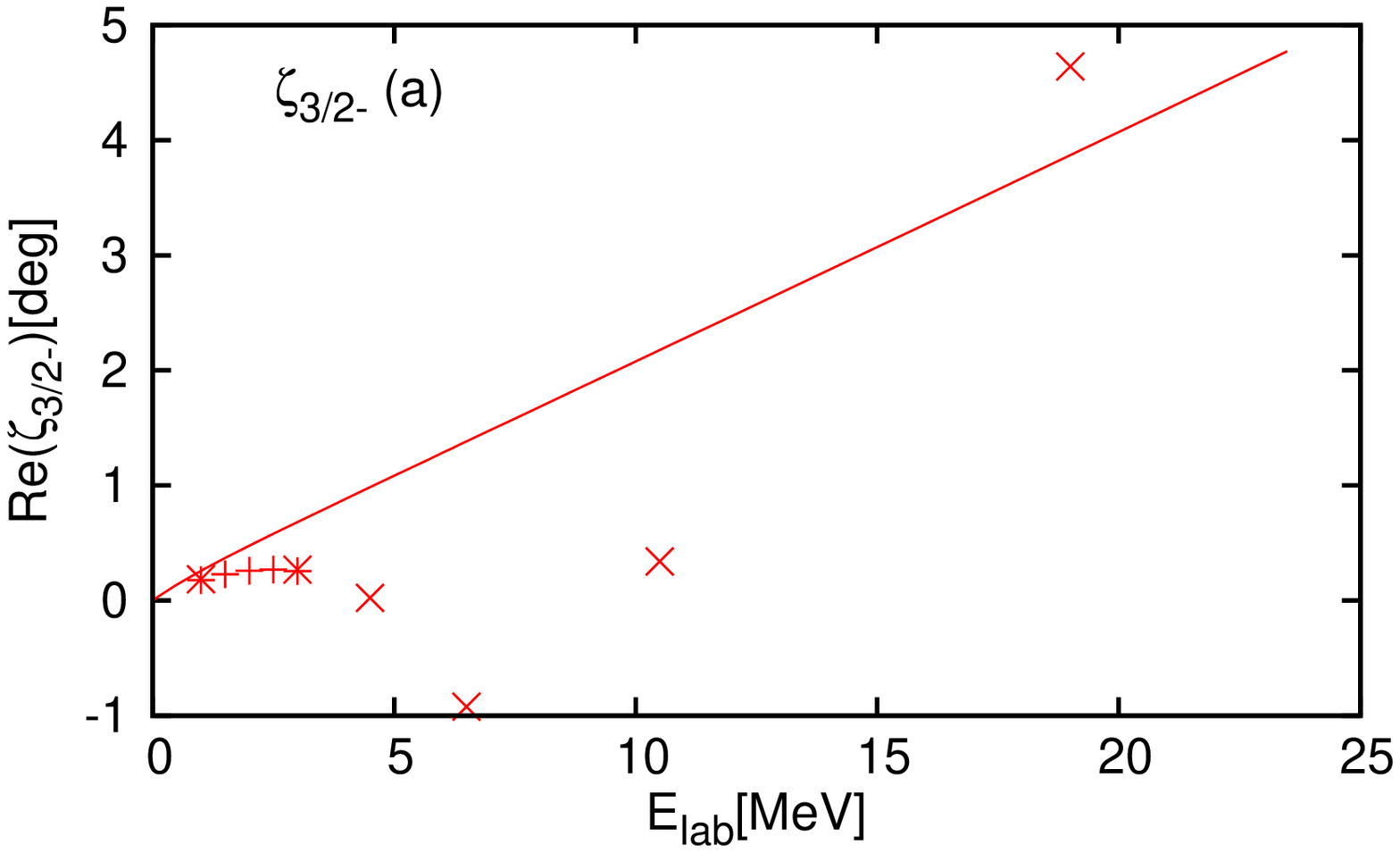}}
	\subfloat{\includegraphics[angle=-90,width=88mm]{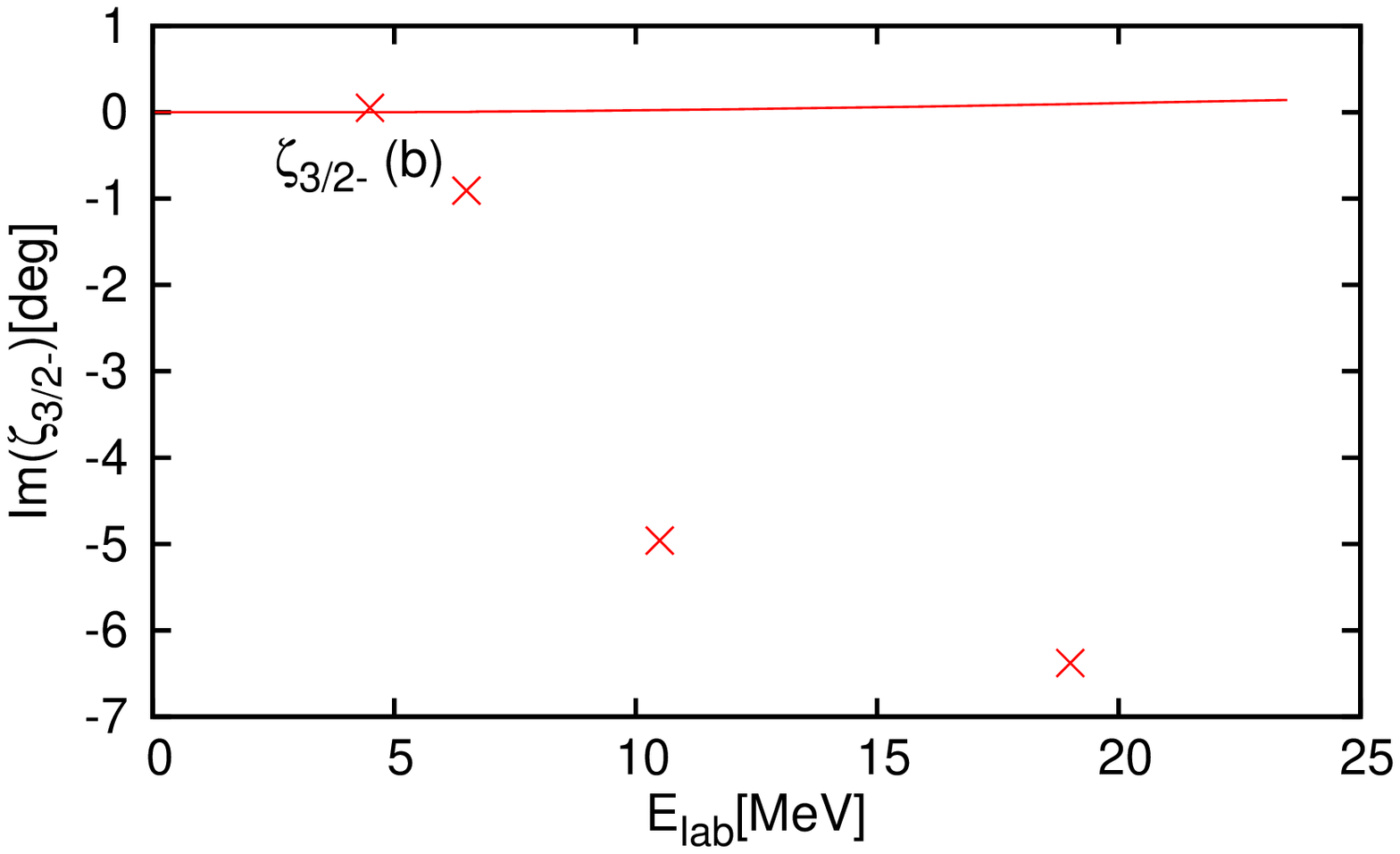}}
	\vspace{-.9cm}

	\subfloat{\includegraphics[angle=-90,width=88mm]{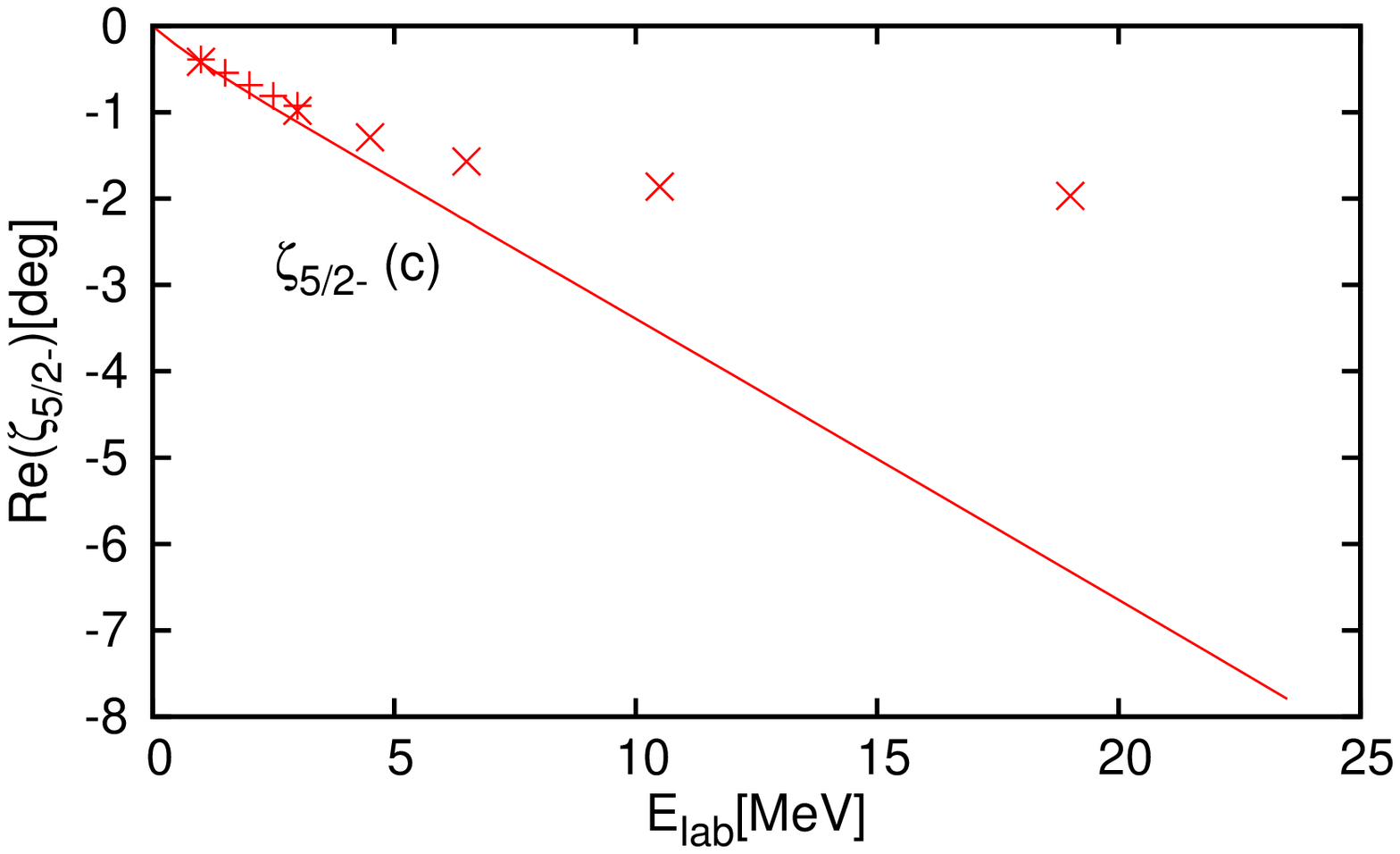}}
	\subfloat{\includegraphics[angle=-90,width=88mm]{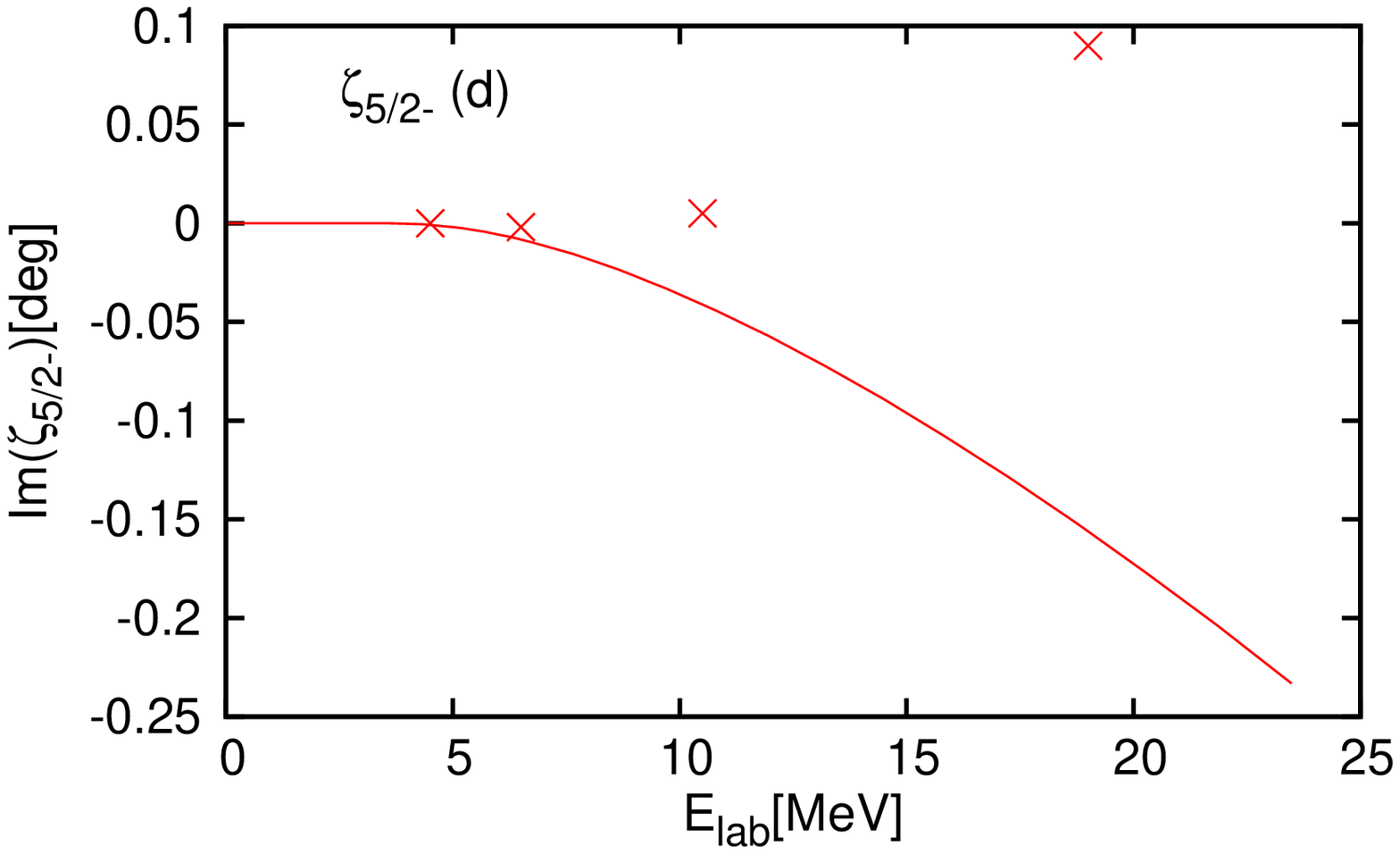}}
	
	\end{center}

	\caption{\label{fig:zetan}(Color online)All $\zeta^{J-}$ for $\Lambda=1600$ MeV left (right) real part (imaginary part).  Data below DBT (crosses) is AV18+UIX \cite{Kievsky:1996ca}, above and below DBT (stars) is Bonn-B \cite{Glockle:1995fb}}
	
\end{figure}

The mixing angles $\zeta^{J+}$ shown in Fig. \ref{fig:zetap} mix states of different angular momentum but the same total spin.  Good qualitative agreement for the real part of the mixing parameters $\zeta^{J+}$ is  seen with the PMC and at low energies the agreement is within $3\%$.  However, the imaginary part of $\zeta^{J+}$ is roughly an order of magnitude smaller than the Bonn-B data at higher energies.  In Fig \ref{fig:zetan} the mixing angles $\zeta^{J-}$ are shown.  For the real part of $\zeta^{\frac{5}{2}-}$ the agreement with PMC at low energies is apparent and at higher energies this breaks down, in addition the imaginary part of $\zeta^{\frac{5}{2}-}$ has an opposite sign compared to the PMC.  For the real part of $\zeta^{\frac{3}{2}-}$ the qualitative dip seen in the potential model is not followed.  However at low energies there is still rough agreement.  The imaginary part of $\zeta^{\frac{3}{2}-}$ like $\zeta^{\frac{3}{2}+}$ seems to be an order of magnitude smaller than the PMC.  Analogously to the $\eta^{J+}$ and $\eta^{J-}$ mixing angles it appears that the real part of $\zeta^{J+}$ matches well to PMC while $\zeta^{J-}$ is qualitatively worse.  This again is likely due the fact that $\zeta^{J-}$ depends on P-waves.

\begin{figure}[hbt]

	\begin{center}

	\subfloat{\includegraphics[angle=-90,width=88mm]{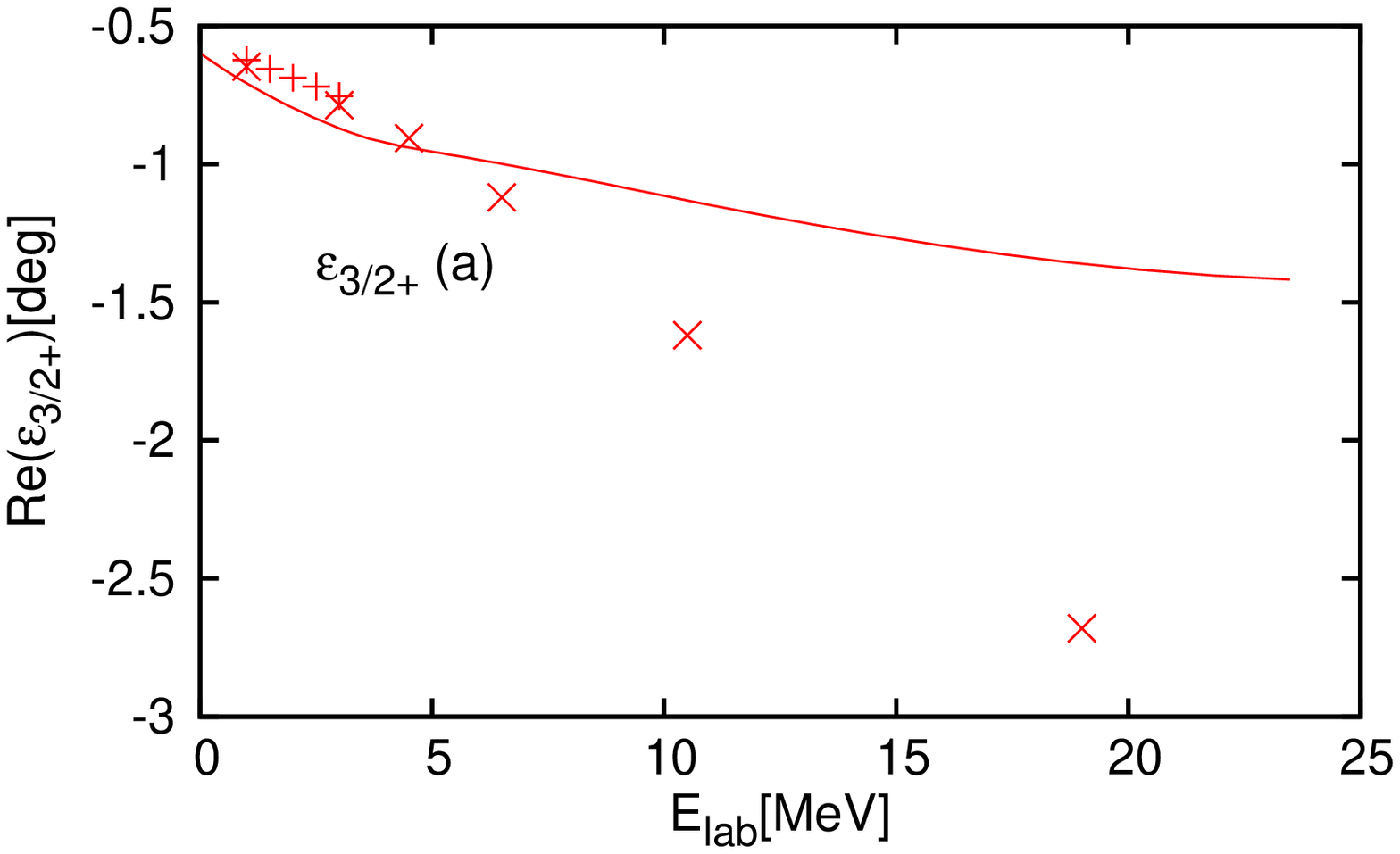}}
	\subfloat{\includegraphics[angle=-90,width=88mm]{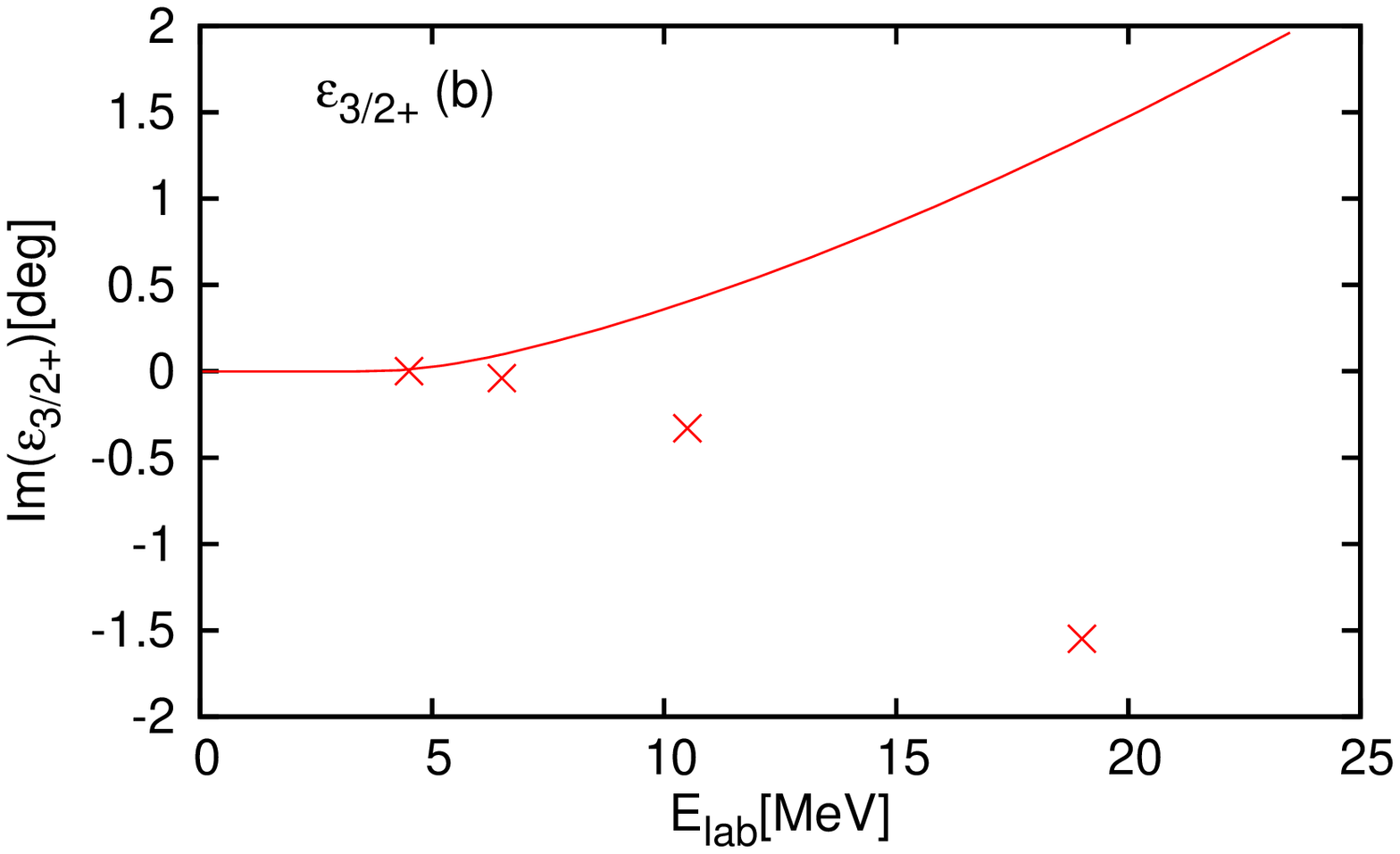}}
	\vspace{-.9cm}

	\subfloat{\includegraphics[angle=-90,width=88mm]{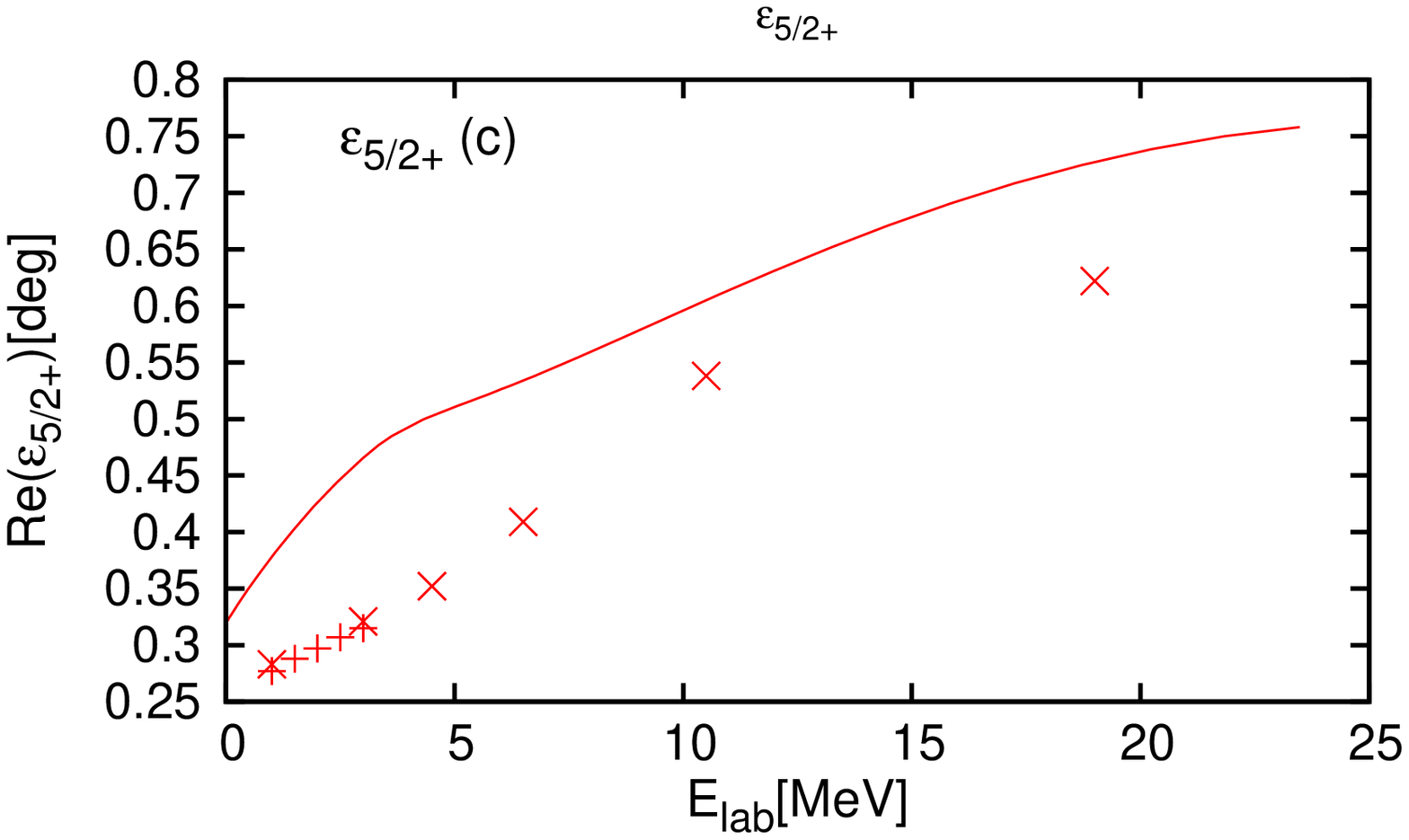}}
	\subfloat{\includegraphics[angle=-90,width=88mm]{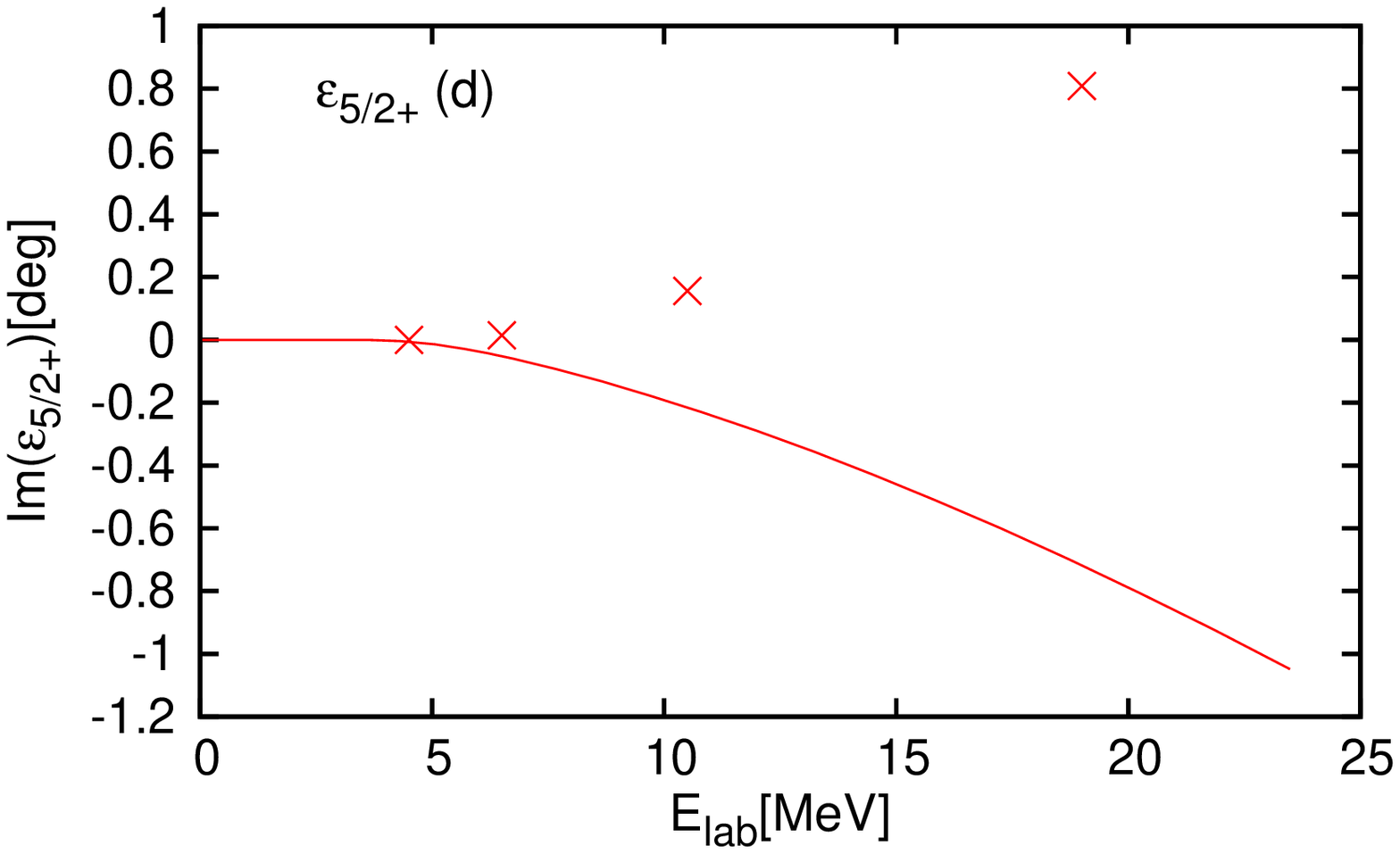}}
	\vspace{-.9cm}

	\subfloat{\includegraphics[angle=-90,width=88mm]{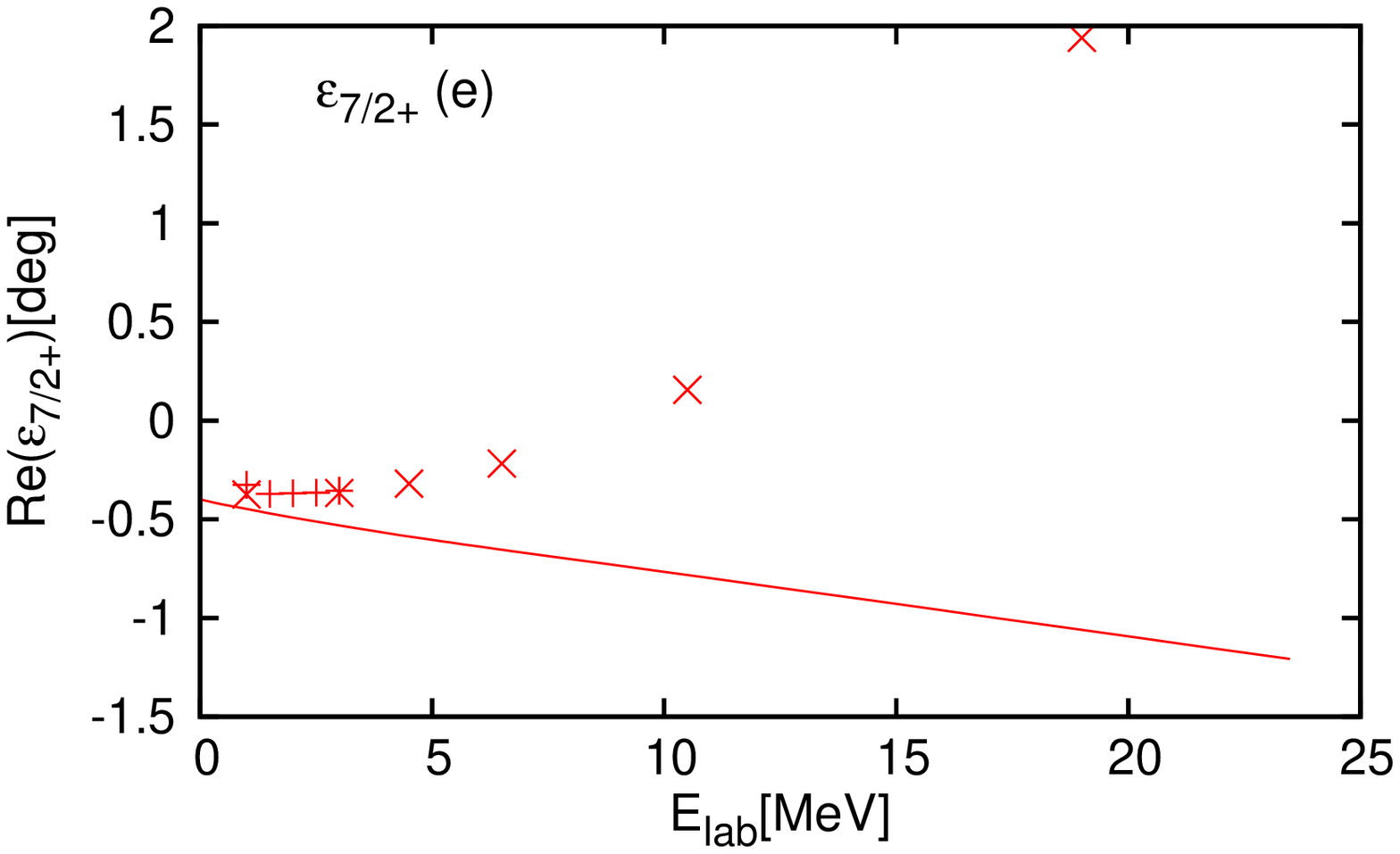}}
	\subfloat{\includegraphics[angle=-90,width=88mm]{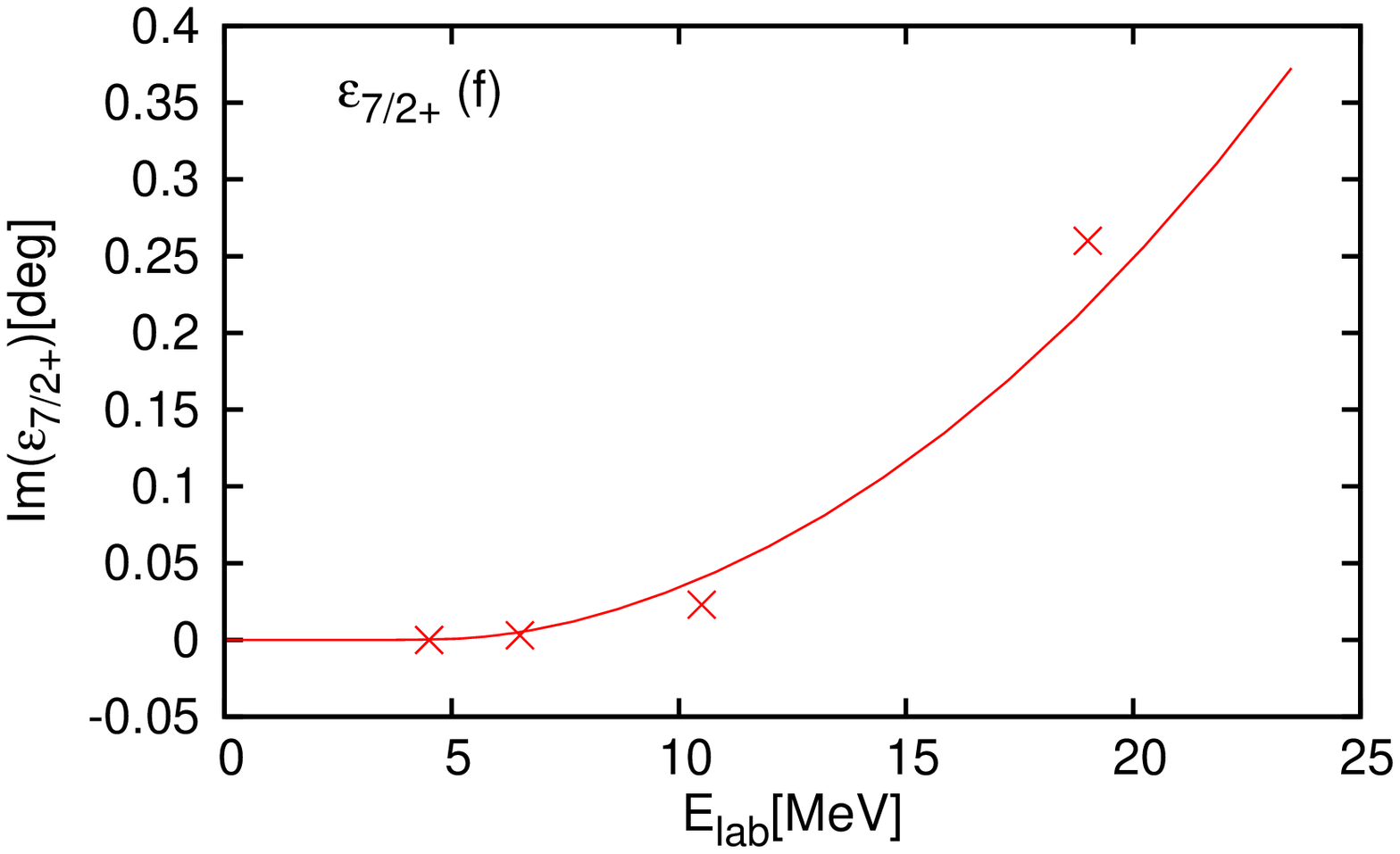}}
	\vspace{-.9cm}
	
	\subfloat{\includegraphics[angle=-90,width=88mm]{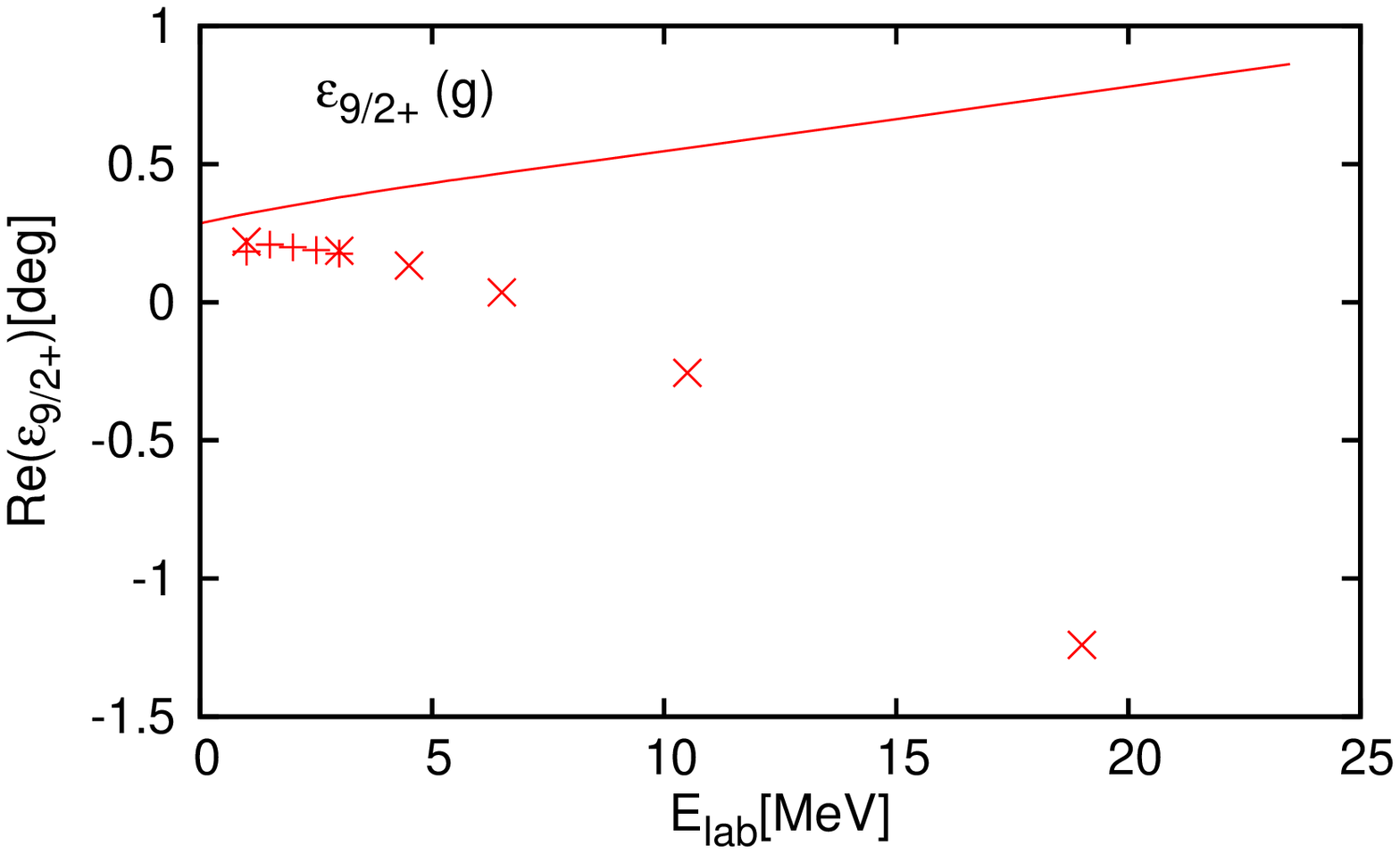}}
	\subfloat{\includegraphics[angle=-90,width=88mm]{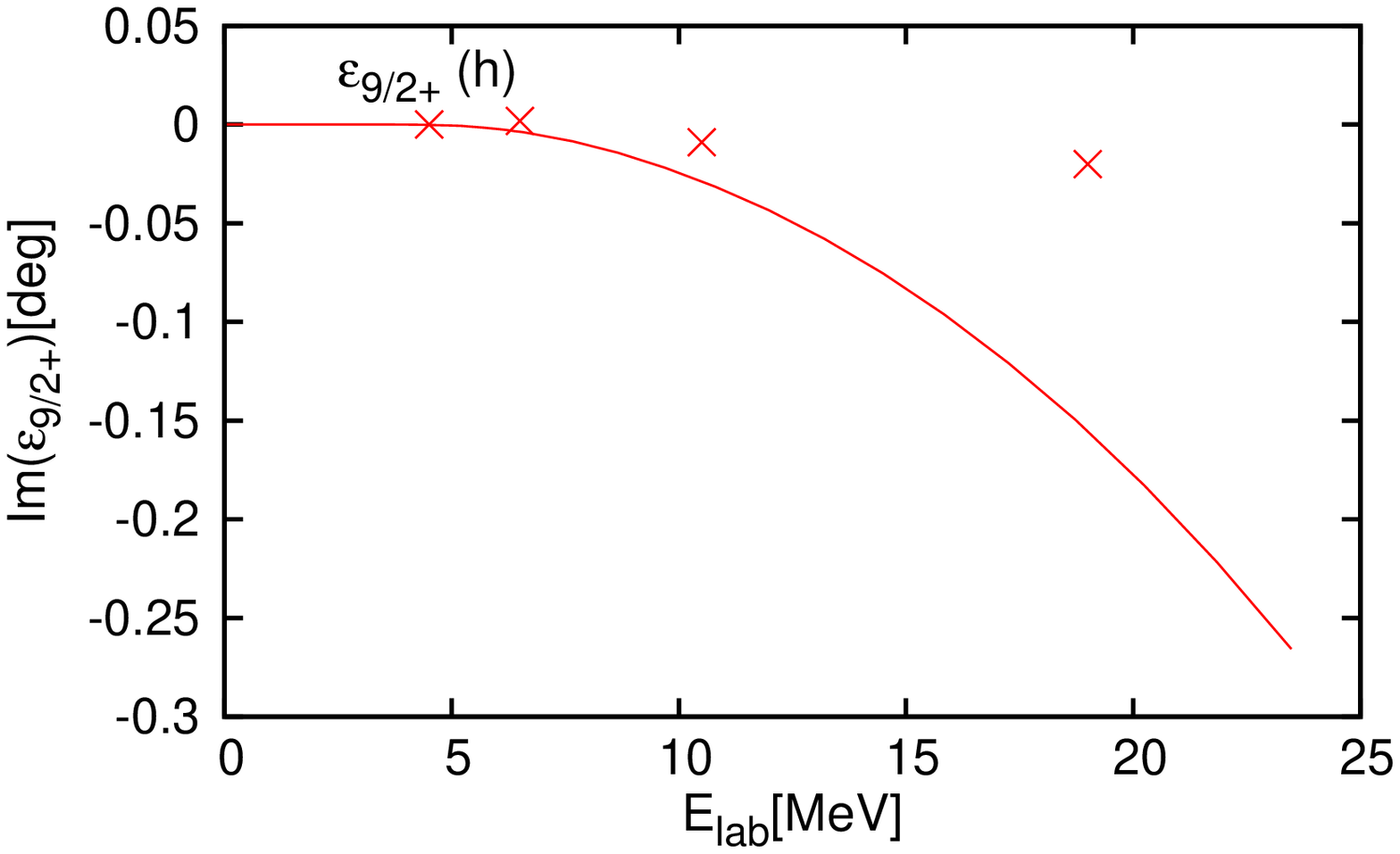}}
	
	\end{center}

	\caption{\label{fig:epsp}(Color online)All $\epsilon^{J+}$ for $\Lambda=1600$ MeV left (right) real part (imaginary part).  Data below DBT (crosses) is AV18+UIX \cite{Kievsky:1996ca}, above and below DBT (stars) is Bonn-B \cite{Glockle:1995fb}}

\end{figure}

\begin{figure}[hbt]

	\begin{center}

	\subfloat{\includegraphics[angle=-90,width=88mm]{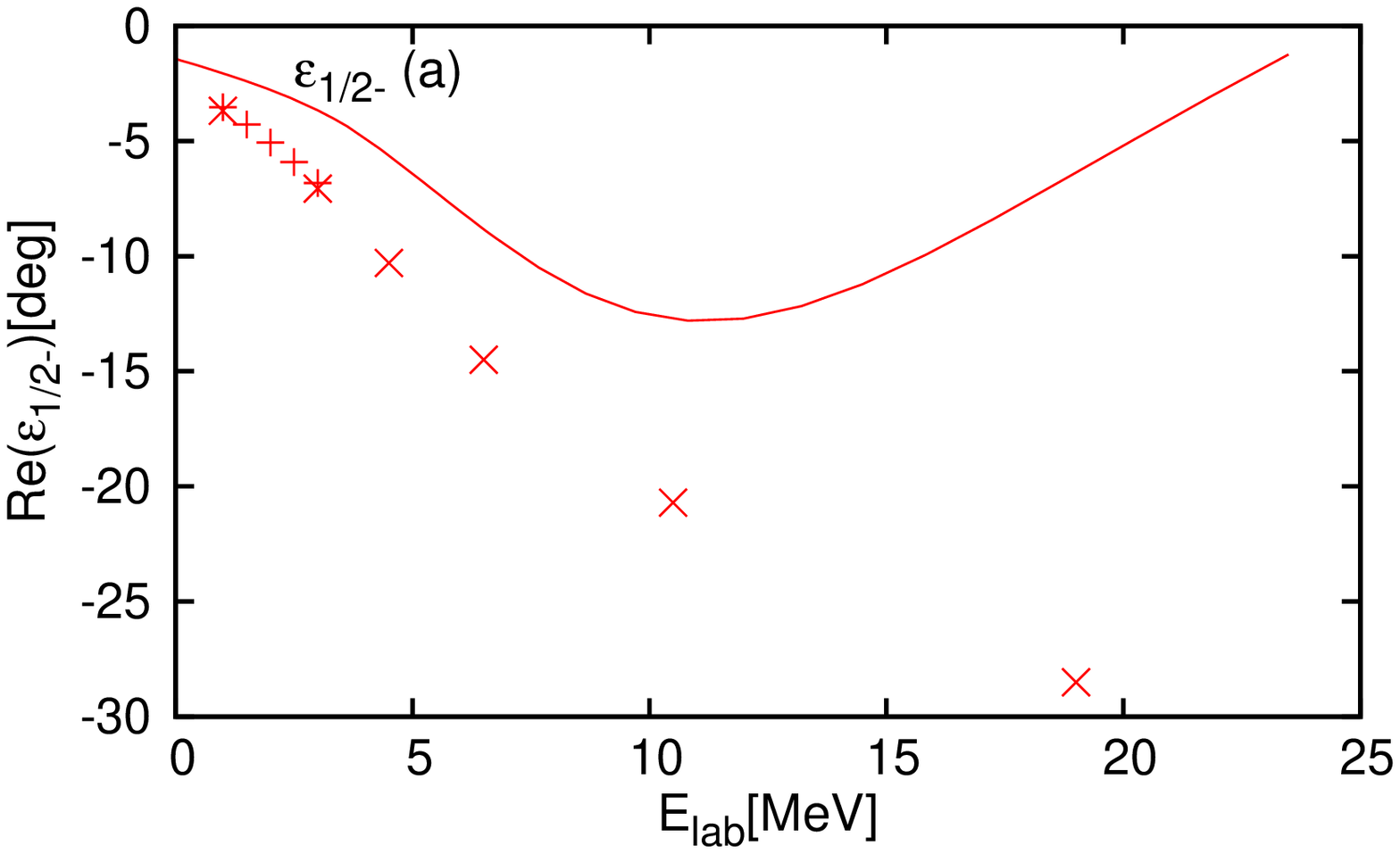}}
	\subfloat{\includegraphics[angle=-90,width=88mm]{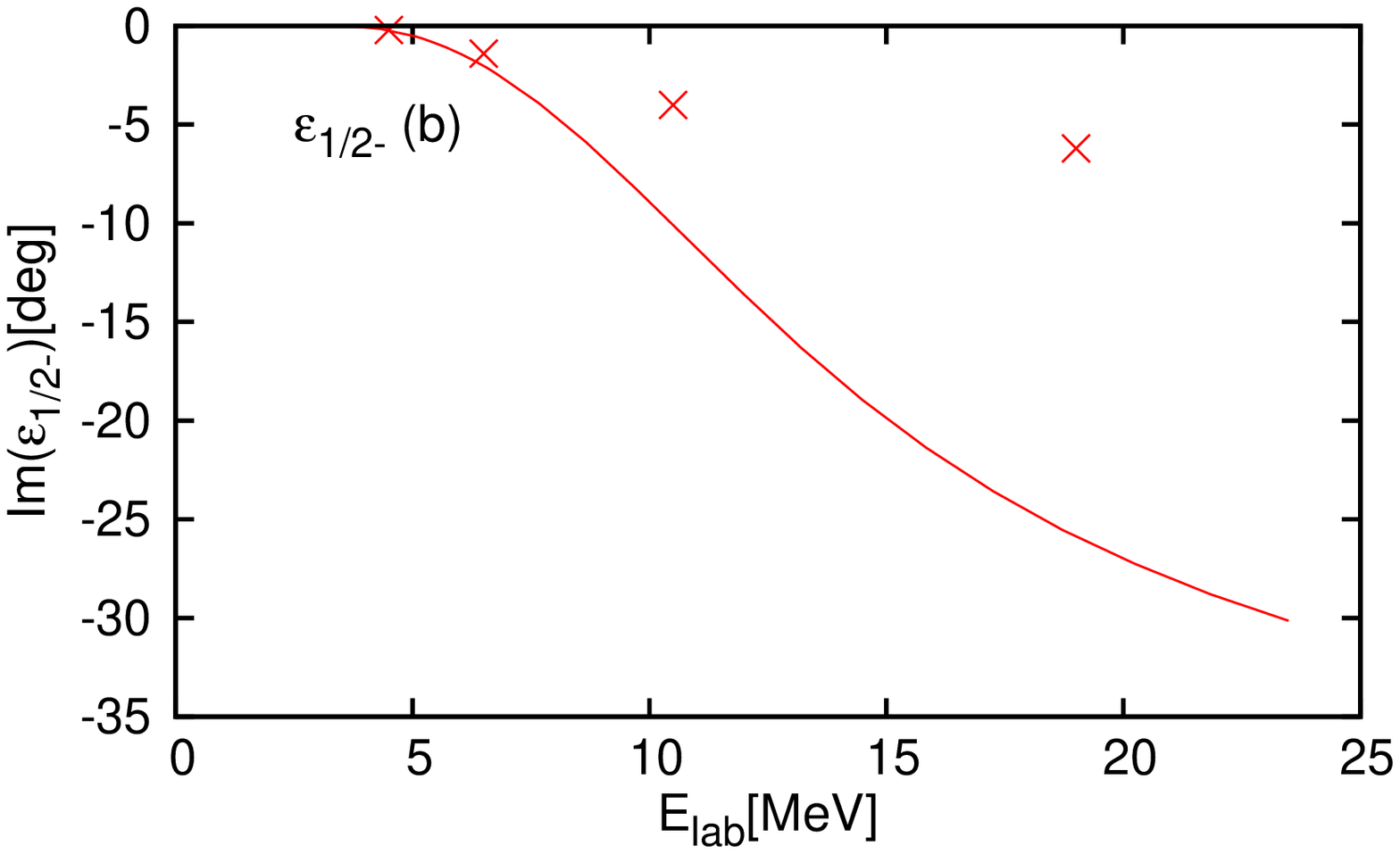}}
	\vspace{-.9cm}

	\subfloat{\includegraphics[angle=-90,width=88mm]{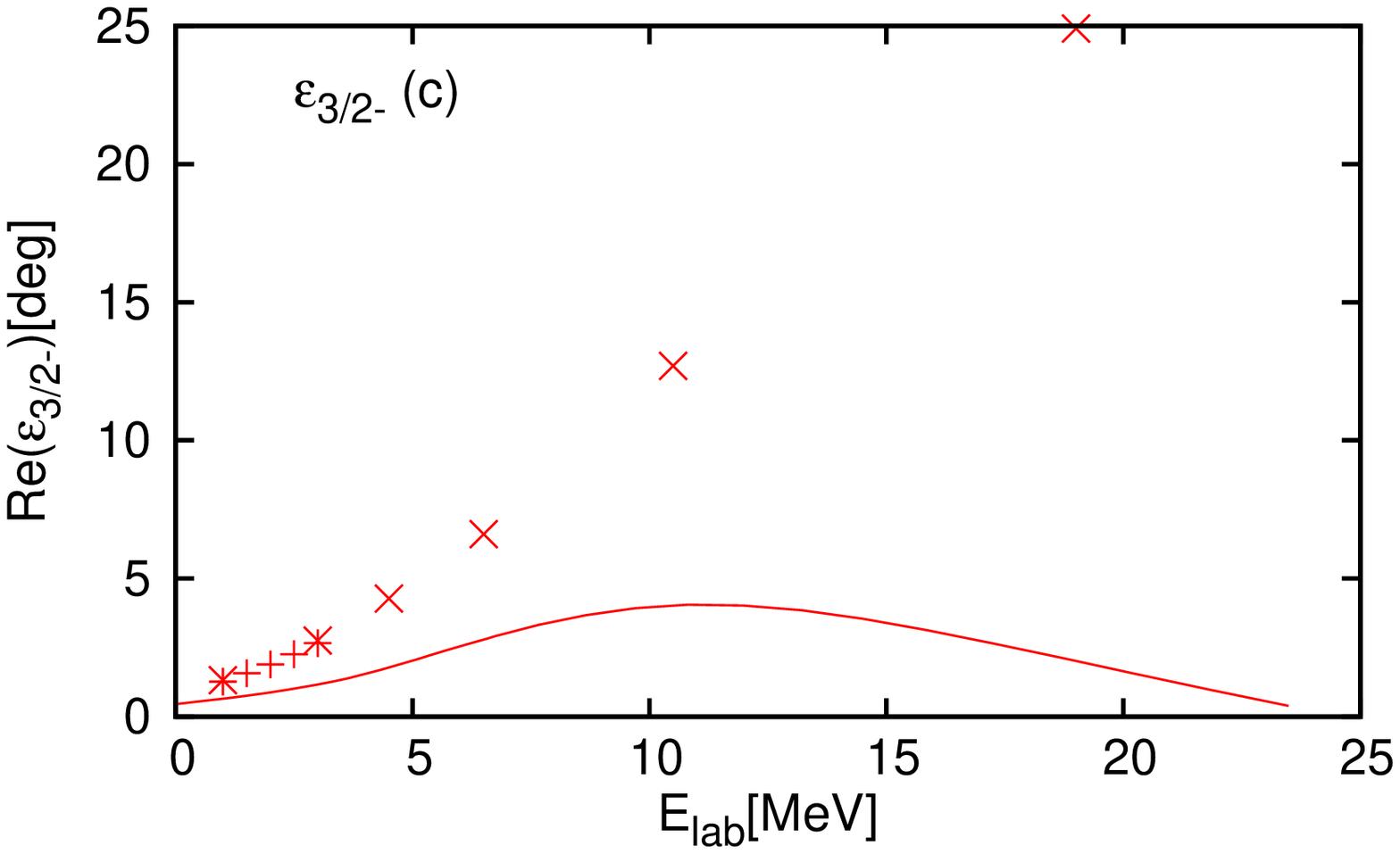}}
	\subfloat{\includegraphics[angle=-90,width=88mm]{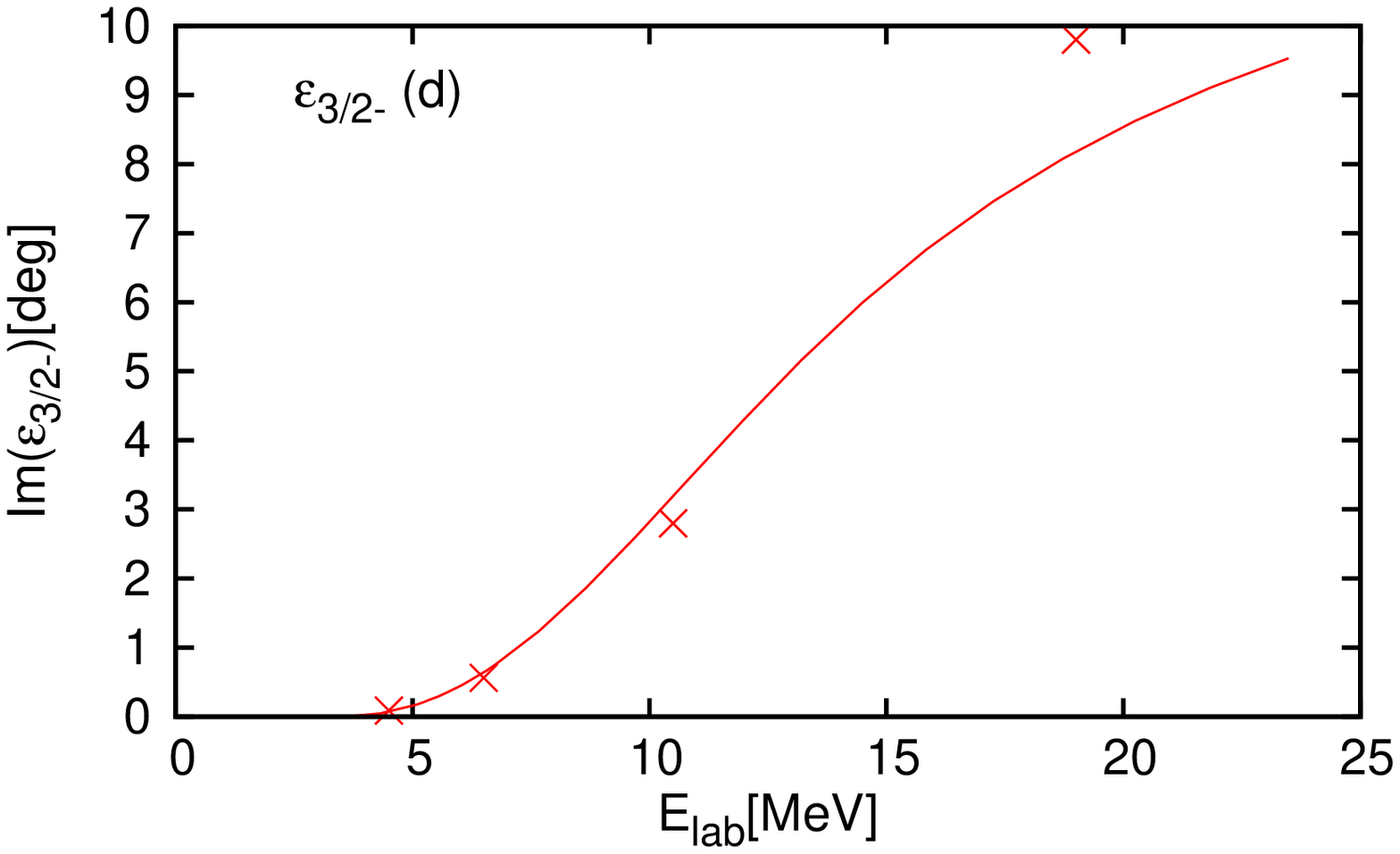}}
	\vspace{-.9cm}

	\subfloat{\includegraphics[angle=-90,width=88mm]{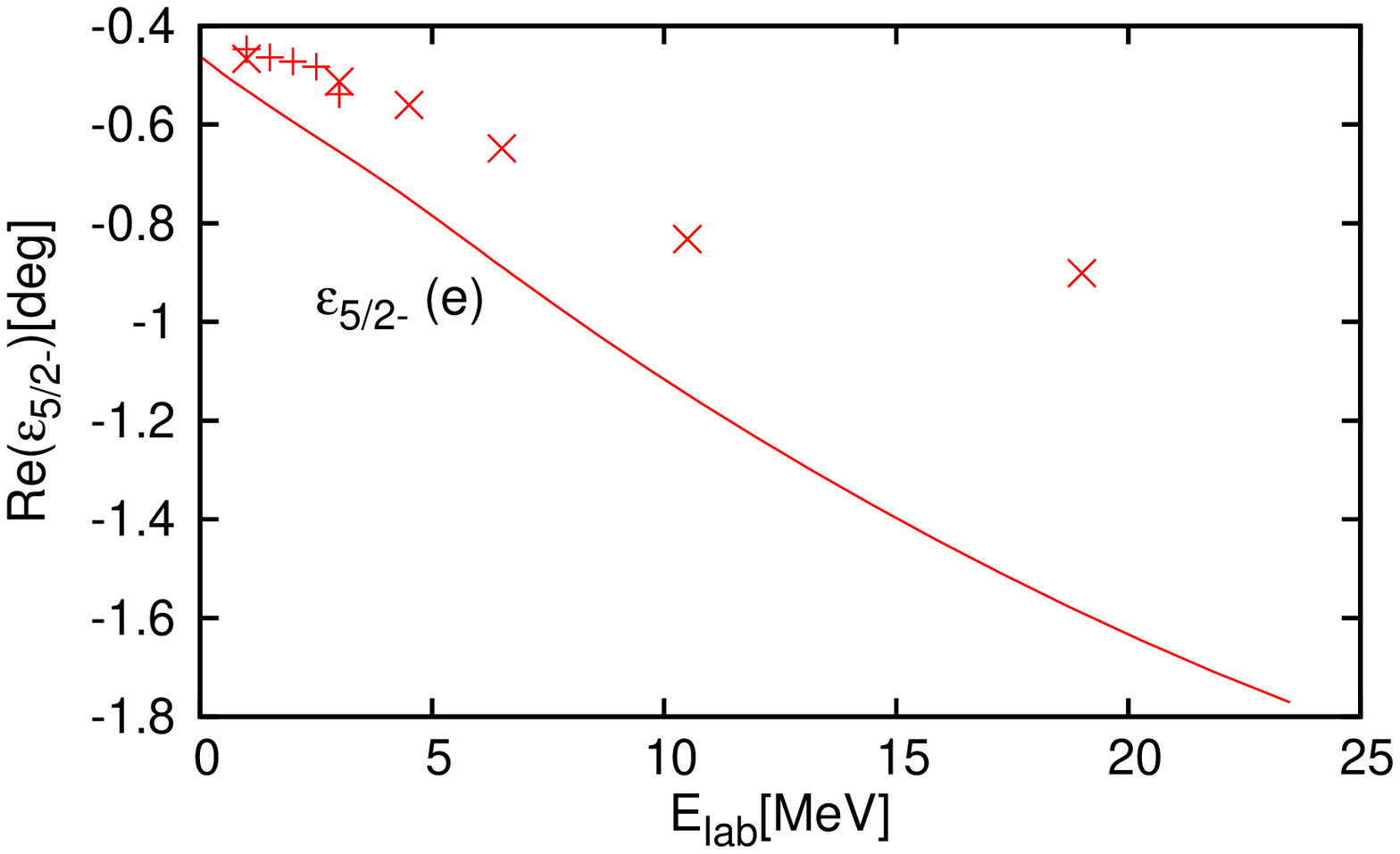}}
	\subfloat{\includegraphics[angle=-90,width=88mm]{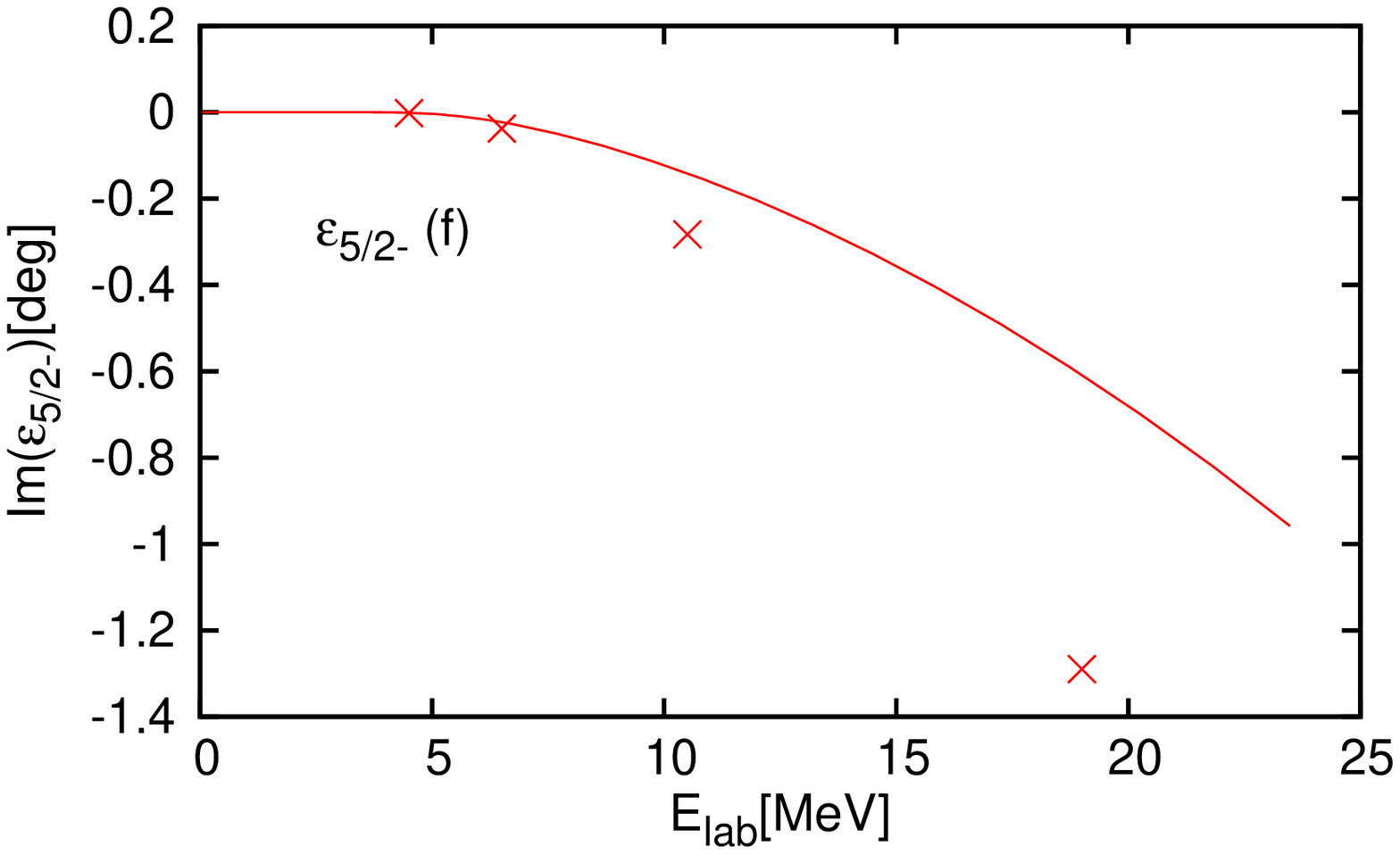}}
	\vspace{-.9cm}

	\subfloat{\includegraphics[angle=-90,width=88mm]{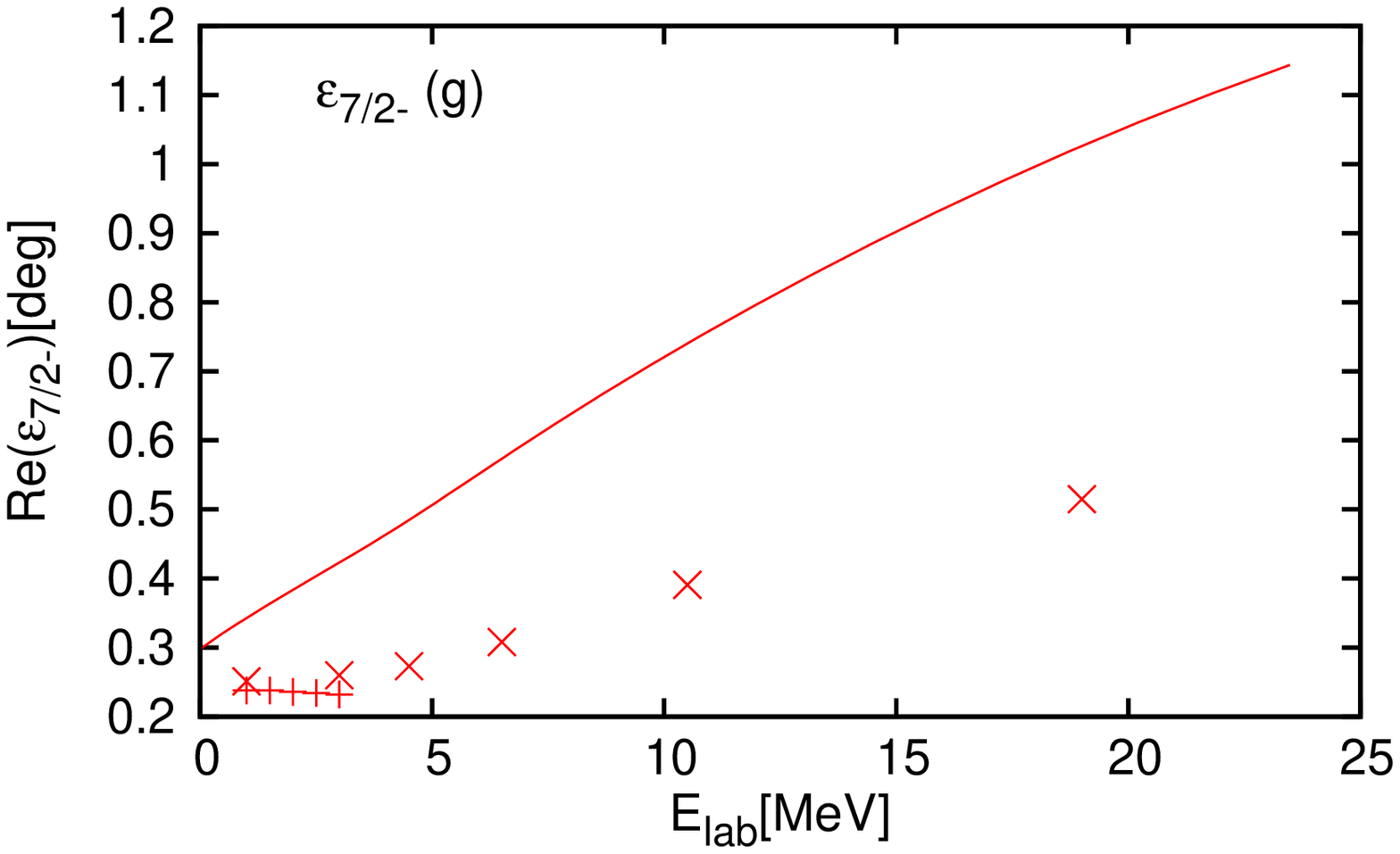}}
	\subfloat{\includegraphics[angle=-90,width=88mm]{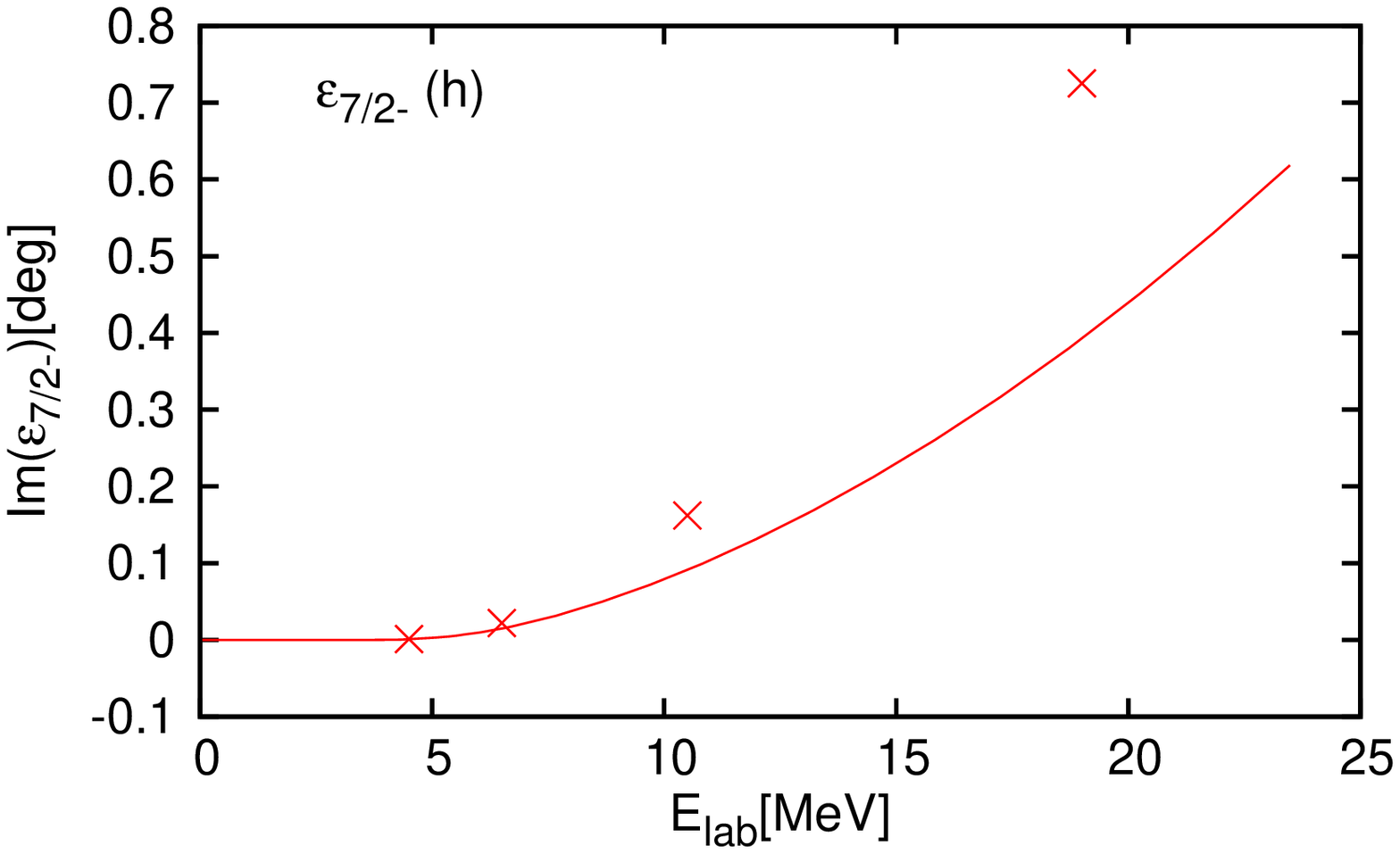}}
	
	\end{center}

	\caption{\label{fig:epsn}(Color online)All $\epsilon^{J-}$ for $\Lambda=1600$ MeV left (right) real part (imaginary part).  Data below DBT (crosses) is AV18+UIX \cite{Kievsky:1996ca}, above and below DBT (stars) is Bonn-B \cite{Glockle:1995fb}}

\end{figure}

Finally we examine the $\epsilon^{J+}$ mixing angles in Fig. \ref{fig:epsp} as well the mixing angles $\epsilon^{J-}$ in Fig. \ref{fig:epsn} which mix different partial waves but the same total spin.  For both sets of mixing angles agreement with the PMC is poor and the qualitative behavior of the PMC is generally not followed.  Even at the lowest energies there is poor agreement with the PMC and the PMC are either over or under predicted.  There is much need for improvement in these mixing parameters and thus a higher order calculation of these mixing parameters is necessary.  An accurate determination of all mixing angles and eigenphase shifts will allow the prediction of polarization observables in nd scattering and in particular the $A_{y}$ asymmetry.

\section{Conclusion}
The technique outlined here has general applicability in including perturbative corrections to integral equations.  It has all of the numerical savings of the partial-resummation technique but with the added benefit that it is strictly perturbative.  The partial waves obtained using this technique are very similar to those produced in the partial-resummation technique and similar issues are found in the doublet P-wave.  As was shown, the quartet scattering length in the partial-resummation technique is closer to the experimental value at NNLO than ours.  However, this is not surprising since the partial-resummation technique does contain certain higher order terms.  More importantly the imaginary part of the quartet S-wave phase shift has the correct sign at NLO and NNLO in this technique, unlike  the partial-resummation technique.  In order to obtain the correct sign in the partial-resummation technique one must resum the effective range to all orders in the deuteron propagator and this introduces spurious poles which complicates the numerics.  Note in this technique the SD-mixing terms could have been included in the integral equation for the NNLO amplitude and will be in future when higher orders are calculated.  Also this technique allows one to calculate diagrams that contain external currents with full off-shell scattering amplitudes, thus in principle enabling one to calculate the three-body process ${}^{3}He+\gamma\to p+d$ , ${}^{3}H+\gamma\to n+d$, or even Compton scattering off of ${}^{3}H$ or ${}^{3}He$.  In short this technique makes the perturbative calculation of scattering in three-body systems numerically simpler.  However, the generalization of this technique to perturbative three-body bound state calculations does not work and different techniques will need to be developed for such calculations.

The calculation of nd scattering up to NNLO including the SD-mixing terms yields good agreement with PMC for the eigenphase shifts.  However, as noted in the quartet P-wave the splittings are larger than expected and in the doublet P-wave where no splittings occur at this order in \EFT the potential models diverge from our results above the DBT.  These discrepancies are likely due to the absence of the rather strong two-body P-wave terms that occur at $\mathrm{N}^{3}\mathrm{LO}$ in \EFT.  Thus a higher order calculation to resolve this issue is in order.  For the real part of the mixing parameters $\eta^{J\pi}$ and $\zeta^{J\pi}$, good agreement with the PMC at low energy has also been shown.  As for the imaginary part of $\eta^{J\pi}$ and $\zeta^{J\pi}$ there are noticeable discrepancies with the PMC, in particular the imaginary part of $\zeta^{J\pi}$ seems to be an order of magnitude smaller than the potential model prediction.  Also for negative parity $\eta^{J-}$ and $\zeta^{J-}$ there is more noticeable discrepancy at higher energies with PMC and this is again likely related to the fact that these mixing angles depend on P-waves, thus further warranting the inclusion of higher order two-body P-wave corrections in \EFT.  Finally the mixing parameters $\epsilon^{J\pi}$ seem to match very poorly to the potential model calculation and could potentially benefit the most from a higher order calculation.

In order to calculate polarization observables it is important to accurately determine the mixing parameters since in the absence of these there are no polarization observables.  The technique outlined here will ease the calculation of higher order contributions necessary for an accurate determination of the mixing parameters.  Of particular interest is to carry out a calculation for the mixing parameters in $pd$ scattering as there are far more experimental measurements to compare with.  However, this will be complicated by the inclusion of Coulomb interactions.  The technique of Hammer and K\"onig \cite{Konig:2011yq} can be combined straightforwardly with the technique in this paper to calculate $pd$ scattering strictly perturbatively in both Coulomb interactions and \EFT.  However, at low energies it is necessary to include the Coulomb interactions nonperturbatively and such a calculation is complicated by the numerical singularities introduced by the full off-shell Coulomb propagator, and new numerical techniques will need to be developed in order to deal with it.


\vspace{-.2cm}
\begin{acknowledgments}
I would like to thank Chen Ji and Harald W. Griesshammer as well as Thomas Mehen and Roxanne Springer for useful discussions during the course of this work.  This work is supported in part by the Department of Energy under Grant No. DE-FG02-05ER41368
\end{acknowledgments}


\end{thebibliography}

\end{document}